\documentclass[aps,pra,twocolumn,superscriptaddress,longbibliography]{revtex4-2}
\usepackage{xcolor}
\usepackage[colorlinks=true,
            citecolor=blue,
            linkcolor=blue,
            urlcolor=blue]{hyperref}

\usepackage{amsmath,amssymb,bm,braket,graphicx}
\graphicspath{{manuscript_figures/}}

\usepackage{orcidlink}
\usepackage{booktabs}
\usepackage{multirow}

\begin{document}

\title{From quantum reservoirs to quantum extreme learning machines through a nearest-neighbor spin chain with tunable quantum memory}

\author{Carlos Ramon-Escandell\orcidlink{0009-0001-8673-3568}}
\email{carlos.ramon@qilimanjaro.tech}
\affiliation{Qilimanjaro Quantum Tech, 08019 Barcelona, Spain}

\author{Arnau Riera\orcidlink{0000-0002-3271-7802}}
\affiliation{Qilimanjaro Quantum Tech, 08019 Barcelona, Spain}

\author{Marcin Płodzień\orcidlink{0000-0002-0835-1644}}
\email{marcin.plodzien@qilimanjaro.tech}
\affiliation{Qilimanjaro Quantum Tech, 08019 Barcelona, Spain}

\date{\today}

\begin{abstract}
    Quantum Reservoir Computing (QRC) processes temporal data by retaining a memory of past inputs in the recurrent state of a quantum system, whereas a Quantum Extreme-Learning Machine (QELM) discards that memory, resetting the system at every step so that only the most recent input shapes the response. The two are usually treated as separate computational paradigms. We show that they are the two limits of a single architecture, connected by the input-encoding length, that is, the number of qubits overwritten with fresh data at each step. When a single qubit is re-encoded the system operates as a standard QRC, when the whole register is re-encoded it operates as a QELM, and intermediate lengths interpolate between them. The overwritten qubits hold the recent past in an explicit register, while the remaining qubits are never reset and carry older inputs forward in their evolving quantum state, so the encoding length redistributes memory between explicit and recurrent storage at fixed system size. Tuning the reservoir Hamiltonian and the evolution time with Bayesian optimization at each encoding length, we find that recurrent quantum memory is essential when a task must reach far into the past, and dispensable when the relevant history is short, where the memoryless reset limit already suffices. For every task the best reservoirs operate at the edge of chaos, where they perform as well as a densely connected reservoir with random all-to-all couplings of the same size, indicating that what temporal processing requires is the dynamical regime rather than the connectivity.
\end{abstract}

\maketitle

\section{Introduction}
\label{sec:intro}

Quantum Machine Learning (QML) has emerged as an active research direction at the intersection of quantum information processing and artificial intelligence~\cite{Biamonte_2017,schuld2021machine,Cerezo_2022}. A central question in the field is whether the large state space and rich many-body dynamics of quantum systems can be exploited as computational resources for learning tasks. Most current approaches pursue this by making the quantum system itself trainable, so that every optimization step must be executed on the quantum device, adjusting its parameters with high precision and estimating each update from repeated measurements~\cite{Cerezo_2021,Bharti_2022}. Beyond this experimental overhead, the training loop faces obstacles of its own. Gradients typically vanish exponentially with the system size, the so-called barren plateaus, so that estimating them from measurements requires exponentially many repetitions and the training becomes impractical beyond a few qubits~\cite{McClean_2018,Cerezo_2021_BP,Zoe_BP,Larocca_2025}. Strategies that avoid the loop entirely, keeping the quantum system fixed and training only a classical readout on the measured outputs, avoid both obstacles at once. Among them, Quantum Extreme Learning Machines (QELMs)~\cite{HUANG2006489,Ghosh2019,Chen_2020,Innocenti2023,Kawanabe2026_TDQELM,Dao2026_QELM} and Quantum Reservoir Computing (QRC)~\cite{Fujii2017,Mujal2021,Mujal2024_QuEra,Hou_2026} have recently gained attention in both theory and experiment.

The latter derives from classical Reservoir Computing (RC), which uses the dynamics of a physical system to map input streams into a high-dimensional representation. The reservoir supplies the nonlinearity through its intrinsic dynamical response, and training reduces to a linear regression solved in closed form~\cite{Jaeger2001,Maass2002}. QRC extends this to quantum many-body systems, where the exponentially large Hilbert space provides a rich feature space for temporal information processing~\cite{Fujii2017,Nakajima2021,Martinez2021,Mujal2021}. Since the unitary evolution of a closed quantum system is reversible, the gradual forgetting of past inputs that temporal processing requires~\cite{Boyd1985} must be introduced by a non-unitary map. Depending on the protocol, this map can enter at the input through an erase-and-write step~\cite{Fujii2017,Martinez2021}, in the evolution through dissipation~\cite{Suzuki_2022,domingo2023takingadvantagenoisequantum,Sannia_2024}, or at the measurement stage via monitoring schemes~\cite{Franceschetto_2026,morgusancho2026generaltheorymonitoredquantum}. In this work we adopt the first of these, the standard choice of an erase-and-write map.

\begin{figure*}[t]
    \centering
    \includegraphics[width=1.0\linewidth]{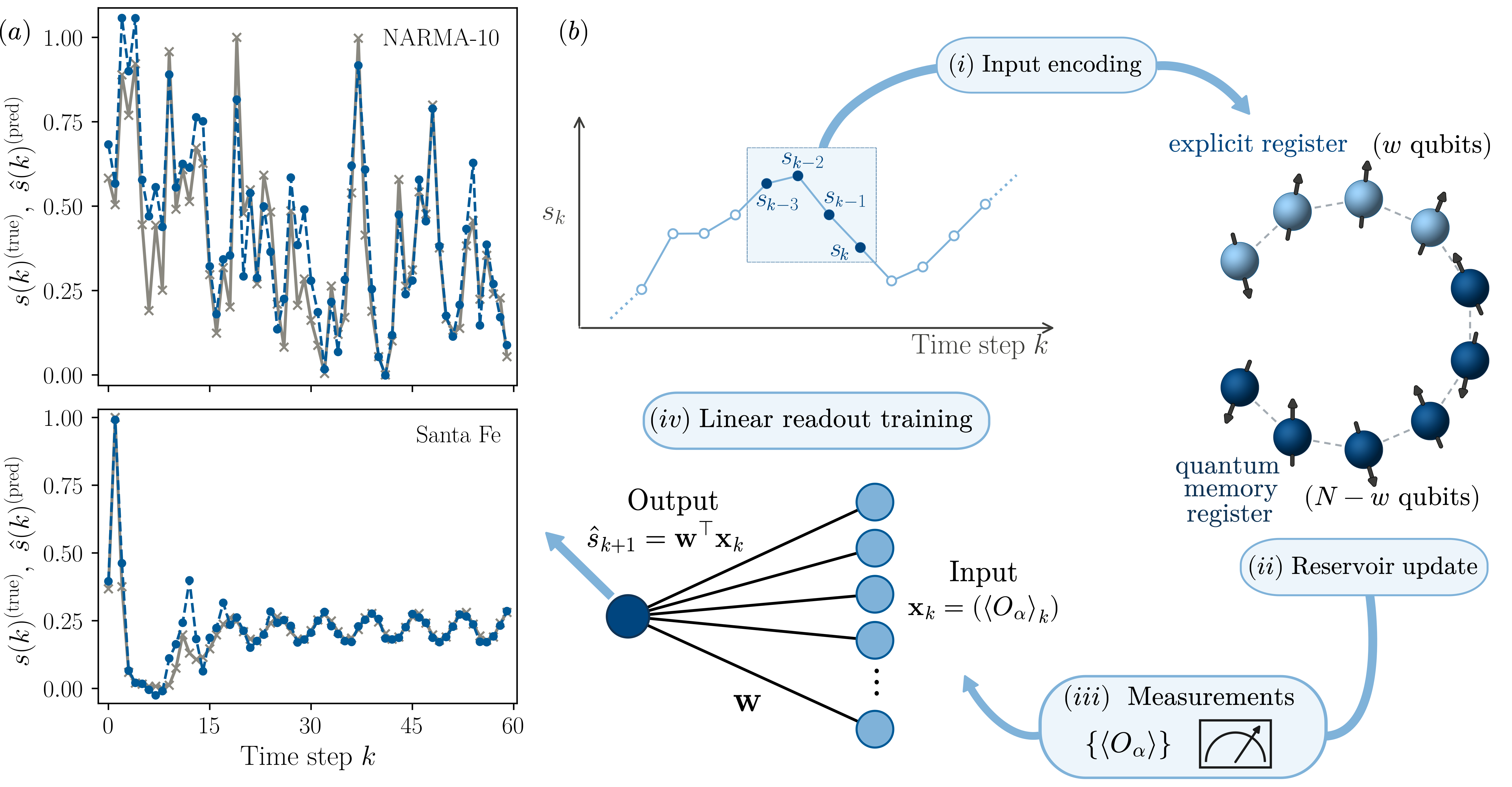}
    \caption{Quantum Reservoir Computing with a tunable memory register. (a) Example of the reservoir prediction $\hat s(k)^{(\mathrm{pred})}$ (dashed blue) against the target $s(k)^{(\mathrm{true})}$ (solid grey) over a subset of the test set, for the NARMA-10 task (top) and the Santa Fe laser series (bottom), both obtained with the intermediate $w = 4$ reservoir sketched in (b). The Hamiltonian is that of Eq.~\eqref{eq:mfim}, with $(J, h_x, h_z, \Delta t) = (1, 0.5, 1.05, 3.14)$ for NARMA-10 and $(1, 6.42, 3.84, 70)$ for Santa Fe. (b) Pipeline of one QRC step. (i) The $w$ most recent inputs $(s_k, s_{k-1}, \dots)$ are encoded into the register qubits (light blue), giving $\rho_{\mathrm{in}}^{(k)}$; these are reset and rewritten at every step and hold the recent inputs explicitly (the explicit register), while the $N-w$ memory qubits (dark blue) are never reset and hold older inputs in their evolving state (the quantum memory register), in the state $\rho_{\mathrm{mem}}^{(k-1)} = \mathrm{Tr}_w[\rho^{(k-1)}]$ inherited from the previous step. (ii) The whole chain then evolves for a time $\Delta t$ under $U = e^{-i H \Delta t}$, so that the reservoir state is updated as $\rho^{(k)} = U\,(\rho_{\mathrm{in}}^{(k)} \otimes \rho_{\mathrm{mem}}^{(k-1)})\,U^{\dagger}$. (iii) The evolved state is measured on a set of observables $\{O_\alpha\}$, collected into the feature vector $\mathbf{x}_k$. (iv) A linear readout $\mathbf{w}$, the only trained part of the protocol, maps these features onto the output.}
    \label{fig:setup}
\end{figure*}

With this choice, the non-unitary step is restricted to the input injection, while the evolution between inputs remains unitary. Over the evolution between inputs, the reservoir Hamiltonian mixes each newly encoded input with the memory stored in the remaining qubits. This scrambling makes combinations of inputs from many past steps accessible to the readout. However, this mixing cannot be pushed indefinitely, since in a high-dimensional Hilbert space almost all states are locally indistinguishable from thermal equilibrium, a manifestation of canonical typicality~\cite{Goldstein_2006, Popescu_2006}. Once the dynamics scramble strongly enough the input delocalizes into global many-body correlations, and few-body observables lose their dependence on it. A useful reservoir therefore operates between two extremes, with too little evolution leaving the map nearly linear and too much erasing the local imprint of the inputs, the best performance falling in between at moderate scrambling~\cite{Bertschinger2004,Mart_nez_Pe_a_2021, Kobayashi_2026,tnfv-lzfx}.

In the Heisenberg picture, both the reach into the past and the nonlinearity of the readout come from one effect, the spreading of few-body observables into many-body operators under the evolution, which lets a local measurement reach inputs injected at earlier steps and combine them into nonlinear features~\cite{Mujal_2021,Cindrak2026}. The architecture provides another knob, namely the number of qubits devoted to the input, which sets how much of the past is stored explicitly rather than left to the recurrent quantum state. We investigate this trade-off in a linear Ising chain as illustrated in Fig.~\ref{fig:setup}(b) in which a register holding the most recent inputs is re-encoded at every step. These qubits store the recent past as an explicit record and form the  \emph{explicit register}. The remaining qubits are never reset, carry older inputs forward in their evolving state, and form the \emph{quantum memory register}. Widening the former trades quantum memory for explicit encoding at fixed system size and observable budget. We ask how this balance should be set, and in particular whether the recurrent quantum memory is always worth keeping. At one extreme, a single input qubit leaves the entire past to that memory; at the other, the nonrecurrent QELM resets every qubit at each step so that only the explicit register survives; intermediate windows interpolate between the two.

To answer how this balance should be set, we benchmark the reservoir on two families of tasks that place complementary demands on its memory and are summarized in Table~\ref{tab:benchmark_tasks}. The first family asks it to reconstruct the past. The Short-Term Memory (STM) task measures the linear memory the reservoir retains about earlier inputs, while the NARMA family couples memory of the recent past with nonlinear processing~\cite{Atiya2000}. The second family asks it to forecast the future of a chaotic signal from its history. We use the Mackey-Glass system~\cite{Mackey1977}, whose attractor ranges from quasi-periodic to high-dimensional chaos, and the Santa Fe laser series~\cite{Weigend1994}, a measured signal with short-ranged correlations. The two signals are deliberately opposite, the Mackey-Glass attractor developing long-range structure that should reward a recurrent memory while the short correlations of the Santa Fe series may not need one.

To compare windows fairly, we evaluate each encoding window at its best-performing dynamics, optimizing the evolution time and the Hamiltonian parameters with Bayesian Optimization (BO) through the Tree-structured Parzen Estimator (TPE)~\cite{Bergstra2011,Akiba2019}. We also characterize how chaotic the best-performing regime is for every window. Earlier work placed this optimum at the edge of many-body quantum chaos, but mostly at a single input site in densely connected models~\cite{Mart_nez_Pe_a_2021, Kobayashi_2026, llodra2024}. Quantum reservoirs are more generally built from randomly and densely coupled systems~\cite{Fujii2017,Nakajima2021,Nakajima_2019,Kutvonen2020,PhysRevResearch.3.013077,Martinez2021} on the premise that complex dynamics call for dense connectivity. We test whether a native nearest-neighbor chain with only uniform local fields already reaches the moderate scrambling regime that temporal processing needs, and benchmark it directly against the densely connected model at equal size. We sweep the full range of encoding windows in the 1D Mixed-Field Ising Model (MFIM), tracking how the optimal regime shifts as the reservoir moves from relying on its quantum memory to relying on an explicit register.

\begin{table}[t]
    \begin{ruledtabular}
        \begin{tabular}{llll}
            \textbf{Task} & \textbf{Target} & \textbf{Task parameter} & \textbf{Score} \\
            \hline
            STM & $s_{k-d}$ & --- & $C_{\mathrm{tot}}$, $d \le 50$ \\
            NARMA-$n$ & Eq.~(\ref{eq:narma}) & $n \in \{1,\dots,10\}$ & $1-\mathrm{NRMSE}$ \\
            Mackey--Glass & $s_{k+d}$ & $\tau_{\mathrm{MG}} \in \{7,\dots,50\}$ & $C^{\mathrm{fut}}_{\mathrm{tot}}$, $d \le 150$ \\
            Santa Fe & $s_{k+d}$ & --- & $C^{\mathrm{fut}}_{\mathrm{tot}}$, $d \le 70$ \\
        \end{tabular}
    \end{ruledtabular}
    \caption{\label{tab:benchmark_tasks} Benchmark tasks. The memory tasks reconstruct a function of past inputs and the forecasting tasks predict the future value of the signal. The third column gives the parameter that tunes the demand within each family, the order $n$ of the recurrence and the delay $\tau_{\mathrm{MG}}$ controlling the attractor dimension, with STM and Santa Fe having no such dial. The last column gives the score, a capacity summed over delays or horizons up to $d_{\mathrm{max}}$, or $1-\mathrm{NRMSE}$ for NARMA.}
\end{table}

In Sec.~\ref{sec:The 1D Mixed-Field Ising Model as Quantum Reservoir}, we introduce the mixed-field Ising chain, characterize its chaotic dynamics, and define the encoding-window protocol. In Sec.~\ref{sec:results} we benchmark the reservoir on the memory and forecasting tasks and compare it against a densely connected reservoir of equal size. We summarize our findings in Sec.~\ref{sec:conclusions}.

\section{The 1D Mixed-Field Ising Model as Quantum Reservoir}
\label{sec:The 1D Mixed-Field Ising Model as Quantum Reservoir}

The quantum reservoir that we use consists of a one-dimensional Mixed-Field Ising Model (MFIM) of $N$ qubits with open boundary conditions. Its Hamiltonian is given by
\begin{equation}
\label{eq:mfim}
    H = J \sum_{i=1}^{N-1} X_{i}X_{i+1}+ h_z \sum_{i=1}^N Z_i+ h_x\sum_{i=1}^N X_i ,
\end{equation}
where $J$ sets the nearest-neighbor coupling along $x$, and $h_x$ and $h_z$ are uniform longitudinal and transverse fields. The MFIM is among the simplest interacting quantum systems realizable on quantum hardware, and a single pair of fields tunes it continuously from an integrable to a fully chaotic regime. Switching off the longitudinal field, $h_x = 0$, reduces Eq.~\eqref{eq:mfim} to the integrable Transverse-Field Ising Model (TFIM), which maps to free, non-interacting fermions through the Jordan-Wigner transformation~\cite{PFEUTY197079}. A nonzero $h_x$ breaks this integrability and drives the system toward quantum chaos, so that $(h_x,h_z)$ alone interpolates between the two regimes at fixed coupling.

This tunability makes the MFIM a suitable reservoir, since the degree of chaos in its dynamics largely controls the memory available for computation, while the nonlinearity of the feature map is set by the encoding and the measured observables. Its static level statistics (Sec.~\ref{subsec:static}) locate the parametric crossover from integrable to chaotic dynamics across the field plane, and its dynamical properties (Sec.~\ref{subsec:dynamic}) set the characteristic timescales of a fixed chaotic reservoir. Sec.~\ref{subsec:reservoir} then describes how the model is operated as a quantum reservoir, and Sec.~\ref{subsec:BO_method} the strategy used to select the best-performing ones.

\subsection{Static properties: level statistics}
\label{subsec:static}

Quantum chaos manifests in the spectrum as repulsion between energy levels, absent in the integrable case, and we use this to characterize the reservoir through its level statistics. To quantify it, rather than the bare level spacing, which depends on the local density of states and requires unfolding, we use the dimensionless ratio of consecutive gaps \cite{Oganesyan_2007,Mart_nez_Pe_a_2021,Palacios_2024}. Ordering the eigenvalues $\{E_n\}$ within a symmetry sector and writing $s_n = E_{n+1}-E_n$, the gap ratio is
\begin{equation}
    \tilde{r}_n = \frac{\min(s_n,s_{n-1})}{\max(s_n,s_{n-1})} \in [0,1],
\end{equation}
whose spectral average $\braket{\tilde{r}}$ separates the two regimes. In the integrable limit the levels are uncorrelated and may cross freely, so the spacing obeys Poisson statistics as conjectured by Berry and Tabor~\cite{Berry1977LevelCI}, giving $P_{\text{Poiss}}(\tilde{r}) = \frac{2}{(1+\tilde{r})^2}$ with mean $\braket{\tilde{r}} = 2\ln 2-1 \approx 0.386$. In the chaotic regime the levels repel, and by the Bohigas-Giannoni-Schmit conjecture their fluctuations follow Random Matrix Theory (RMT)~\cite{PhysRevLett.52.1}. The gap-ratio distribution is then well described by the surmise
distribution $P_\beta(\tilde r) = \frac{2}{Z_\beta}\,\frac{(\tilde r + \tilde r^{2})^{\beta}}{(1 + \tilde r + \tilde r^{2})^{1+3\beta/2}}$~\cite{Atas_2013}, with $Z_{\beta}$ a normalization constant and the Dyson index $\beta$ labeling the symmetry class. Since the Hamiltonian from Eq.\eqref{eq:mfim} is real and symmetric, its chaotic regime belongs to the Gaussian Orthogonal Ensemble (GOE) $\beta = 1$, with $Z_1 = 8/27$ and numerical mean $\braket{\tilde{r}} \approx 0.531$ in the large-matrix limit. Therefore, a value of $\braket{\tilde{r}}$ near $0.386$ indicates integrable dynamics, one near $0.531$ a chaotic regime, and intermediate values track the crossover.

\begin{figure}[t]
    \centering
    \includegraphics[width=1\linewidth]{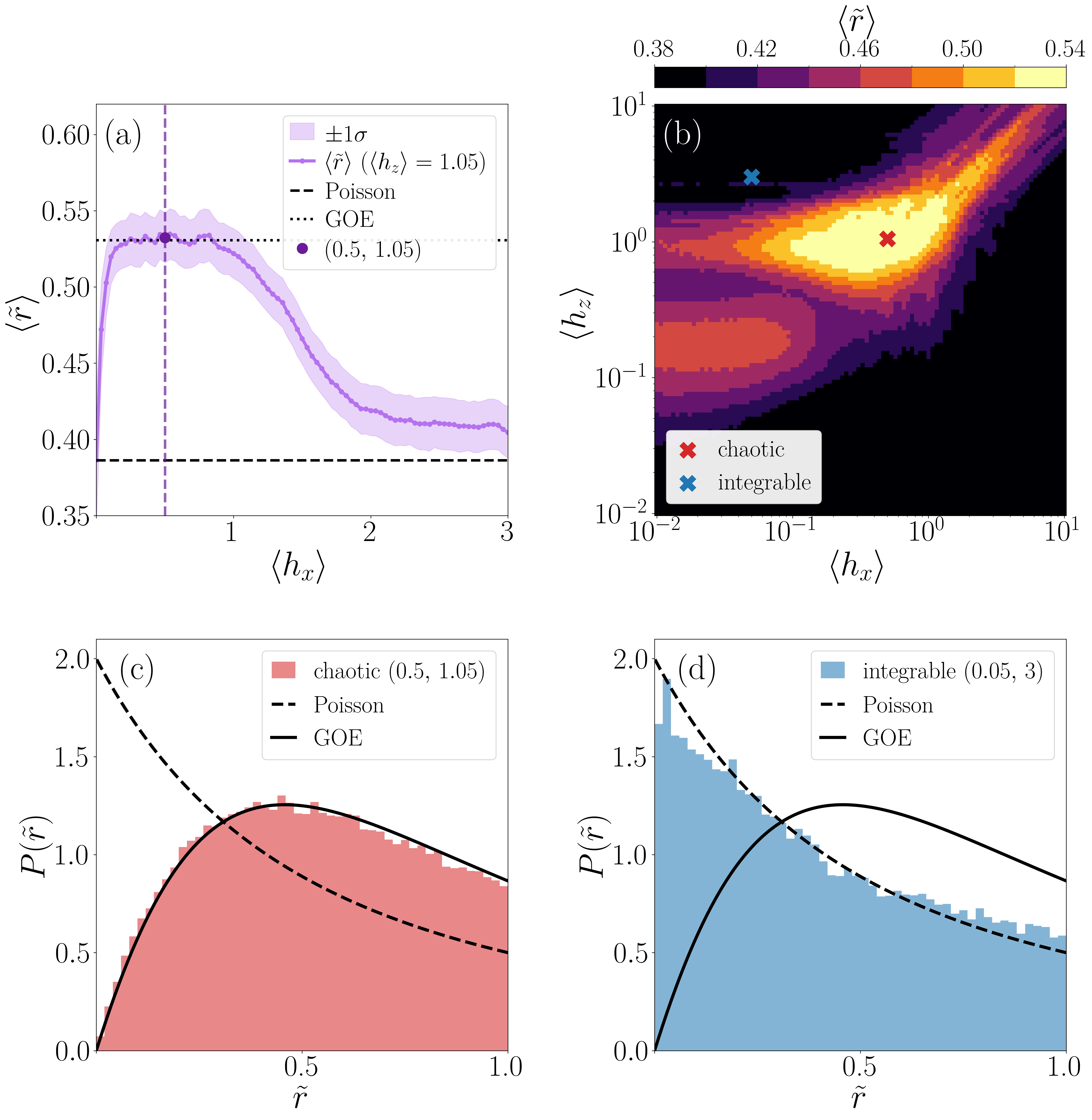}
    \caption{Spectral characterization of the MFIM reservoir for $N=10$ at coupling $J=1$, with fields drawn from Gaussian distributions of width $\epsilon=0.05$ around their nominal values $\langle h_\bullet\rangle$ and averaged over $n_r=1000$ realizations in (a,\,b) and $n_r=200$ in (c,\,d). (a) Average gap ratio $\langle\tilde r\rangle$ versus the longitudinal field $\langle h_x\rangle$ at fixed $\langle h_z\rangle=1.05$, with the shaded band showing $\pm 1\sigma$. The dashed and dotted lines mark the Poisson ($0.386$) and GOE ($0.531$) values, and the dot marks the chaotic point used in this work, $(\langle h_x\rangle,\langle h_z\rangle)=(0.5,1.05)$. (b) Heatmap of $\langle\tilde r\rangle$ over the field plane, with the chaotic (red) and integrable (blue) reference points marked. (c-d) Gap-ratio distributions $P(\tilde r)$ at the chaotic point $(0.5,1.05)$ and the integrable point $(0.05,3)$, compared with the Poisson and GOE predictions. 
    }
\label{fig:levelstats}
\end{figure}

We compute $\braket{\tilde{r}}$ for $N = 10$, fixing $J=1 $ throughout unless stated otherwise. Some care is needed with the symmetry sectors, since superposing the uncorrelated spectra of different sectors washes out the level repulsion and biases the statistics towards Poisson even in the chaotic regime \cite{PhysRevX.12.011006}. With uniform fields, Eq.~\eqref{eq:mfim} is invariant under the spatial reflection $i\to N+1-i$ about the center of the chain, whose parity operator commutes with $H$ and splits the Hilbert space into symmetric and antisymmetric sectors. We therefore diagonalize $H$ in each sector separately and evaluate the gap ratio within it. To smooth the strong sample-to-sample fluctuations of $\braket{\tilde{r}}$ in a finite chain, we average over $n_r = 1000$ realizations whose uniform fields are drawn with Gaussians of width $\epsilon = 0.05$ centered at the nominal values $\braket{h_x}$ and $\braket{h_z}$. Finally, since the gap ratio is only meaningful where the density of states is smooth we discard the lowest and highest $25\%$ of the spectrum, where boundary effects dominate, and evaluate $\braket{\tilde{r}}$ over the central bulk. Along a cut at fixed $\braket{h_z}=1.05$ (Fig.~\ref{fig:levelstats}(a)), $\braket{\tilde{r}}$ sits at the Poisson value at $\braket{h_x}=0$, where the model reduces to the integrable TFIM, rises to the GOE value as $\braket{h_x}$ grows and chaos sets in, and falls back toward Poisson at large fields, where the coupling and the dominant longitudinal field share an eigenbasis and the dynamics become nearly classical. The dot marks the chaotic point $(\braket{h_x},\braket{h_z})=(0.5,1.05)$, benchmarked in earlier works~\cite{Ba_uls_2011,Balasubramanian_2021}, which we adopt as the chaotic reservoir Hamiltonian below. The same structure recurs over the full field plane (Fig.~\ref{fig:levelstats}(b)), the ratio dropping to Poisson at very low fields or wherever one field dominates the other and the coupling, and reaching GOE across a broad region where both fields are comparable to the coupling; the red and blue crosses mark the chaotic Hamiltonian and an integrable reference point $(0.05,3)$. The gap-ratio distributions at these two points (Fig.~\ref{fig:levelstats}(c,d)) confirm the distinction, the chaotic one following the GOE surmise and vanishing as $\tilde{r}\to0$ in a clear signature of level repulsion, the integrable one following the Poisson form with none.

\subsection{Dynamical properties: scrambling timescales}
\label{subsec:dynamic}

How far the encoded information spreads depends not only on whether the spectrum is chaotic but also on the evolution time $\Delta t$, since the chaotic correlations that drive the scrambling build up only beyond a characteristic timescale. Two such scales govern the dynamics of a finite chaotic quantum system, one timescale marking the point at which the discreteness of the spectrum is resolved and a shorter one marking the onset of the chaotic correlations. Before determining them, we fix the unit of time by normalizing the Hamiltonian by its root-mean-square spectral width, $\sigma_H = \sqrt{\mathrm{Tr}(H^2)/\mathcal{D}}$, which coincides with the standard deviation of the eigenvalues since $H$ is traceless, so that the rescaled $\tilde{H} = H/\sigma_H$ has unit spectral variance. This removes the dependence on the overall energy scale set by $(J,h_x,h_z)$, ensuring that a given evolution time corresponds to comparable dynamics across parameters, and sets a common dimensionless time unit.  All evolution times of this work, including the two scales above, refer to this rescaled evolution $e^{-i\tilde{H}\Delta t}$.

\begin{figure*}[t!]
    \centering
    \includegraphics[width=1\linewidth]{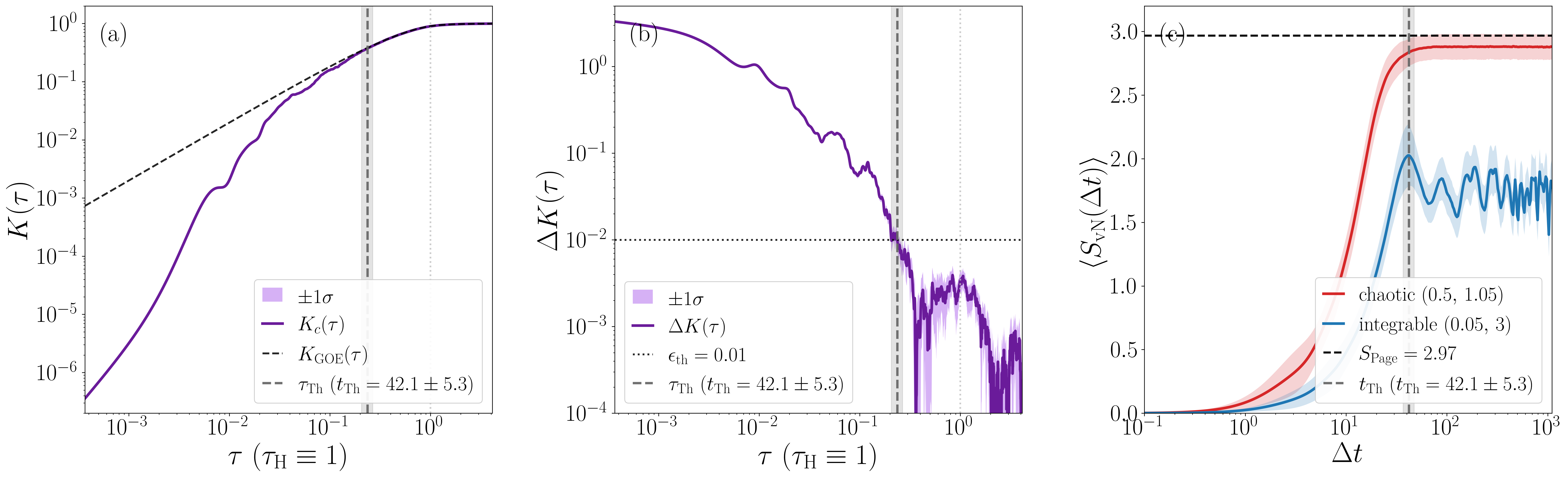}
    \caption{Dynamical signatures of quantum chaos in the MFIM reservoir, at the chaotic point $(h_x, h_z) = (0.5, 1.05)$ and, for comparison in (c), also at the integrable point $(0.05, 3)$. (a) Connected spectral form factor (SFF) $K_c(\tau)$ as a function of the rescaled time $\tau = t/t_H$ (solid), compared with the GOE prediction $K_{\mathrm{GOE}}(\tau)$ (dashed), averaged over $n_r = 2000$ realizations with fields drawn uniformly from a window of half-width $\epsilon_{\mathrm{dis}} = 0.15$ around the chaotic point. (b) Deviation $\Delta K(\tau) = |\log_{10}[K_c(\tau)/K_{\mathrm{GOE}}(\tau)]|$, with the threshold $\epsilon_{\mathrm{th}} = 0.01$ (dotted horizontal line) whose crossing gives $t_{\mathrm{Th}} = 42.1 \pm 5.3$ in physical units. (c) Half-chain von Neumann entropy $\langle S_{\mathrm{vN}}(t)\rangle$, averaged over $200$ initial random product states evolved under the chaotic (red) and the integrable (blue) Hamiltonians, with the Page value $S_{\mathrm{Page}} \approx 2.97$ marked (dashed horizontal line). In all panels the vertical band marks $t_{\mathrm{Th}}$, and shaded bands show $\pm 1\sigma$, over disorder realizations in (a,\,b) and over initial states in (c).}
    \label{fig:ThoulesTime_SFF}
\end{figure*}

The longest timescale of the two, present in any system with a discrete spectrum, is the Heisenberg time $t_H$, set by the inverse mean level spacing, which marks the scale at which individual levels are resolved. For the quantum reservoir at the selected chaotic point $(J,h_x,h_z) = (1,0.5,1.05)$, it takes the value $t_H \simeq 178$, and the details of how it is computed can be found in Appendix~\ref{Appendix_sec:Calculation of the Heisenberg time}. The other timescale, specific to the chaotic regime, is the Thouless time $t_{\mathrm{Th}}$, which following Ref.~\cite{Schiulaz_2019} can be interpreted as the time that an initially localized many-body state takes to spread, through local interactions, over the part of the Hilbert space compatible with its energy.

The Thouless time is read from the Spectral Form Factor (SFF), the Fourier transform of the spectral two-point correlations, which measures how those correlations manifest dynamically. In a chaotic system its connected version $K_c(\tau)$ follows the universal GOE ramp $K_{\mathrm{GOE}}(\tau)$ only beyond $t_{\mathrm{Th}}$: as Fig.~\ref{fig:ThoulesTime_SFF}(a) shows, $K_c(\tau)$ departs from the GOE prediction at short times, joins it at $\tau_{Th}$, and saturates to the plateau at $\tau_H=1$. We locate the onset through the deviation $\Delta K(\tau)=|\log_{10}[K_c(\tau)/K_{\mathrm{GOE}}(\tau)]|$ of Fig.~\ref{fig:ThoulesTime_SFF}(b), moving inward from large $\tau$ to the first crossing of a threshold $\epsilon_{th}=0.01$. A single finite spectrum yields a noisy $K_c(\tau)$, since the SFF does not self-average and only converges to the universal ramp once averaged over an ensemble~\cite{_untajs_2020}. Our model carries no intrinsic disorder, so we build that ensemble from a small neighborhood of the chaotic point, drawing the fields uniformly from a window of half-width $\epsilon_{\mathrm{dis}} = 0.15$ around $(h_x, h_z) = (0.5, 1.05)$ and averaging over $n_r = 2000$ realizations. This gives $t_{\mathrm{Th}} = 42.1 \pm 5.3$, representative of the chaotic region rather than of one fine-tuned point. The connected SFF, the GOE ramp, and the full extraction pipeline, including the spectral unfolding and filtering, are given in Appendix~\ref{Appendix_sec:Detailed calculation of the Thouless time}.

Figure~\ref{fig:ThoulesTime_SFF}(c) gives an intuitive picture of the Thouless timescale. We prepare $n_s = 200$ random product states $\ket{\psi_0} = \bigotimes_{i=1}^N \ket{\varphi_i}$, with $\ket{\varphi_i} = \cos(\theta_i/2)\ket{0}+e^{i\phi_i}\sin(\theta_i/2)\ket{1}$, $\cos\theta_i \sim \mathcal{U}[-1,1]$ and $\phi_i \sim \mathcal{U}[0,2\pi)$, evolve each under $e^{-i\tilde{H}\Delta t}$, and compute the half-chain von Neumann entropy
\begin{equation}
    S_{vN}(\Delta t) = -\text{Tr}\left(\rho_A(\Delta t) \ln\rho_A(\Delta t) \right),
\end{equation}
with $\rho_A(\Delta t) = \text{Tr}_B(\ket{\psi(\Delta t)}\bra{\psi(\Delta t)})$, averaged over the initial states. At the chaotic point $(J,h_x,h_z) = (1,0.5,1.05)$ the entanglement grows ballistically from zero and saturates, around $t_{\mathrm{Th}}$, close to the Page value $S_{\text{Page}}  \simeq \ln \mathcal{D}_A - \frac{\mathcal{D}_A}{2\mathcal{D}_B} \approx 2.97$ for $N=10$~\cite{PhysRevLett.71.1291}. At the integrable point $(h_x,h_z) = (0.05,3)$ it also grows ballistically but saturates well below the Page value and oscillates, the conserved quantities of the integrable model preventing the state from spreading uniformly over the Hilbert space and instead returning it periodically toward less entangled configurations, a signature of the absence of thermalization.

\subsection{The MFIM as a quantum reservoir}
\label{subsec:reservoir}

We operate the MFIM as a reservoir through the noiseless erase-and-write map, in which the dissipation comes from the active reset of the input qubits at each step. The full protocol, sketched in Fig.~\ref{fig:setup}(b), repeats at every step the encoding of the inputs (i), the unitary evolution of the reservoir (ii), and the measurement of a set of observables (iii), with the linear readout (iv) trained only once at the end. The reservoir is initialized in a reference state that is easy to prepare, here $\rho^{(0)} = \ket{0}\bra{0}^{\otimes N}$. In step (i), the $w$ input qubits are reset and the $w$ most recent inputs of the sequence $\{s_k\}$, with $s_k \in [0,1]$, are injected into the explicit register through single-qubit rotations $R_y(\pi s)$, so that the reservoir state becomes
\begin{equation}
    \label{eq:reservoir_update}
    \rho^{(k)} = \rho_{\text{in}}^{(k)} \otimes \text{Tr}_w[\rho^{(k-1)}],
\end{equation}
where $\rho_{\text{in}}^{(k)} = \bigotimes_{j=0}^{w-1} \rho(s_{k-j})$, with $\rho(s) = R_y(\pi s)\ket{0}\bra{0}R_y^{\dagger}(\pi s)$, and $\text{Tr}_w$ the partial trace over the input qubits, which leaves the $N-w$ memory qubits untouched. In step (ii), the whole chain evolves under the normalized Hamiltonian for a time $\Delta t$,
\begin{equation}
\label{eq:reservoir_evolution}
    \rho^{(k)}(\Delta t) = e^{-i\tilde{H}\Delta t} \left(\rho_{\text{in}}^{(k)} \otimes \text{Tr}_w\left[\rho^{(k-1)}\right] \right) e^{i\tilde{H}\Delta t},
\end{equation}
with $\tilde{H} = H/\sigma_H$. 

For a reservoir to process information reliably it must satisfy the Echo State Property (ESP), by which its state, after a sufficient number of steps, becomes a function of the input history alone, independent of the initial state $\rho^{(0)}$~\cite{Jaeger2001,YILDIZ20121,Kobayashi_2024}. In our protocol, the erase-and-write map drives the reservoir toward the ESP through the partial reset of Eq.~\eqref{eq:reservoir_update}, which discards part of the reservoir state at every step and progressively washes out the memory of $\rho^{(0)}$. On its own, the reset only removes the information stored in the input qubits. Combined with the scrambling evolution of Eq.~\eqref{eq:reservoir_evolution}, which feeds the rest of the state into those qubits before they are overwritten, it eventually erases the initial state entirely, as we verify numerically in Appendix~\ref{Appendix_sec:Convergence_property}.

The same convergence underlies the Fading Memory Property (FMP), by which the reservoir forgets each past input as more recent ones are injected, so that the recent past weighs more than the distant one~\cite{gonon2020fadingmemoryechostate, Mart_nez_Pe_a_2023}. An input stays written in the explicit register for its first $w$ steps, where it is held in the explicit register, and once it leaves this register it survives only in the recurrent memory of the $N-w$ qubits, fading under the repeated resets. The reservoir thus holds the last $w$ inputs explicitly and a decaying memory of everything older, a balance we study in Sec.~\ref{sec:results}.

The reservoir must also be input-separable, meaning that distinct input histories drive it to distinguishable states~\cite{morgusancho2026generaltheorymonitoredquantum,Mart_nez_Pe_a_2023}. This sets a second demand on the contraction behind the washout, which has to erase the initial condition without erasing the differences between input histories. The input-dependent reset of the erase-and-write map, which writes a distinct pure state into the explicit register at each step, provides this separation. We verify it directly in Appendix~\ref{Appendix_sec:Separability}, initializing two reservoirs in the same state and driving them with different input sequences. Once the initial state is washed out, the two trajectories do not merge but settle at a finite distance, confirming that the stationary trajectory remains input-driven. That distance is however small, of the order of $10^{-2}-10^{-1}$ in Frobenius norm, so that resolving it from measured statistics would demand a large number of shots. This limitation does not affect the present work, where all features are exact expectation values, but it sets the measurement budget of any experimental implementation of the protocol~\cite{Mujal_2023,Ahmed2024}.

Once the washout is complete, we begin recording the features. These are collected in step (iii), after the evolution of Eq.~\eqref{eq:reservoir_evolution} and before the next reset. On the evolved state, we measure the single-qubit expectations $\braket{X_i}$, $\braket{Y_i}$, $\braket{Z_i}$ and the two-body correlators $\braket{X_iX_j}$, $\braket{Y_iY_j}$, $\braket{Z_iZ_j}$, a total of $3N + 3\binom{N}{2} = 165$ observables for $N = 10$ that we collect into the feature vector $\mathbf{x}_k \in \mathbb{R}^{165}$.
 
The readout, step (iv), is a linear map from the features to the target, $\hat{y}_k = \mathbf{w}^{\top}\mathbf{x}_k$, and it is the only part of the protocol that is trained. We collect the recorded feature vectors as rows of a design matrix $X$, augmented with a constant column for the bias, and the corresponding targets of the training set into a vector $\boldsymbol{s}^{(\text{true})}_{\text{train}}$. The weights then follow from ridge regression, which admits the closed-form solution $\mathbf{w} = (X^{\top}X + \lambda R)^{-1} X^{\top}
\boldsymbol{s}^{(\text{true})}_{\text{train}}$, with $R = \mathrm{diag}(0,1,\dots,1)$. The regularization $\lambda$ penalizes large weights and helps prevent overfitting, while the zero first entry of $R$ leaves the bias unpenalized, so that a constant offset in the target is not artificially suppressed.

We split the post-washout data into a training and a test set in an $80/20$ ratio, preserving the test set for the final performance estimate alone. The regularization strength $\lambda$ is not fixed beforehand but selected on the training data through a further $80/20$ split into a fit and validation set. For each $\lambda$ on a logarithmically spaced grid of ten values from $10^{-8}$ to $10^1$, we solve the ridge equation on the fit and evaluate the Mean Square Error (MSE) on the validation set, retaining the value $\lambda^*$ that minimizes it. We then retrain the readout on the full training set with $\lambda^*$ and evaluate once on the test set.

\subsection{Optimal reservoir search}
\label{subsec:BO_method}

A fair comparison between the different encoding windows requires evaluating each $w$ at its own best-performing dynamics. Since each evaluation is expensive, involving a full reservoir
trajectory and the training of the readout, we locate these optima with Bayesian
Optimization (BO), using the Tree-structured Parzen Estimator
(TPE)~\cite{Bergstra2011,watanabe2026treestructuredparzenestimatorunderstanding} as
implemented in Optuna~\cite{Akiba2019}. TPE builds a probabilistic model of where the
good-scoring configurations concentrate relative to the bad ones and proposes the next
trial where this ratio is largest, so that the costly simulations are spent on the most
promising regions of the search space, as we detail in
Appendix~\ref{appendix_sec:tpe}. Each optimization in our simulations comprises $150$
trials, the first $50$ drawn at random to seed the model and the remaining $100$ guided by
TPE, and is run independently for every $w$.

\begin{figure*}[t]
    \centering
    \includegraphics[width=1\linewidth]{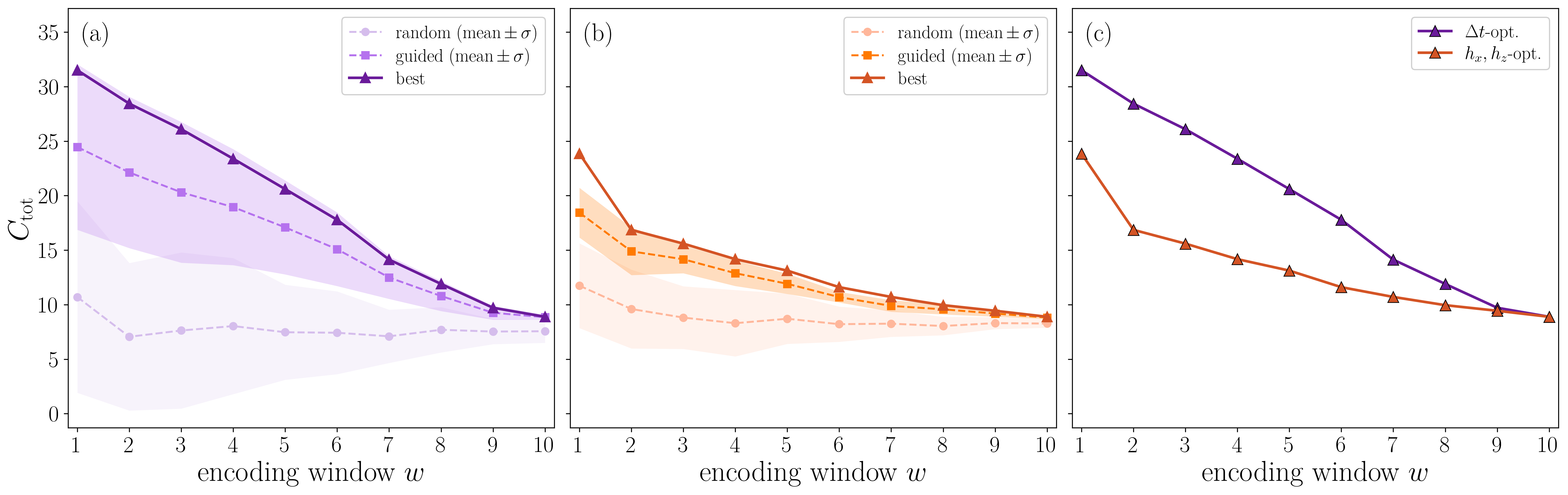}
    \caption{Total short-term memory capacity $C_{\mathrm{tot}}=\sum_{d=1}^{50} C(d)$ of the $N=10$ MFIM reservoir versus encoding window $w$, under the two strategies of Sec.~\ref{subsec:BO_method}, optimizing $\Delta t$ at the fixed chaotic point $(h_x,h_z)=(0.5,1.05)$ in (a) and $(h_x,h_z)$ at fixed $\Delta t=70$ in (b). Light markers with band show the mean $\pm\sigma$ over the $50$ random configurations, dark markers the $100$ guided (TPE) trials, and the solid line the single best configuration. Panel (c) overlays the two best curves. All collapse to a common value at the QELM limit $w\to N$.}
    \label{fig:STM_results}
\end{figure*}

We deploy this strategy on two separate experiments for each encoding window $w$ and each task. The first strategy is the parametric case, where the evolution time is fixed and the fields are optimized over $(h_x,h_z)\in[0,10]^2$ at fixed $J=1$. We set $\Delta t = 70$, inside the window $t_{\mathrm{Th}} < \Delta t < t_H$ established in Sec.~\ref{subsec:dynamic} for the chaotic point marked in Fig.~\ref{fig:levelstats}(b), so that the chaotic correlations of the dynamics are fully developed. This range covers the entire field plane of Fig.~\ref{fig:levelstats}(b), from the integrable boundaries to the chaotic plateau, so the optimizer is free to place the reservoir anywhere across the integrable-to-chaotic crossover rather than being confined to a regime chosen by hand. The second strategy is the temporal case, where we fix the Hamiltonian to the chaotic point $(h_x,h_z)=(0.5,1.05)$ characterized in Secs.~\ref{subsec:static} and~\ref{subsec:dynamic} and optimize the evolution time over $\Delta t \in [10^{-1},10^3]$, sampled on a logarithmic scale to probe all timescales evenly, from evolutions much shorter than $t_{\mathrm{Th}}$ to well beyond $t_H$. Separating the two searches keeps each optimization low-dimensional and isolates the two handles on the dynamics, asking which Hamiltonian performs best at a fixed evolution time, and how much evolution the fixed chaotic Hamiltonian needs.

\section{Results}
\label{sec:results}

We evaluate every encoding window $w$ at its own optimum under the two strategies of Sec.~\ref{subsec:BO_method}.

\subsection{Memory tasks}
\label{sec:memory}

\subsubsection{Short-Term Memory (STM) task}
\label{subsubsec:stm}

The STM task tests the linear memory of the reservoir. It is driven with independent random inputs $s_k \sim \mathcal{U}[0,1]$, and the readout is trained to reconstruct the input injected $d$ steps in the past, $y_k = s_{k-d}$. The target is the input itself, so the task involves no processing and measures only how much information about $s_{k-d}$ remains linearly accessible after $d$ further injections. For each delay we quantify it through the capacity $C(d) = \frac{\mathrm{cov}^2\left(\hat{y}_k, s_{k-d}\right)}{\mathrm{var}\left(\hat{y}_k\right)\mathrm{var}\left(s_{k-d}\right)} \in
[0,1]$, the squared Pearson correlation between prediction and target on the test set. Summing over delays gives the total memory capacity $C_{\text{tot}} = \sum_{d=1}^{d_{\max}} C(d)$, the score maximized by the BO, with $d_{\max}=50$. We use $8000$ steps in total, discarding the first $N_{\text{wo}} = 3000$ as a washout to erase the initial condition (fixed also for the remaining tasks) and splitting the remaining $5000$ steps $80/20$ into a $4000$-step training set and a $1000$-step test set, on which the capacities $C(d)$ are evaluated.

\begin{figure}[h!]
    \centering
    \includegraphics[width=1\linewidth]{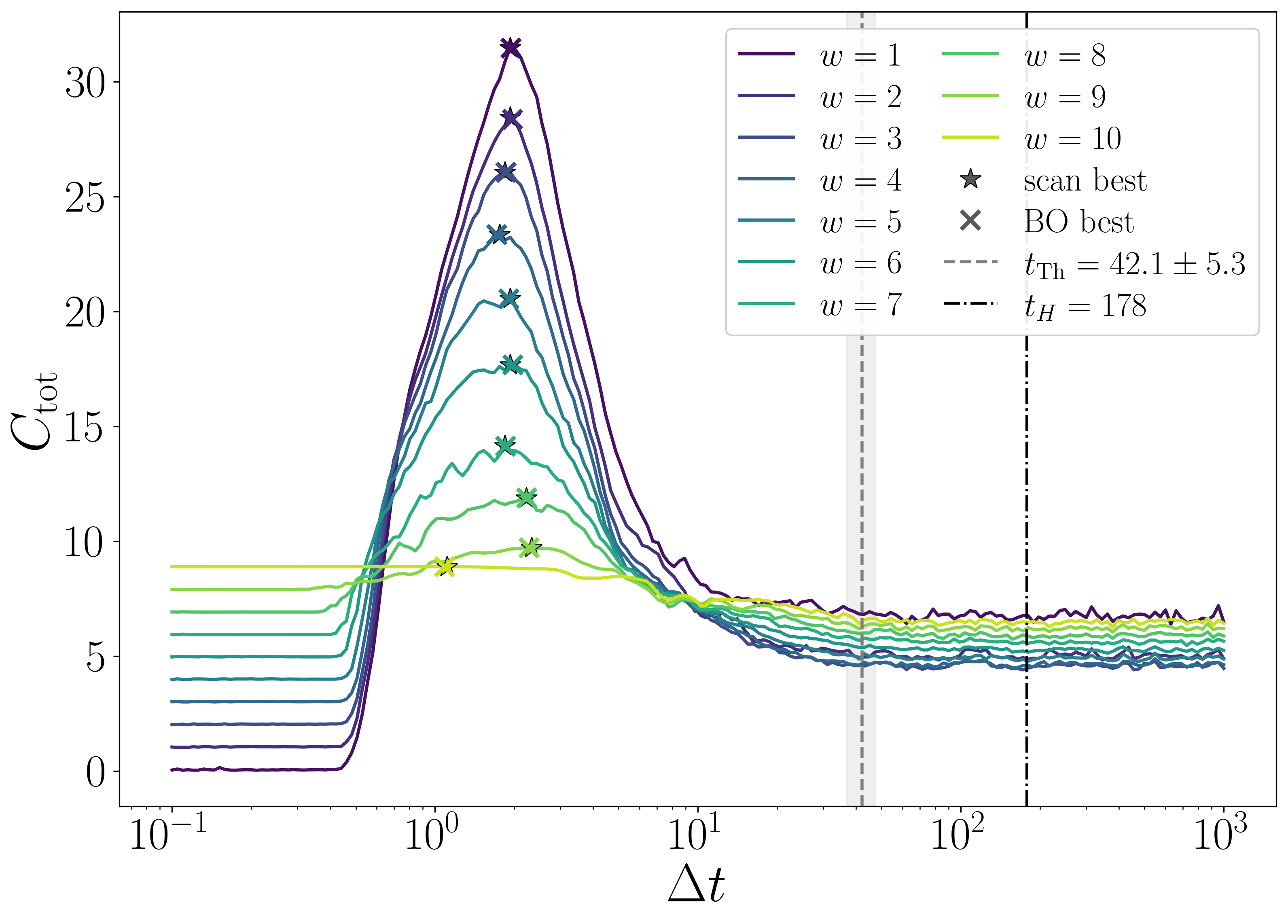}
    \caption{Total memory capacity $C_{\mathrm{tot}}=\sum_{d=1}^{50} C(d)$ versus evolution time $\Delta t$ at the chaotic point $(h_x,h_z)=(0.5,1.05)$, scanned for each window $w=1$ to $10$. For every window the maximum of the scan (star) coincides with the BO optimum (cross). The capacity peaks at intermediate $\Delta t$, well before the Thouless time $t_{\mathrm{Th}}=42.1\pm5.3$ (dashed) and the Heisenberg time $t_H=178$ (dash-dotted), and settles onto a low plateau at long $\Delta t$. The peak height decreases monotonically with $w$, from $w=1$ to the QELM limit $w=N$.}
    \label{fig:STM_scan}
\end{figure}

In panels (a) and (b) of Fig.~\ref{fig:STM_results}, the guided bands sit well above the random one, so the search is finding better reservoirs than random sampling would. The temporal scan of Fig.~\ref{fig:STM_scan} corroborates the search, since the optimum found by BO (crosses) coincides with the true maximum of the scanned curve (stars) at every window. For the parametric search we report no analogous scan, since a grid over the two-dimensional field plane is far more costly and requires averaging over several instances; instead, we analyze the optimal fields $(h_x,h_z)$ in Appendix~\ref{appendix_subsec:STM_expansion}.

The highest capacity is reached at $w=1$ in both panels, with $C_{\text{tot}}$ decreasing monotonically as $w$ grows. At $w=1$ the accessible past is stored in the $N-1$ memory qubits, whose correlations hold information about many past inputs at once. Increasing $w$ trades these qubits for encoding qubits, which recall the current input and its $w-1$ predecessors perfectly but shorten how long any input is retained (analyzed further in Appendix~\ref{appendix_subsec:STM_expansion}). Finally, the dispersion of both the random and the guided distributions shrinks as $w$ increases. At small $w$ the capacity depends delicately on the configuration, so tuning matters most, whereas at larger $w$ the encoding window guarantees recall of the recent inputs and the dynamics matter less. At the QELM limit $w=N$, all curves collapse to a common value set by the register alone, since no memory qubit remains and the reservoir reduces to a nonrecurrent time-delayed quantum feature map fed with an explicit input window. In this limit, the feature map depends only on the externally supplied delay vector, and while its features still depend on $H$ and $\Delta t$, empirically the STM score becomes weakly sensitive to these parameters for this linear recall task~\cite{delorenzis2026entanglementclassicalsimulabilityquantum}.

\begin{figure*}[t!]
    \centering
    \includegraphics[width=1\linewidth]{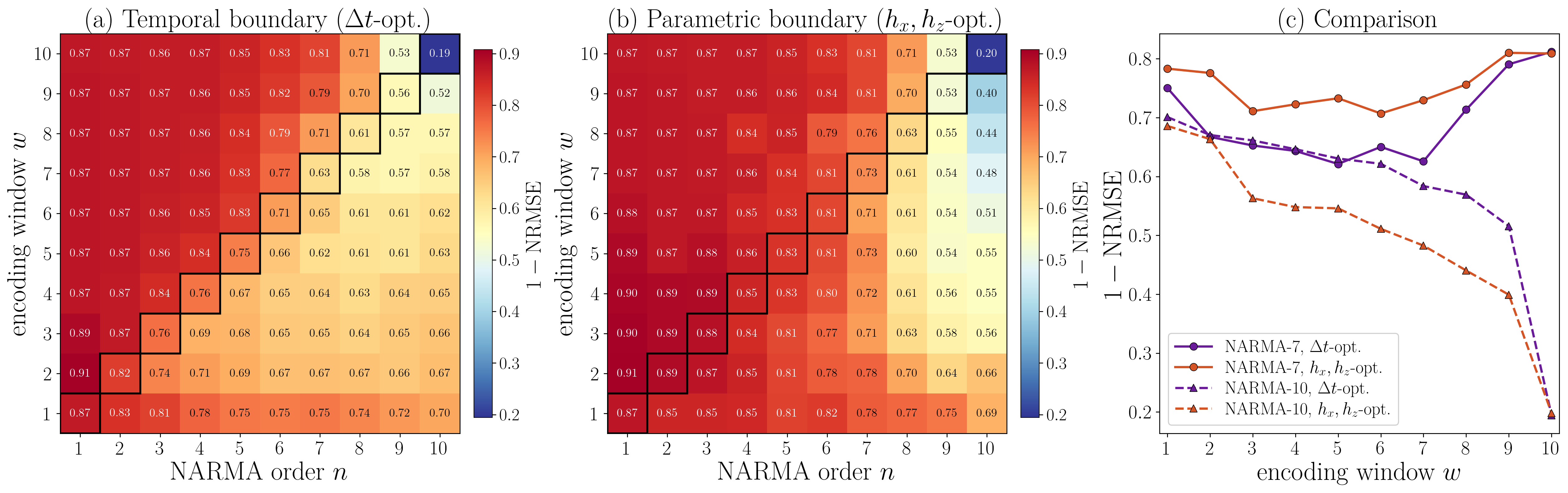}
    \caption{NARMA-$n$ score $1-\mathrm{NRMSE}$ over the plane of order $n$ and encoding window $w$ for the $N=10$ reservoir, showing the best score reached by optimizing $\Delta t$ at the fixed chaotic Hamiltonian (a) and $(h_x,h_z)$ at fixed $\Delta t=70$ (b). The black staircase marks the diagonal $n=w$, at and below which the required lag exceeds the register and must be supplied by the memory qubits. Panel (c) compares the two strategies for NARMA-$7$ and NARMA-$10$ versus $w$; the ordering inverts between them, with the field-optimized reservoir ($h_x,h_z$-opt., solid) leading for NARMA-$7$ and the temporally optimized one ($\Delta t$-opt., dashed with triangles) for NARMA-$10$, the two collapsing as $w\to N$.}
    \label{fig:NARMA_results}
\end{figure*}

Comparing the two strategies, the temporal search shows wider random and guided bands than the parametric one (Fig.~\ref{fig:STM_results}(a,\,b)), so the capacity depends more sensitively on the evolution time than on the fields. This sensitivity rewards the tuning, and panel (c) shows the payoff, with the best temporal configuration exceeding the best parametric one at every window $w<N$, the two meeting only at the QELM limit. The scan of Fig.~\ref{fig:STM_scan} explains this spread. At the shortest evolution times the dynamics have not yet mixed the input into the chain, so the reservoir recalls only what is written explicitly in its register, and the curves are ordered by $w$, with $w=N$ highest. A moderate evolution reverses this, spreading the input into the memory qubits so that the capacity peaks, now largest at $w=1$ where the recurrent memory is deepest, before a longer evolution flattens it to a low plateau, reached already before the Thouless time, where the input has delocalized beyond the reach of the local readout and every window performs alike, as under a Haar-random unitary. The peak marks the moderate scrambling that works best, and appears for every encoding window with $w<N$.

\subsubsection{NARMA-$n$ task}
\label{subsubsec:narma}

The Nonlinear Auto-Regressive Moving Average of order $n$ (NARMA-$n$)~\cite{Atiya2000} extends the demand from pure reminiscence to memory combined with nonlinear processing. The reservoir is driven with independent random inputs $s_k \sim \mathcal{U}[0,0.5]$, and the target is generated by the recurrence
\begin{equation}
    y_k = 0.3y_{k-1} + 0.05y_{k-1}\!\sum_{i=0}^{n-1} y_{k-1-i} + 1.5s_{k-n}s_{k} + 0.1.
    \label{eq:narma}
\end{equation}
The cross-term $s_{k-n}\,s_k$ is the critical ingredient, since it requires the reservoir both to recall the input from $n$ steps in the past and to compute its nonlinear product with the current input. The order $n$ therefore acts as a tunable memory demand, with larger $n$ requiring progressively longer recall and more complex nonlinear transformations. During the optimization the objective is $1-\mathrm{NRMSE}$, with $\mathrm{NRMSE} = \sqrt{\langle(\hat{y}_k - y_k)^2\rangle/\mathrm{var}(y_k)}$. The dataset follows the same protocol as the STM task, except that the first $n$ steps are additionally dropped, since the recurrence of Eq.~\eqref{eq:narma} is undefined before the input history reaches the order.

\begin{figure*}[t!]
    \centering
    \includegraphics[width=1\linewidth]{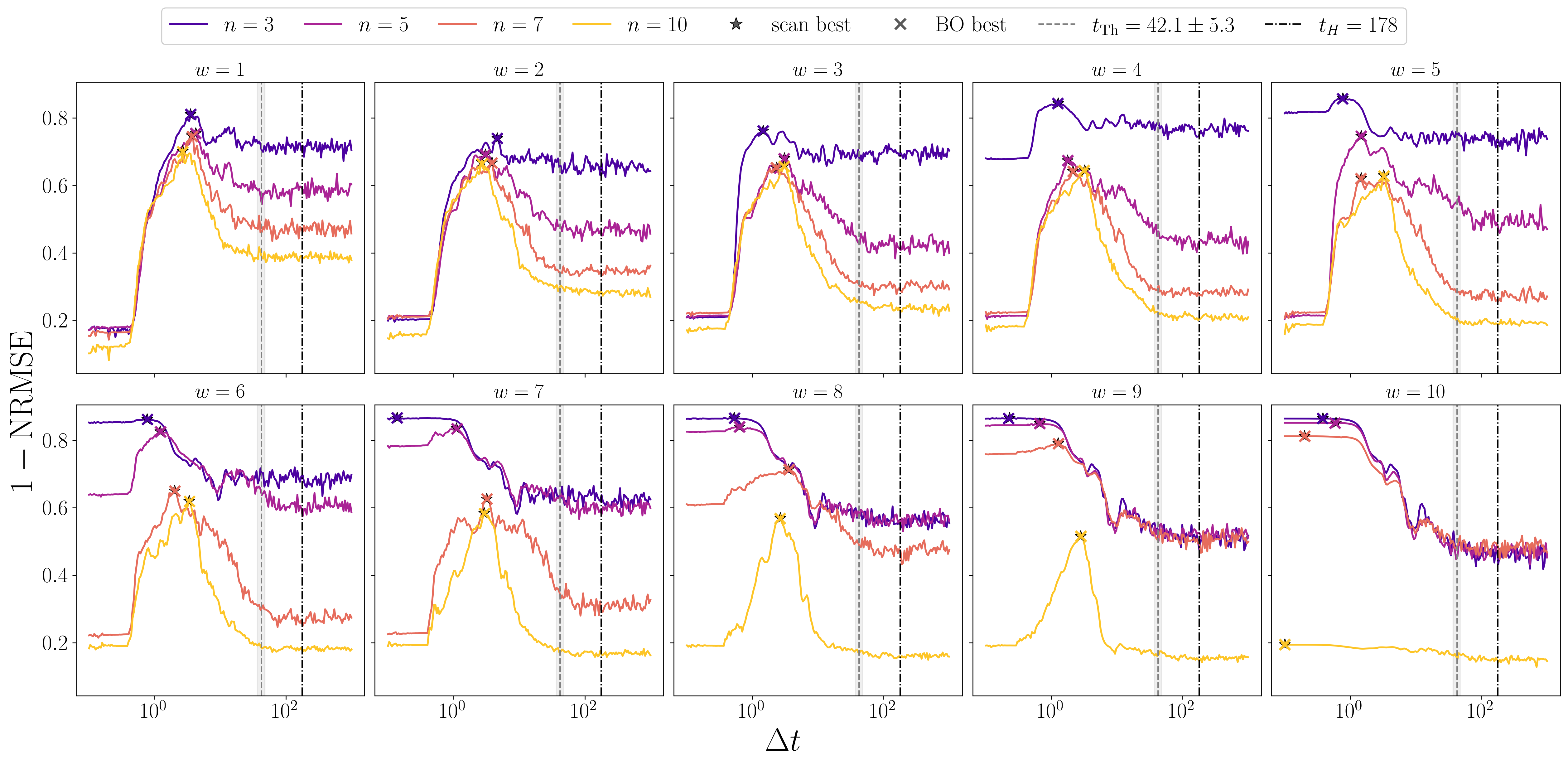}
    \caption{NARMA-$n$ score $1-\mathrm{NRMSE}$ versus evolution time $\Delta t$ at the chaotic point $(h_x,h_z)=(0.5,1.05)$, scanned for each window $w=1$ to $10$ (panels) and orders $n=3,5,7,10$ (colors). Stars mark the maximum of each scan and crosses the BO optimum, which coincide throughout. The score peaks at intermediate $\Delta t$ when $n\geq w$ (sharpening with order) and is already high at short $\Delta t$ when $n<w$, where both inputs of the cross-term lie inside the window. Vertical lines mark the Thouless time $t_{\mathrm{Th}}=42.1\pm5.3$ (dashed) and Heisenberg time $t_H=178$ (dash-dotted).}
    \label{fig:NARMA_scan}
\end{figure*}

The best score at each order $n$ and encoding window $w$, under both strategies, is organized by the diagonal $n=w$ marked with the black staircase (Fig.~\ref{fig:NARMA_results}). Since the window holds the inputs at delays $0,\dots,w-1$, the lag $s_{k-n}$ required by the cross-term of Eq.~\eqref{eq:narma} is physically present in the reservoir whenever $n<w$. In this region the recall is guaranteed by construction, and the nonlinear products the target requires are already exposed by the encoding and the two-body observables, so both panels show uniformly high scores. The degradation at higher orders reflects the growing nonlinear demand of the target, whose autoregressive sum couples an increasing number of past values. The corner $n=10$, $w=N$ collapses entirely, since the required lag falls outside even the full register and no memory qubits remain to carry it, consistent with the QELM collapse seen for the STM task.

Below the diagonal the required lag leaves the encoding register and its recall must be supplied by the memory qubits, and here the two strategies behave differently. In the temporal case the best score at each order is consistently reached at $w=1$, where the memory register is largest and the evolution time can be tuned to balance the two competing demands of the task, keeping the required lag alive in the memory register while mixing it with the recent inputs so that their product becomes accessible to the readout. The temporal scans of Fig.~\ref{fig:NARMA_scan} show this balance explicitly. Whenever $n\geq w$ the score peaks at intermediate evolution times, where the dynamics have mixed the inputs enough to build the nonlinear cross-term but not so much as to erase the required lag. The peak is sharpest at small windows and high orders, where the lag left the encoding register the longest ago and the range of useful evolution times, long enough to build the cross-term yet short enough to preserve the lag, is narrowest. When $n<w$, by contrast, the score is high already at vanishing evolution times, since both inputs entering the cross-term sit inside the encoding register and their product is directly available to the readout through the two-body correlators. In the parametric case the optimum sits at intermediate windows for some orders, although the differences are small, and at the highest orders it returns to $w=1$.

\begin{figure*}[t!]
    \centering
    \includegraphics[width=1\linewidth]{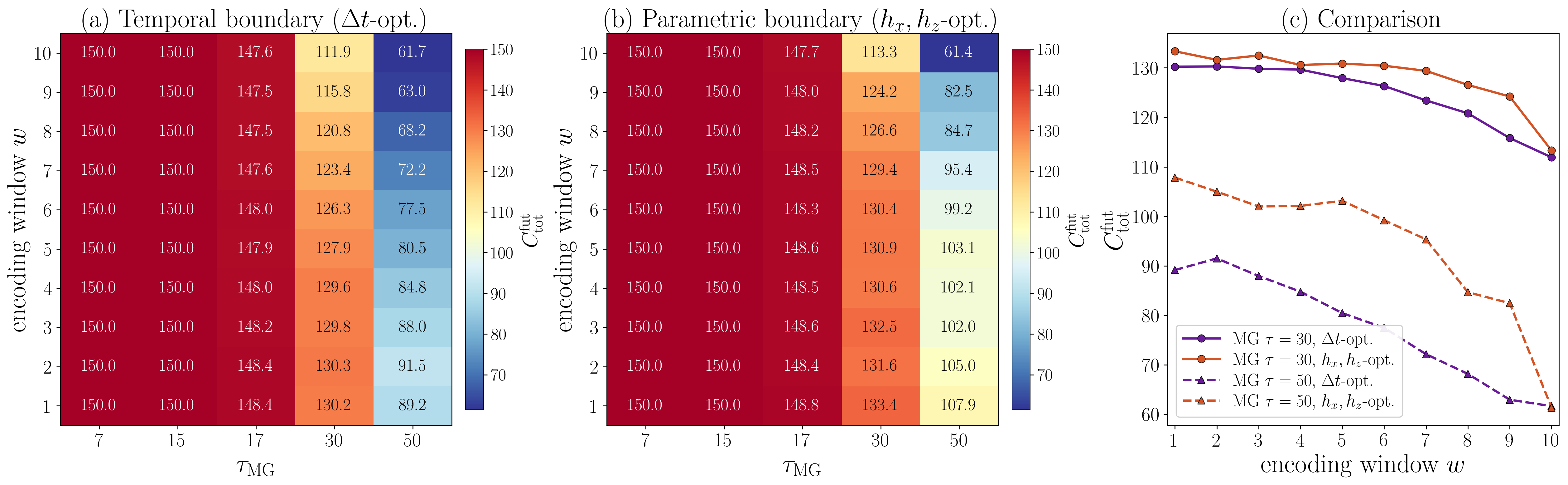}
    \caption{Mackey-Glass forward-prediction capacity $C_{\rm tot}^{\rm fut}$ over the plane of delay $\tau_{\rm MG}$ and encoding window $w$ for the $N=10$ reservoir, optimizing $\Delta t$ at the fixed chaotic Hamiltonian (a) and $(h_x,h_z)$ at fixed $\Delta t = 70$ (b). The score is bounded by $d_{\max} = 150$, reached for a perfectly predicted horizon. Panel (c) compares the two strategies versus $w$ for the chaotic delays $\tau_{\rm MG} = 30$ and $50$, where the field-optimized reservoir ($h_x,h_z$-opt.) exceeds the temporally optimized one ($\Delta t$-opt.), the two converging at the QELM limit $w = N$.}
    \label{fig:Mackey-Glass_results}
\end{figure*}

Which of the two strategies performs better changes with the order (Fig.~\ref{fig:NARMA_results}(c)). For NARMA-$7$ the parametric search obtains higher scores over most of the window range, while for NARMA-$10$ the ordering reverses and the temporally optimized reservoir performs best, with both strategies collapsing together at $w=N$. The comparison between the two strategies thus depends on how the task balances its two demands, recalling the lag and processing it nonlinearly. At moderate orders both demands are low, and the parametric search benefits from its larger freedom, as it can explore the whole field plane. At higher orders, however, the long evolution time fixed in the parametric case becomes a bottleneck, and no choice of fields compensates for it, so optimizing the evolution time of the fixed chaotic Hamiltonian reaches better reservoirs. Extended results across all orders and windows, together with the corresponding best configurations found, are collected in Appendix~\ref{appendix_subsec:NARMA_expansion}.

\subsection{Forecasting tasks}
\label{sec:forecasting}

The second family asks the reservoir to predict the future of a temporal signal from its history. Now the reservoir must build, from the input history, the nonlinear functions from which the future of the signal can be linearly extrapolated. We quantify the performance through the forward counterpart of the STM capacity,
\begin{equation}
    C^{\mathrm{fut}}(d) = \mathrm{cor}^2\!\left(\hat{y}_k, s_{k+d}\right),
    \qquad
    C_{\mathrm{tot}}^{\mathrm{fut}} = \sum_{d=1}^{d_{\max}} C_{\mathrm{fut}}(d),
    \label{eq:forward_capacity}
\end{equation}
where each future horizon $d$ is served by its own independently trained readout, so that the reservoir always receives the true signal as input and no prediction is fed back. The total capacity is the score maximized by the BO, and it aggregates the squared predictive correlation across horizons, with each fully correlated horizon contributing one unit to the sum.

\subsubsection{Mackey-Glass task}
\label{subsubsec:mackeyglass}

The Mackey-Glass system is a delay-differential equation introduced to model physiological processes such as blood-cell regulation~\cite{Mackey1977}, governed by
\begin{equation}
    \frac{dx}{dt} = \beta\,\frac{x(t-\tau_{\rm MG})}{1+x(t-\tau_{\rm MG})^{n}} - \gamma\, x(t),
    \label{eq:mackey_glass}
\end{equation}
with $n=10$, and the standard parameters $\beta = 0.2$, $\gamma = 0.1$. The first term is a nonlinear production driven by the past state and the second a linear decay of the present one, so the variable is pushed by its own history while relaxing in the present. The delay sets how much of that history takes part in the dynamics and thus the complexity of the signal. For small delays the system settles on a stable limit cycle, and chaos develops beyond $\tau_{\rm MG} \gtrsim 17$, with the attractor growing in dimension as the delay increases, so that a single parameter tunes the difficulty of the forecasting task. We generate datasets at $\tau_{\rm MG} \in \{7, 15, 17, 30, 50\}$, spanning from quasi-periodic motion, at $\tau_{\rm MG}=7$ and $15$, through the onset of chaos at $\tau_{\rm MG}=17$, to the moderate and high-dimensional chaotic attractors at $\tau_{\rm MG}=30$ and $50$. The series are rescaled to $[0,1]$ before encoding, and the dataset follows the same protocol as the memory tasks. We evaluate the forward capacity of Eq.~\eqref{eq:forward_capacity} up to a horizon of $d_{\max} = 150$ steps, with $C_{\text{tot}}^{\text{fut}}$ as the objective to maximize with BO.

\begin{figure*}[t!]
    \centering
    \includegraphics[width=1\linewidth]{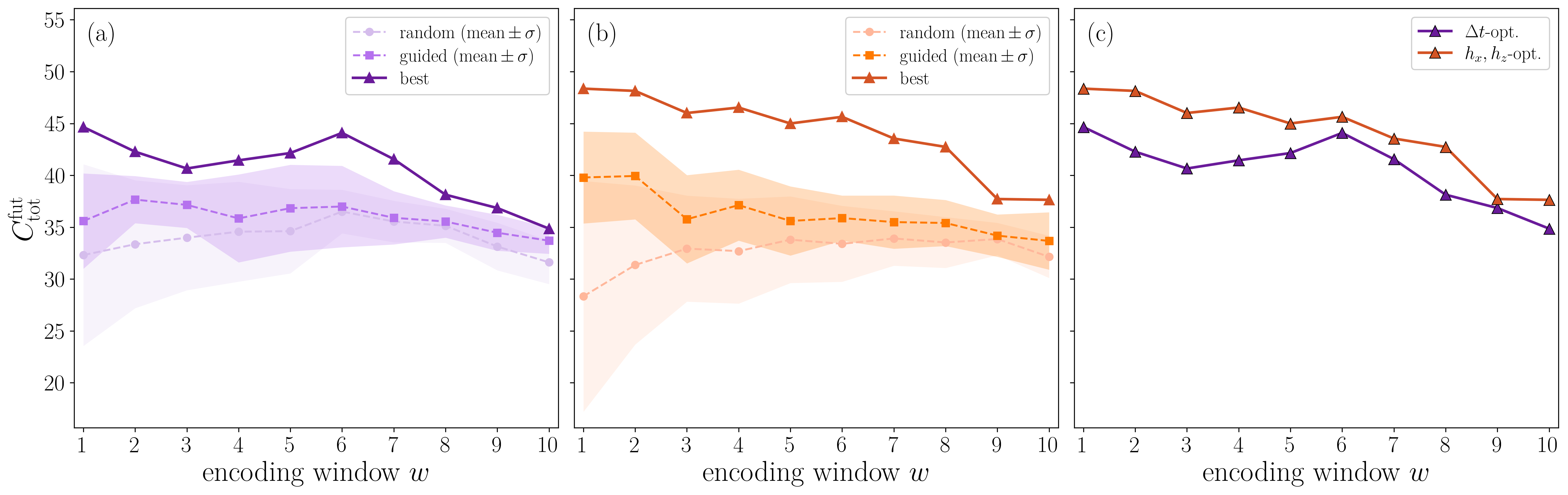}
    \caption{Santa Fe forward-prediction capacity $C_{\rm tot}^{\rm fut}$ versus encoding window $w$ for the $N=10$ reservoir, optimizing $\Delta t$ at the fixed chaotic Hamiltonian (a) and $(h_x,h_z)$ at fixed $\Delta t = 70$ (b). Markers follow Fig.~\ref{fig:STM_results}, with light markers for the random trials, dark for the guided (TPE) trials, and the solid line for the best configuration. Panel (c) overlays the two best curves. The capacity is highest at small windows and decreases toward the QELM limit, with the field-optimized search higher throughout.}
    \label{fig:SantaFe_results}
\end{figure*}

Across delays and windows, the forward capacity tracks the attractor complexity directly (Fig.~\ref{fig:Mackey-Glass_results}). In the quasi-periodic regime, at $\tau_{\rm MG} = 7$ and $15$, the reservoir reaches the maximum $C_{\rm tot}^{\rm fut} = d_{\max} = 150$ for every window, reproducing the full horizon with near-unit correlation even in the QELM limit. The signal is regular enough that the instantaneous feature map alone reconstructs it, and no quantum memory is required. At the onset of chaos, $\tau_{\rm MG} = 17$, the capacity remains close to this maximum value, around $C_{\rm tot}^{\rm fut} \simeq 148$ for both strategies and all windows, with only a slight decrease as $w$ grows. Once the attractor gains dimension, at $\tau_{\rm MG} = 30$ and $50$, the capacity degrades, and more pronouncedly toward large $w$. In both strategies the score is highest at small windows and decreases toward the QELM limit, which is the lowest at each delay. Still, the decline is much lower than in the memory tasks. At $\tau_{\rm MG} = 30$ the QELM retains most of the small-window capacity, whereas in the STM task it kept only the recall of the encoding register itself, and it is only at $\tau_{\rm MG} = 50$ that the collapse becomes comparable, with the QELM falling to $C_{\rm tot}^{\rm fut} \simeq 61$, less than half of the perfect value, so that every window below $N$ outperforms it and the margin widens with the complexity of the signal. The STM task queries lags of up to $d_{\max}=50$ steps directly, which the QELM cannot reach by construction, while forecasting draws on the history only through what it contributes to the future of the signal, a demand that grows with the attractor dimension. Reconstructing a higher-dimensional attractor requires a longer stretch of that history, and it is the memory register that supplies it, a temporal channel the QELM loses once the chain is rewritten in full.

For the two chaotic delays (Fig.~\ref{fig:Mackey-Glass_results}(c)), both strategies degrade with $w$ and approximately meet at the QELM limit, where no quantum memory register remains and the reservoir reduces to a nonrecurrent feature map whose score depends only weakly on how its dynamics were selected. Below that limit, the field-optimized search reaches higher capacities than the temporal one across the windows. On this task, fixing the evolution time and optimizing the fields lets the search explore the whole field plane, and this larger freedom reaches dynamics better matched to the long memory that forecasting the attractor demands. Extended results across all delays and windows, together with the parameters of the best configurations found, are collected in Appendix~\ref{appendix_subsec:MG_expansion}.

\subsubsection{Santa Fe task}
\label{subsubsec:santafe}

The second forecasting signal is the Santa Fe laser series~\cite{Weigend1994}, a measured record of the output intensity of a far-infrared NH$_3$ laser and a classic benchmark from the 1991 Santa Fe prediction competition. Unlike Mackey-Glass, it is recorded from an experiment rather than generated from a known equation, so it carries measurement noise on top of its deterministic dynamics, and its bursts of intensity decorrelate over a few steps. It therefore complements the previous task, testing the reservoir on a real, short-memory chaotic signal rather than a synthetic one with tunable long-range structure. We follow the same protocol as in the other tasks, now with a horizon of $d_{\max} = 70$.

The Santa Fe capacity (Fig.~\ref{fig:SantaFe_results}) follows the same pattern as the chaotic Mackey-Glass, with the score higher at small windows and decreasing toward the QELM limit, but the decline is somewhat smaller, since the QELM retains close to $80\%$ of the small-window capacity in both strategies, against roughly $60\%$ at the most complex Mackey-Glass delay $\tau_{\text{MG}} = 50$. This is consistent with the correlations of the signal, which decay over a few steps, so that the recent inputs held explicitly in the register carry most of the information needed for the prediction. Forecasting the high-dimensional Mackey-Glass attractor instead requires reconstructing a long stretch of the signal history, with relevant lags extending well beyond any register, and it is there that removing the memory qubits is most costly. As in the previous tasks, the guided bands lie above the random ones throughout, with the largest gains at small windows, where the performance depends more delicately on the configuration. The field-optimized reservoir again reaches higher capacities than the temporal one across the whole window range (Fig.~\ref{fig:SantaFe_results}(c)), with a residual gap remaining even at $w=N$. This gap, however, should be read with care, since near the QELM limit the score barely depends on the evolution time and the temporal optimum becomes ill-defined, as the analysis of the best configurations in Appendix~\ref{Appendix_subsec:results_Santa_Fe} shows. Extended results, including the per-horizon capacities and the parameters of the best configurations found, are collected there.

\subsection{Comparison with a densely connected model}
\label{subsec:comparison_models}

\begin{figure}[t!]
    \centering
    \includegraphics[width=\linewidth]{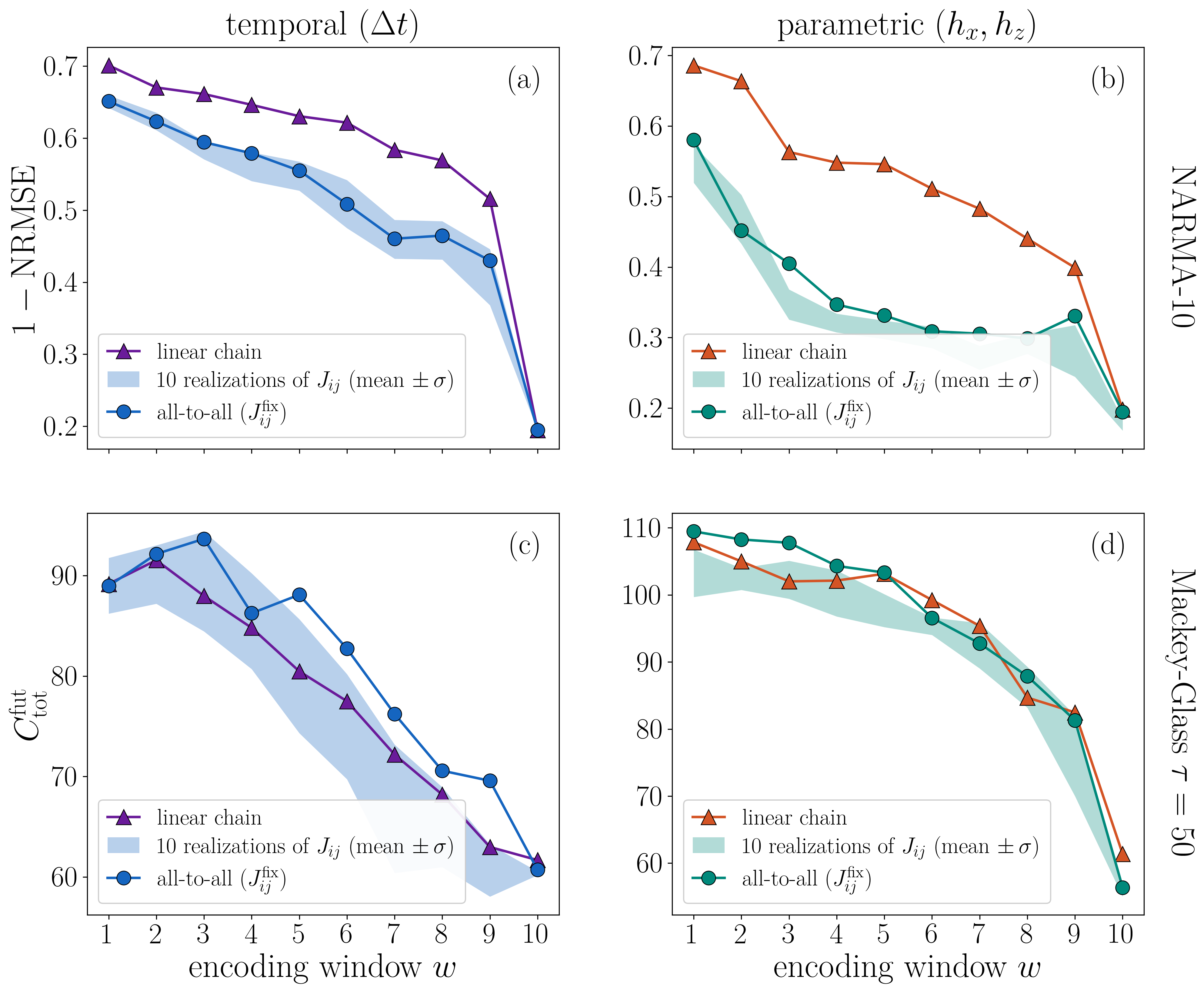}
    \caption{Comparison between the linear chain and the densely connected model on NARMA-10 (a,\,b) and Mackey-Glass at $\tau_{\rm MG} = 50$ (c,\,d), under the temporal (left) and parametric (right) optimization strategies of Sec.~\ref{subsec:BO_method}. Triangles show the best configuration of the chain at each encoding window $w$, reproduced from Figs.~\ref{fig:NARMA_results} and~\ref{fig:Mackey-Glass_results}, and circles the best configuration of the dense model optimized on the fixed realization of the couplings $J_{ij}^{\rm fix}$. The shaded band spans the mean $\pm 1\sigma$ of that configuration re-evaluated on $10$ realizations of the couplings, quantifying the sensitivity to the particular draw. On NARMA-10 the chain outperforms the dense model at every window below the QELM limit under both strategies, while on Mackey-Glass the two models perform similarly, with neither model showing a systematic advantage.}
    \label{fig:dense_comparison}
\end{figure}

The results so far show that a properly tuned nearest-neighbor chain performs well across different tasks. Whether a densely connected model would do better at the same system size is a separate question, and we address it by repeating the optimization for the model
\begin{equation}
    \label{eq:H_dense_main}
    H_{\rm dense} = \sum_{i<j} J_{ij}\, X_i X_j + h_x \sum_i X_i
    + h_z \sum_i Z_i ,
\end{equation}
with the same $N = 10$ qubits and uniform local fields, and couplings $J_{ij} \sim \mathcal{U}[-1,1]$ drawn independently for every pair. The only difference with the chain is thus the interaction term, where the nearest-neighbor coupling is replaced by an all-to-all connectivity, the construction on which quantum reservoirs are usually built~\cite{Fujii2017,Nakajima_2019,Kutvonen2020,Martinez2021}. We compare the two models on the most demanding task of each family, NARMA-10 for the memory tasks and Mackey-Glass at $\tau_{\rm MG} = 50$ for the forecasting ones. The couplings are not optimization parameters. We fix one realization ($J_{ij}^{\text{fix}}$) and run the two optimization strategies of Sec.~\ref{subsec:BO_method} exactly as for the chain. Each best configuration found is then re-evaluated on $10$ further realizations of the couplings, and the standard deviation over them, shown as the band in Fig.~\ref{fig:dense_comparison}, quantifies how sensitive the result is to the particular draw. The specific realization employed, together with the spectral characterization of the model, its timescales, and the verification of the reservoir properties of Sec.~\ref{subsec:reservoir}, are collected in Appendix~\ref{Appendix_sec:dense_model}.

Fig.~\ref{fig:dense_comparison} shows the comparison. On NARMA-10 (top panels) the chain outperforms the dense model at every window below the QELM limit, under both the temporal (a) and the parametric (b) strategies. The task requires keeping the lag $s_{k-n}$ accessible to the local readout while it is combined with the recent inputs, and the dense model, even at its own optimal evolution time, holds it there less effectively. The gap is widest under the parametric strategy, where the fixed $\Delta t = 70$ sits less than a factor of two above the Thouless time of the chain but an order of magnitude above that of the dense model (Appendix~\ref{Appendix_sec:dense_model}), so the latter is evaluated much deeper in the scrambled regime and no choice of fields recovers what the dynamics have already delocalized. On Mackey-Glass (bottom panels) the two models perform alike, with the chain lying within the band of the dense model at nearly every window. Forecasting draws on the history only through what it contributes to the future of the signal, rather than through the recall of one specific input, and the faster mixing is then no handicap. In all four panels the performance decreases as the explicit register grows and collapses at $w = N$, so the trade-off between explicit and recurrent storage is a property of the protocol rather than of the connectivity. Finally, the fixed realization generally sits at the upper edge of its own band, so the tuning is specific to the draw it was performed on, although no conclusion of the comparison depends on it. The optimal parameters found for the dense model are collected in Appendix~\ref{app:extended_results_dense_model}.

The dense connectivity thus brings no advantage for the tasks studied here. The two models perform alike on the forecasting task, and on NARMA-10, where keeping a specific past input accessible is essential, the locality of the chain becomes an asset. In the chain an input spreads gradually along the lattice. Its imprint remains concentrated on few qubits, within the span of the one- and two-body observables of the readout. On the other hand, in the dense model every coupling distributes it over all ten spins at once, moving it into higher-order correlations that no measured observable captures, regardless of the evolution time. A nearest-neighbor chain with uniform fields is therefore not a compromise but a sufficient architecture for temporal processing at this size.

\section{Conclusions}
\label{sec:conclusions}

In this work we considered a one-dimensional mixed-field Ising chain with nearest-neighbor interactions as a quantum reservoir for time series processing. By varying the length of the input-encoding window we interpolated continuously between the QRC and the QELM paradigms, evaluating every window at its own optimum via Bayesian optimization. We found that the recurrent quantum memory is essential when the target reaches far into the past and dispensable when the relevant history is short, where the QELM limit already suffices. In every case, the best reservoirs operate at moderate scrambling, and a direct comparison at equal system size showed that dense all-to-all connectivity brings no advantage under the measurement scheme employed here. We conclude that quantum reservoir computing for temporal tasks can be realized with a nearest-neighbor spin chain by tuning the local fields or the evolution time, contrary to the common construction based on dense connectivity.

Within this chain, the encoding register of $w$ qubits sets how the past is stored, trading the recurrent memory of the quantum memory register for an explicit record of the recent inputs. Sweeping $w$ from $1$ to $N$ identifies which problems genuinely require quantum memory. Whenever the task reaches past the encoding register, as in recalling distant inputs, matching a high-order NARMA target, or reconstructing a high-dimensional attractor, performance peaks at small windows $w$ and collapses at the QELM limit. When the relevant correlations of the signal are short instead, the dependence on the window flattens. The quasi-periodic Mackey-Glass regimes are forecast with near-unit correlation across all windows, and on the Santa Fe series, a measured chaotic signal that decorrelates over a few steps, the nonrecurrent QELM retains close to $80\%$ of the best capacity. For such problems, a time-delayed quantum feature map fed with a short
explicit record suffices, and quantum memory is no longer the critical
resource.

Across the tasks studied, the two optimization strategies considered converge independently on the same regime. The temporal search selects evolution times an order of magnitude below the Thouless time of the chaotic dynamics, and the parametric search settles on weakly chaotic Hamiltonians, both realizing a moderate scrambling that mixes each new input with the stored history without delocalizing it into many-body correlations inaccessible to the local readout. Because the two strategies produce reservoirs of comparable quality across most tasks, one can fix whichever parameter is harder to adjust on a given platform and optimize the other. On current hardware, the evolution time is typically the easier parameter to tune.

The comparison with the densely connected model tests this conclusion directly. The two models perform alike on forecasting, while on NARMA-10 the chain leads, its slower spreading keeping past inputs within reach of the one- and two-body observables of the readout, whereas the dense couplings move them into higher-order correlations that no measured observable captures. This advantage is thus tied to the measured observable set, and a readout with access to higher-order correlators could in principle recover the delocalized information, at the cost of an observable set growing combinatorially with the correlation order.

A further observation emerging from our results is that the relevance of the precise reservoir dynamics is itself strongly task- and regime-dependent. In some cases, good performance is confined to a relatively narrow range of Hamiltonian parameters or evolution times, whereas in others the performance landscape becomes broad and markedly different reservoir configurations yield essentially the same score. This behavior is visible, for instance, in the widening of the near-optimal regions for several tasks and encoding windows, but its microscopic origin is not yet clear. Related Hamiltonian insensitivity has recently been observed in QELMs for classification,  where a local integrable $XX$ model reaches, beyond a characteristic evolution time, a performance plateau comparable to that obtained with Haar-random unitaries, despite their radically different dynamical complexity \cite{delorenzis2026entanglementclassicalsimulabilityquantum}. Together, these observations suggest that high learning performance does not always require fine control over the microscopic Hamiltonian, while in other regimes such tuning remains important. Identifying which properties of the task, encoding, readout, and dynamics determine this robustness therefore remains an interesting open question, with direct implications for how precisely quantum reservoirs need to be engineered and calibrated.

Our results were obtained under ideal conditions, using exact expectation values computed from the density matrix. Because the input-dependent signal that survives in the stationary trajectory can be small, a critical next step is to quantify how much of this performance survives under finite measurement shots and realistic device noise~\cite{Mujal_2023,Ahmed2024}, a budget that the input-separability analysis already bounds and that the dilution of the per-site signal makes more demanding for the dense model than for the chain. On a broader level, our study identifies the optimal dynamical regimes for temporal processing, either a strongly chaotic Hamiltonian evolved for short times or a weakly chaotic one evolved for long times. Whether such an \textit{a priori} selection, guided by scalable diagnostics~\cite{domingo2026diagnosingquantumreservoirsscale}, can replace the costly task-by-task optimization performed here remains open. This study is finally limited to $N = 10$ qubits, and whether the reported trade-offs persist at larger sizes, where the recurrent state grows exponentially while both the register and the measured observable set grow only polynomially, remains to be established, a question made timely by recent analog realizations of quantum reservoirs~\cite{Mujal2024_QuEra}.

\vspace{-0.5cm}
\begin{acknowledgments}
\vspace{-0.25cm}
The authors thank Ana Palacios for insightful discussions. The authors acknowledge the computational resources provided by the Barcelona Supercomputing Center in MareNostrum 5 through the RES project INNO-2026-1-0004, and funds from MICIU/AEI/10.13039/501100011033/ FEDER, UE.
\end{acknowledgments}

\textbf{Data Availability:} Data are available from the authors upon reasonable request.

\bibliography{references}

\newpage
\appendix
\section{Calculation of the Heisenberg time}
\label{Appendix_sec:Calculation of the Heisenberg time}

The Heisenberg time is the longest dynamical scale of a quantum system, set by the inversemean level spacing,
 \begin{equation}
    t_H = \frac{1}{\overline{\delta E}},
    \label{eq:tH_def}
\end{equation}
with $\hbar = 1$. It marks the time at which the dynamics resolve individual energy levels, beyond which the SFF saturates to its plateau, as shown in Fig.~\ref{fig:ThoulesTime_SFF}(a). Since the level spacing is not uniform across the spectrum but scales as the inverse of the local density of states, the average in Eq.~\eqref{eq:tH_def} must be taken over a definite energy window. We evaluate it in the bulk of the spectrum, where the majority of the levels lie and where all the spectral diagnostics of the main text are computed. In this region the estimate becomes analytic, following the protocol of Ref.~\cite{_untajs_2020}. For many-body systems with few-body interactions the density of states is well approximated by a Gaussian centered at the mean energy, whose width is the standard deviation of the spectrum. We compute this width as the ensemble-averaged variance
\begin{equation}
    \Gamma_0^2 = \frac{1}{\mathcal{D}}\braket{\mathrm{Tr}(\tilde{H}^2)}
    - \frac{1}{\mathcal{D}^2}\braket{\mathrm{Tr}(\tilde{H})}^2,
    \label{eq:gamma0}
\end{equation}
where $\braket{\bullet}$ denotes the average over the same ensemble used for the Thouless time in Sec.~\ref{subsec:dynamic}, namely $n_r=2000$ realizations of the normalized Hamiltonian with fields drawn uniformly from a window of half-width $\epsilon_{\mathrm{dis}}=0.15$ around the chaotic point $(\braket{h_x},\braket{h_z}) = (0.5,1.05)$.

For the Gaussian density of states, centered at $\bar E = 0$ for the traceless normalized Hamiltonian, the fraction of levels lying within one standard deviation above the spectral center is
\begin{equation}
\begin{aligned}
    \chi &= \frac{1}{\sqrt{2\pi}\,\Gamma_0} \int_0^{\Gamma_0}
    e^{-E^2/(2\Gamma_0^2)}\, dE
    = \frac{1}{2}\,\mathrm{erf}\!\left(\frac{1}{\sqrt{2}}\right)
    \approx 0.3413.
    \label{eq:chi}
\end{aligned}
\end{equation}
This is the usual fraction of a Gaussian distribution lying between its mean and one standard deviation. The window thus contains $\chi\mathcal{D}$ levels spread over an energy range $\Gamma_0$, so the bulk mean level spacing and the Heisenberg time read
\begin{equation}
    \overline{\delta E} = \frac{\Gamma_0}{\chi\, \mathcal{D}},
    \qquad
    t_H = \frac{\chi\, \mathcal{D}}{\Gamma_0}.
    \label{eq:tH_final}
\end{equation}
Here $\mathcal{D}$ is the dimension of the symmetry sector in which the spectrum is computed. As in the main text, we work in the reflection-parity sector $+1$, whose dimension exceeds half of the Hilbert space because the reflection-symmetric basis states, those invariant under the reflection $i\to N+1-i$, have no antisymmetric partner and belong entirely to it, giving $\mathcal{D}=(2^N+2^{N/2})/2 = 528$ for $N=10$. The only quantity left to evaluate numerically is the spectral width, for which the ensemble average yields $\Gamma_0 = 1.013 \pm 0.001$, close to unity since it is computed with the normalized Hamiltonian $\tilde{H} = H/\sigma_H$. Inserting these values into Eq.~\eqref{eq:tH_final} gives
\begin{equation}
    t_H = \frac{\chi\,\mathcal{D}}{\Gamma_0} \approx 178,
\end{equation}
with a relative uncertainty below $0.1\%$, negligible for all practical purposes. This value enters the SFF analysis of the main text through the unfolding procedure, which rescales each spectrum to unit mean level spacing so that the scaled Heisenberg time is $\tau_H=1$ by construction, and physical times are recovered as $t=\tau\, t_H$. In particular, the Thouless time quoted in the main text follows as $t_{\mathrm{Th}}=\tau_{\mathrm{Th}}\, t_H$.

\section{Detailed calculation of the Thouless time}
\label{Appendix_sec:Detailed calculation of the Thouless time}

In Sec.~\ref{subsec:dynamic} we defined the Thouless time as the onset beyond which the connected SFF follows the GOE ramp. Here we give their definitions, the full extraction pipeline, and the error analysis behind the reported value.

The SFF is the Fourier transform of the spectral two-point correlations. In the chaotic regime it follows the universal GOE prediction
\begin{equation}
    K_{\text{GOE}}(\tau) =
    \begin{cases}
    \label{eq:SFFGOE}
        2\tau - \tau \ln(1+2\tau), & \tau \le 1, \\
        2 - \tau \ln \left(\frac{2\tau+1}{2\tau-1}\right), & \tau > 1,
    \end{cases}
\end{equation}
where $\tau = t/t_H$ is the rescaled time. The ramp reflects the level repulsion of the chaotic spectrum, which suppresses the correlations that survive at intermediate times, while the plateau sets in once the discrete levels are individually resolved, at $\tau_H = 1$. We use the connected SFF,
\begin{equation}
\begin{aligned}
\label{eq:SFFMFIM}
    K_c(\tau) &= \frac{1}{Z}\Bigg(\Bigg<\left| \sum_{\alpha =1}^{\mathcal{D}} \rho(\varepsilon_{\alpha})e^{-i \varepsilon_{\alpha}\tau}\right|^2 \Bigg> \\
    &- \frac{A}{B} \left| \Bigg<\sum_{\alpha = 1}^{\mathcal{D}}\rho(\varepsilon_{\alpha})e^{-i \varepsilon_{\alpha}\tau}\Bigg>\right|^2\Bigg),
\end{aligned}
\end{equation}
where $\{\varepsilon_{\alpha}\}$ is the unfolded spectrum, $\mathcal{D}$ the dimension of the symmetry sector, $\braket{\bullet}$ the average over disorder realizations, and $\rho(\varepsilon)$ the Gaussian filter of Eq.~\eqref{eq:gaussian_filter} that suppresses the spectral edges. Note that in this convention, where $t_H = 1/\overline{\delta E}$ and $\tau=t/t_H$, the factor of $2\pi$ typically present in the SFF phase is omitted to directly match the physical time units used in the text. The second term subtracts the disconnected part of the SFF, the contribution from the smooth, non-universal shape of the filtered spectral density rather than from genuine correlations between levels; removing it lets the ramp follow the random-matrix prediction cleanly from short times. The constants
\begin{equation}
    A = \Bigg<\left|\sum_{\alpha=1}^{\mathcal{D}}\rho(\varepsilon_{\alpha})\right|^2\Bigg>, \hspace{0.5cm} B = \left|\Bigg<\sum_{\alpha=1}^{\mathcal{D}}\rho(\varepsilon_{\alpha})\Bigg>\right|^2,
\end{equation}
and the normalization $Z = \braket{\sum_{\alpha}|\rho(\varepsilon_{\alpha})|^2}$ ensure that $K_c(\tau)$ vanishes as $\tau\to0$ and saturates to unity at long times, as shown in Fig.~\ref{fig:ThoulesTime_SFF}(a). We locate the Thouless time from the deviation
\begin{equation}
\label{eq:deltaK}
    \Delta K(\tau) = \left|\log_{10} \frac{K_c(\tau)}{K_{\text{GOE}}(\tau)} \right|,
\end{equation}
starting from large $\tau$, where $K_c$ and $K_{\text{GOE}}$ agree, and moving inward to the first time at which $\Delta K(\tau)$ exceeds the threshold $\epsilon_{th} = 0.01$, as in Fig.~\ref{fig:ThoulesTime_SFF}(b).

The spectra belong to the reflection-parity sector $+1$ of the normalized Hamiltonian at the chaotic point, with the two fields $(h_x, h_z)$ drawn uniformly from a window of half-width $\epsilon_{\mathrm{dis}} = 0.15$ around $(0.5, 1.05)$. Each sample is drawn independently for each of the $n_r = 2000$ realizations, the same ensemble used for the Heisenberg time in Appendix~\ref{Appendix_sec:Calculation of the Heisenberg time}.

\begin{figure}[h!]
    \centering
    \includegraphics[width=1\linewidth]{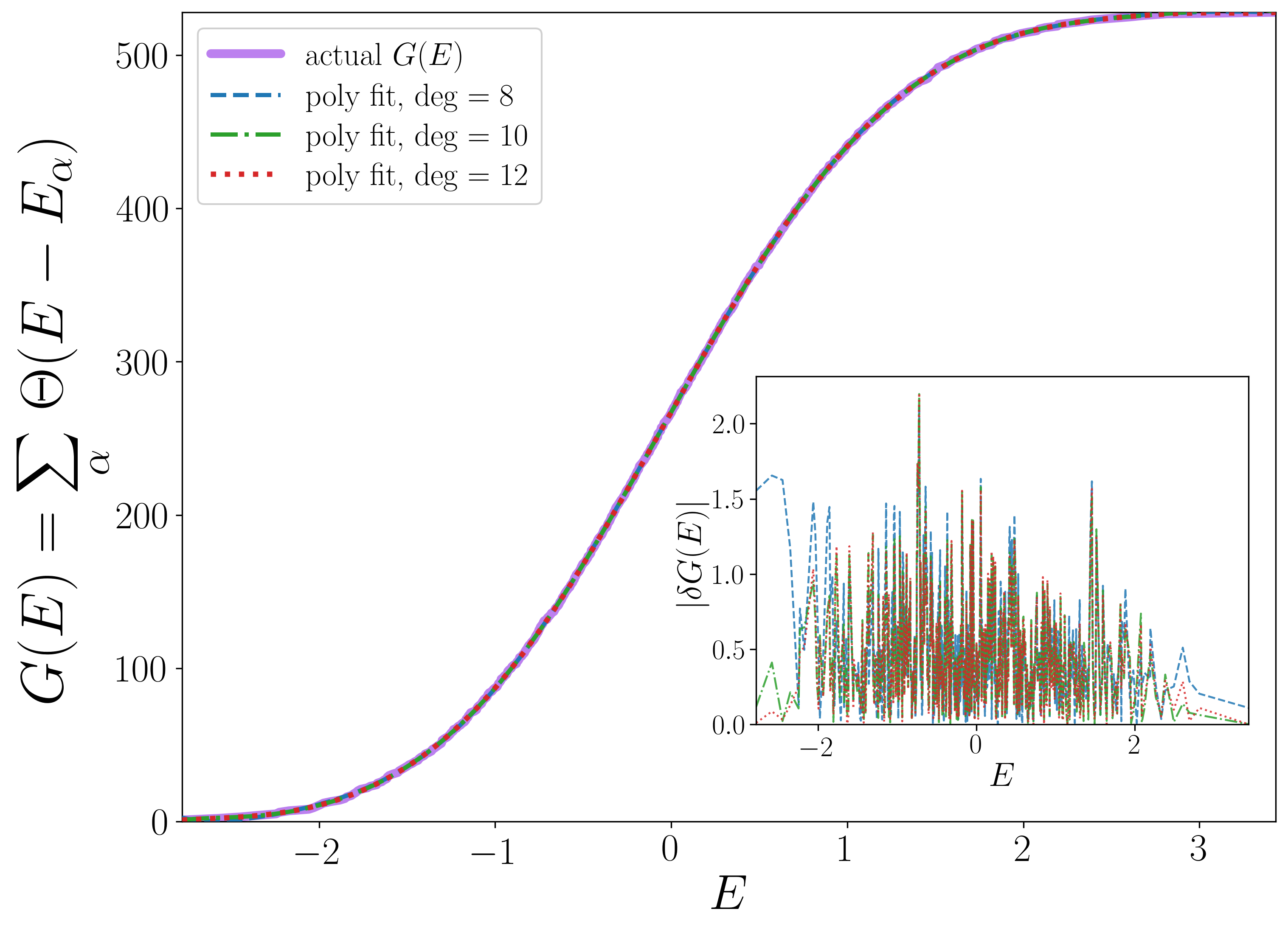}
    \caption{Cumulative spectral function $G(E)$ for a single disorder realization at the chaotic point (solid purple), together with the polynomial fits $\bar g_n(E)$ used for the unfolding, of degree $n=8$ (dashed blue), $n=10$ (dash-dotted green) and $n=12$ (dotted red). The three fits are indistinguishable on this scale. The inset shows the absolute residuals $|\delta G(E)| = |G(E) - \bar g_n(E)|$ for the three degrees, with the same color code, which remain of the order of a single level across the whole spectrum, showing that the fits capture the smooth part of the density of states without following its level-to-level fluctuations. The slightly larger residuals of the lowest degree near the spectral edges fall outside the Gaussian filter of Eq.~\eqref{eq:gaussian_filter} and do not enter the SFF.}
    \label{fig:unfolding_appendix}
\end{figure}

To compute the SFF, the first step is the spectral unfolding of the energy levels, which removes the secular variation of the level spacing discussed in Appendix~\ref{Appendix_sec:Calculation of the Heisenberg time}, since the SFF is only universal once this smooth component is taken out. For each realization we build the cumulative spectral function $G(E) = \sum_{\alpha} \Theta(E-E_{\alpha})$ and fit its smooth part with a polynomial $\overline{g}_n(E)$ of degree $n$, so that $G(E) = \overline{g}_n(E)+\delta G(E)$, where $\delta G(E)$ contains the level-to-level fluctuations. As shown in Fig.~\ref{fig:unfolding_appendix}, the fit captures the full shape of $G(E)$ equally well for all the degrees considered, with residuals $|\delta G(E)|$ of the order of a single level out of the  $\mathcal{D}=528$ in the sector giving a relative deviation below $10^{-2}$. We select $n=12$ and check that this choice is immaterial by repeating the full pipeline for $n\in\{8,10,12\}$, which shifts $t_{\mathrm{Th}}$ by less than $\pm 0.2$.

The unfolding is least reliable near the spectral edges, whose correlations are in any case non-universal, so each level is weighted by the Gaussian filter entering Eq.~\eqref{eq:SFFMFIM},
\begin{equation}
    \rho(\varepsilon_\alpha) = \exp\!\left[-\frac{(\varepsilon_\alpha-\bar{\varepsilon})^2}{2(\eta\Gamma)^2}\right],
    \label{eq:gaussian_filter}
\end{equation}
with $\bar{\varepsilon}$ and $\Gamma$ the mean and standard deviation of the unfolded spectrum. We set $\eta = 0.5$, so that levels within one standard deviation of the spectral center, about $68\%$ of the spectrum, carry a filter weight above $e^{-2}$, while the non-universal edges are smoothly suppressed. This filter plays the same role as the hard cut used for the gap-ratio statistics of Fig.~\ref{fig:levelstats}, where the lowest and highest $25\%$ of each spectrum were simply discarded, but the two situations demand different tools. The gap ratio is a local quantity, built from consecutive spacings, so an abrupt truncation removes edge levels without disturbing the ratios that remain. The SFF is instead a coherent sum of phases over the whole spectrum, a Fourier transform of the filtered level density, and an abrupt rectangular window would imprint spurious oscillations on $K_c(\tau)$ through its slowly decaying Fourier tails. A smooth filter such as the Gaussian introduces no such artifacts~\cite{_untajs_2020}.

With the unfolded and filtered spectra in hand, we evaluate the amplitude $z(\tau)=\sum_\alpha \rho(\varepsilon_\alpha)\,e^{-i \varepsilon_\alpha \tau}$ of each realization on a logarithmically spaced grid of $5000$ points spanning $\tau\in[1/(2\pi\mathcal{D}),\,5]$, and assemble the connected SFF of Eq.~\eqref{eq:SFFMFIM} by averaging over the $n_r=2000$ realizations. The resulting curve is smoothed with a $200$-point centered running mean to remove the residual point-to-point noise of the ramp, giving the result shown in Fig.~\ref{fig:ThoulesTime_SFF}(a). We then form the deviation $\Delta K(\tau)$ of Eq.~\eqref{eq:deltaK} and, starting from the largest $\tau$, where $K_c(\tau)$ and $K_{\text{GOE}}(\tau)$ agree, we move inward and locate $\tau_{\mathrm{Th}}$ at the first crossing of the threshold $\epsilon_{\text{th}}=0.01$, as shown in Fig.~\ref{fig:ThoulesTime_SFF}(b). The physical Thouless time follows as $t_{\mathrm{Th}}=\tau_{\mathrm{Th}}\,t_H$, with $t_H\simeq 178$ from Appendix~\ref{Appendix_sec:Calculation of the Heisenberg time}.

Unlike the Heisenberg time, which is a closed-form function of the spectral moments and carries a negligible uncertainty, the Thouless time is read off the shape of a finite sample average through a threshold crossing, and therefore fluctuates. Two independent sources contribute to its error. The dominant one is statistical. At finite $n_r$ the ramp is noisy, since the SFF only self-averages in the limit of infinitely many realizations, and the threshold crossing sits on a nearly flat portion of $\Delta K(\tau)$, so small vertical fluctuations of the curve translate into a sizeable horizontal scatter of $\tau_{\mathrm{Th}}$. We quantify it with a nonparametric bootstrap over the disorder realizations, resampling the $n_r$ spectra with replacement $200$ times, rebuilding $K_c(\tau)$ and re-extracting $\tau_{\mathrm{Th}}$ for each resample, and taking the standard deviation of the resulting values, which, converted to physical units through $t_H$, gives $\pm 5.3$. The second source is systematic and comes from the unfolding degree, whose spread over $n\in\{8,10,12\}$ is only $\pm 0.2$ in the same units, more than an order of magnitude smaller. We therefore report
\begin{equation}
    t_{\mathrm{Th}} = 42.1 \pm 5.3,
\end{equation}
dominated by the finite-ensemble statistical error.

\section{Echo state property and input separability}
\label{Appendix_sec:ESP_and_input_dependence}
In this Appendix we verify numerically the dynamical properties of the reservoir map established in Sec.~\ref{subsec:reservoir}. First we check the ESP, tracking how two copies of the reservoir prepared in different initial states converge to the same trajectory when driven by the same input sequence, and quantifying how the number of steps needed to erase the initial state depends on the Hamiltonian and the evolution time. We then check input separability by exchanging the roles in that test, so that the two copies now start from the same initial state and are driven by different input sequences, and we confirm that their trajectories remain apart instead of merging. Together, the two tests show that after the washout the reservoir trajectory is determined by the input history alone.

\subsection{Verification of the Echo State Property}
\label{Appendix_sec:Convergence_property}

To test the ESP we prepare one copy of the reservoir in the reference state $\rho_A^{(0)}
= \ket{0}\!\bra{0}^{\otimes N}$ used throughout this work and a second copy in a random product state $\rho_B^{(0)}$, with the single-qubit states drawn as in Sec.~\ref{subsec:dynamic}, and drive both with the same input sequence, the same normalized Hamiltonian $\tilde{H}$ and the same evolution time $\Delta t$. At each step, after the evolution of Eq.~\eqref{eq:reservoir_evolution}, we track the distance between them,
\begin{equation}
    \label{eq:esp_distance}
    D_k = \lVert \rho_A^{(k)} - \rho_B^{(k)} \rVert_F,
\end{equation}
with $\lVert \bullet \rVert_F$ the Frobenius norm, averaged over $200$ random draws of $\rho_B^{(0)}$. Both copies are propagated by the same reservoir map, composed of the reset of Eq.~\eqref{eq:reservoir_update} and the unitary evolution of Eq.~\eqref{eq:reservoir_evolution}. While only the trace distance is mathematically guaranteed to contract under arbitrary channels, across all configurations tested we observe the Frobenius distance to decrease monotonically,
\begin{equation}
    D_{k+1} \leq D_k,
    \label{eq:esp_contraction}
\end{equation}
and the ESP is fulfilled when it becomes negligible. How fast it does is set by the composition with the scrambling dynamics, as discussed in Sec.~\ref{subsec:reservoir}. In practice we declare the initial state washed out once $D_k$ falls below $\epsilon = 10^{-3}$, after which it keeps decreasing down to machine precision. Throughout this Appendix we consider the encoding window $w=1$, which is the worst case scenario and therefore the slowest to converge, since tracing out more qubits at each step erases more information from the reservoir. The washout length estimated at $w=1$ thus upper-bounds the requirement for any $w>1$.

\begin{figure}[t]
    \centering
    \includegraphics[width=1\linewidth]{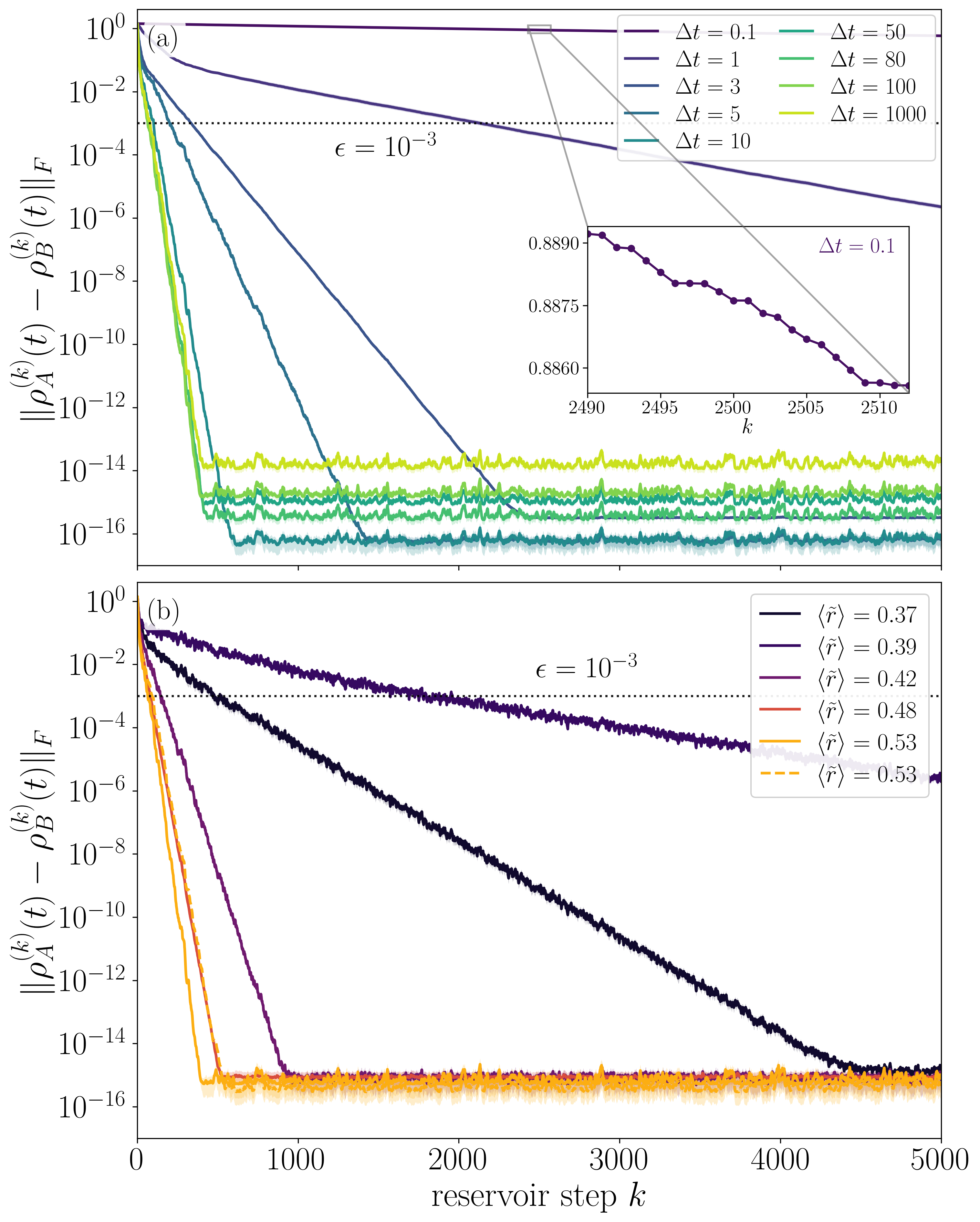}
    \caption{Convergence to the echo-state property, measured by the Frobenius distance $D_k = \lVert \rho_A^{(k)} - \rho_B^{(k)} \rVert_F$ between two reservoir copies initialized in the reference state and in a random product state, driven by the same input sequence at $w = 1$ and averaged over $200$ draws of $\rho_B^{(0)}$. The dotted line marks the washout threshold $\epsilon = 10^{-3}$. (a) Fixed chaotic Hamiltonian $(h_x, h_z) = (0.5, 1.05)$ for different evolution times $\Delta t$, with the inset zooming into $\Delta t = 0.1$, where the distance still decreases at every step but at a vanishingly small rate. (b) Fixed $\Delta t = 70$ across the integrable-to-chaotic crossover, labeled by the mean gap ratio $\langle\tilde{r}\rangle$.}
    \label{fig:esp_appendix}
\end{figure}

Figure~\ref{fig:esp_appendix}(a) shows the convergence at the chaotic point $(h_x,h_z)=(0.5,1.05)$ for evolution times spanning four orders of magnitude. For $\Delta t \gtrsim 10$ the two copies meet machine precision within a few hundred steps, while shorter evolution times slow the convergence down dramatically. The reason is that between resets the dynamics must propagate the erasure from the input qubit into the memory qubits. When $\Delta t$ is too short, the evolution barely correlates the input qubit with the rest of the chain before the next reset, and the information stored in the memory qubits survives almost untouched, so the initial condition is erased only after many more steps. In the extreme case $\Delta t = 0.1$ the distance appears flat on the scale of the main panel, but the inset confirms that it still decreases at every step, only at a rate so small that the washout becomes impractically long. Panel (b) fixes $\Delta t = 70$ and varies the Hamiltonian across the integrable-to-chaotic crossover, labeled by its mean gap ratio $\langle \tilde{r} \rangle$. The convergence rate broadly increases with the degree of chaos. The two GOE Hamiltonians ($\langle\tilde{r}\rangle=0.53$), with equal gap ratio but different fields, converge fastest, confirming that the rate is set by $\langle\tilde{r}\rangle$ rather than by the microscopic parameters, and both are erased within a few hundred steps. The near-Poissonian Hamiltonians are the slowest, lagging by orders of magnitude, although the ordering is no longer strict near the integrable regime, where $\langle\tilde{r}\rangle=0.37$ converges somewhat faster than $\langle\tilde{r}\rangle=0.39$. There the gap ratio has essentially saturated at its Poisson value and no longer discriminates between Hamiltonians, so the residual ordering reflects the particular fields rather than a difference in the degree of chaos. In every case the distance keeps decreasing steadily and eventually reaches machine precision, confirming that the echo-state property holds across the whole crossover, with the dynamics setting only how fast it is attained.

The results of Fig.~\ref{fig:esp_appendix} motivate the washout length used in the main text. At the chaotic point, every evolution time down to $\Delta t = 1$ crosses the washout threshold well before $N_{\mathrm{wo}} = 3000$ steps, the slowest after a few hundred, and only the extreme $\Delta t = 0.1$ fails to converge within the simulated window. At the fixed $\Delta t = 70$ of the parametric strategy, the threshold is met across the whole crossover, the chaotic Hamiltonians crossing it within a few hundred steps and the most integrable ones later but still within the washout. The value $N_{\mathrm{wo}} = 3000$ thus guarantees a fully erased initial condition for both strategies over the configurations the searches explore.

\subsection{Verification of input separability}
\label{Appendix_sec:Separability}

To test input separability we exchange the roles of the previous experiment. Both copies now start from the reference state, $\rho_A^{(0)} = \rho_B^{(0)} = \ket{0}\!\bra{0}^{\otimes N}$, and are driven by two independent input sequences drawn from the same uniform distribution, with the same Hamiltonian and evolution time for both. We track the distance between the reduced states of the memory qubits,
\begin{equation}
    D_k^{\mathrm{mem}} = \lVert \mathrm{Tr}_w[\rho_A^{(k)}]
    - \mathrm{Tr}_w[\rho_B^{(k)}] \rVert_F.
    \label{eq:sep_distance}
\end{equation}
Tracing out the encoding register removes the trivial contribution of the new injected inputs, which differ by construction, so that Eq.~\eqref{eq:sep_distance} examines only the memory carried by the memory register. Input separability demands the complementary behavior, since the contraction that erases the initial condition must not also erase the differences between input histories. The distance starts at zero, since the two copies share the initial state, and instead of decaying it grows and keeps fluctuating around a finite nonzero value, the fluctuations reflecting that each new input continues to drive the two copies apart.

\begin{figure}[t]
    \centering
    \includegraphics[width=1\linewidth]{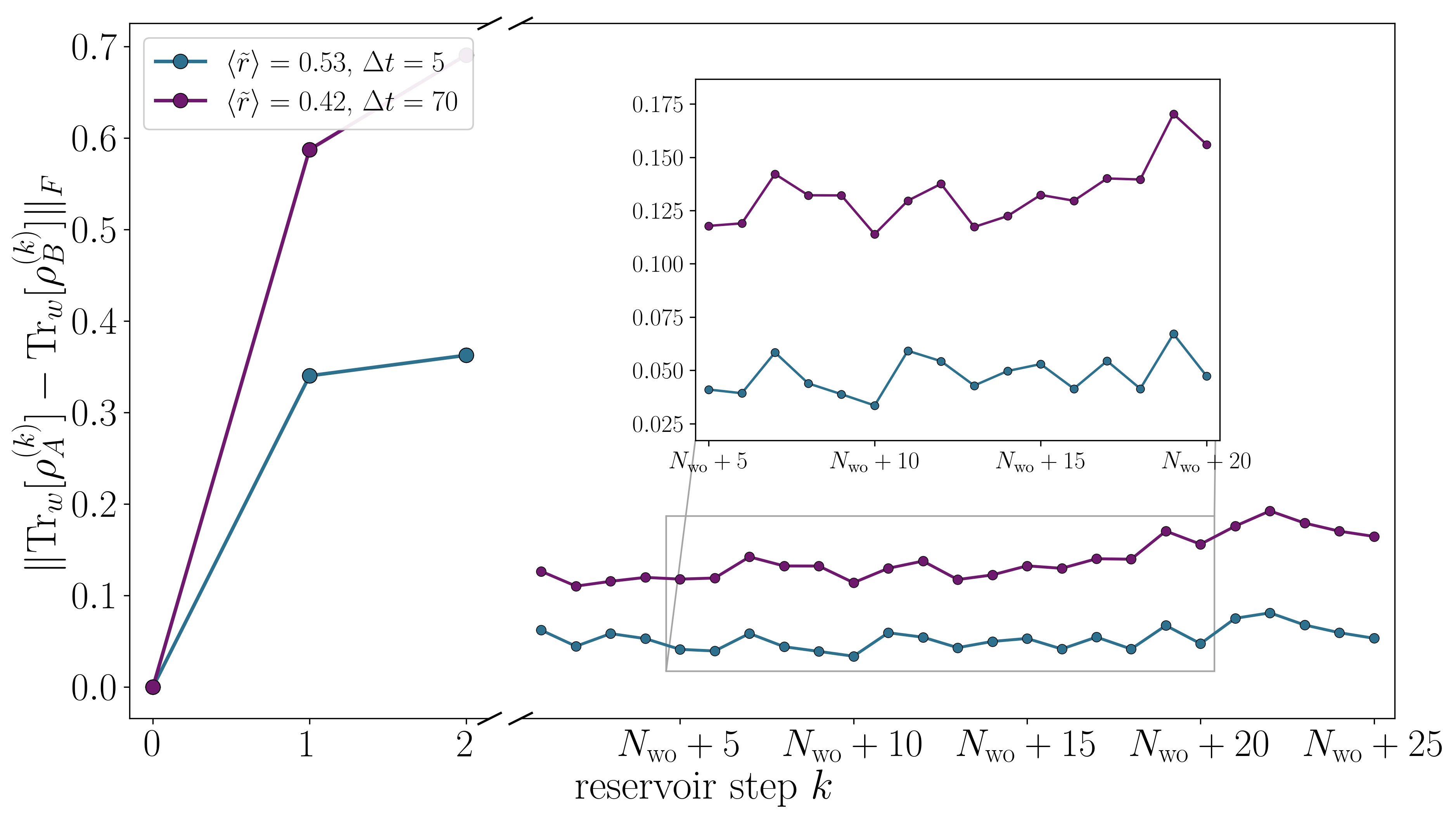}
    \caption{Verification of input separability, measured by the Frobenius distance $D_k^{\mathrm{mem}}$ of Eq.~\eqref{eq:sep_distance} between the memory qubits reduced states of two reservoir copies initialized in the same reference state and driven by different input sequences at $w=1$. Two configurations are shown, the chaotic Hamiltonian $(h_x,h_z)=(0.5,1.05)$ at $\Delta t = 5$ (blue) and a weakly chaotic one with $\langle\tilde r\rangle = 0.42$ at $\Delta t = 70$ (purple). In both cases the distance grows from zero and then fluctuates around a finite value instead of decaying, the signature of a stationary trajectory that remains input-driven, in contrast with the collapse of Fig.~\ref{fig:esp_appendix}.}
    \label{fig:separability_appendix}
\end{figure}

Figure~\ref{fig:separability_appendix} shows this distance for two representative configurations, the chaotic Hamiltonian at $(h_x,h_z)=(0.5,1.05)$ with $\Delta t = 5$, close to the optimal operating point of the memory and forecasting tasks, and a weakly chaotic Hamiltonian with $\langle\tilde r\rangle = 0.42$ at $\Delta t = 70$, representative of the winners of the parametric search. In both cases the distance starts at zero, rises within the first two steps as soon as the two input sequences begin to differ, and thereafter fluctuates around a finite value without decaying. The inset zooms into the post-washout region, where the distance keeps oscillating well above zero rather than settling to a constant, the fluctuations reflecting that each new input continues to drive the two copies apart. This confirms that the stationary trajectory remains input-driven, so that distinct input histories are mapped to distinguishable reservoir states and the map is input separable.

The distance is, however, small. The input-dependent signal that survives in the memory qubits is of the order of $5\times 10^{-2}$ for the chaotic configuration and $10^{-1}$ for the weakly chaotic one, in Frobenius distance. Resolving such differences from measured statistics would require a large number of shots, which adds no difficulty in this work, where all features are exact expectation values computed from the density matrix, but would become a relevant overhead in any finite-shot implementation.

\section{Bayesian Optimization: Tree-Structured Parzen Estimator (TPE) algorithm}
\label{appendix_sec:tpe}

Locating the best-performing dynamics at each encoding window is a black-box optimization problem. Writing $x$ for the configuration, $x=(h_x,h_z)$ in the parametric search and $x=\Delta t$ in the temporal one, and $f(x)$ for the task score, every evaluation of $f$ requires simulating a full reservoir trajectory and training the readout, no gradient is available, and the landscape is nonconvex. With a budget of a few hundred evaluations per study, a grid fine enough to resolve the structure of $f$ is unaffordable, while random search keeps sampling regions that earlier trials have already shown to be poor. Bayesian Optimization (BO)~\cite{jones1998efficient,Shahriari2016} improves on both by letting every completed trial inform the choice of the next one.

Each BO iteration combines two ingredients. A surrogate model, fit to all completed trials, provides a probabilistic estimate of $f$ across the search space, and an acquisition function turns that estimate into a decision. The decision is taken by scoring candidates by how much an evaluation is expected to pay off and balancing the exploitation of regions predicted to score well against the exploration of regions where the surrogate remains uncertain. The candidate maximizing the acquisition is evaluated with the full simulation, the outcome is appended to the record, and the surrogate is refit. Our sampler is built on the Expected Improvement (EI)~\cite{jones1998efficient}, which for a maximization problem reads
\begin{equation}
    \label{eq:Expected_Improvement}
    \mathrm{EI}(x) = \int_{y^*}^{\infty} (y-y^*) p(y|x) dy ,
\end{equation}
the expected margin by which a trial at $x$ would exceed a reference score $y^*$ under the surrogate prediction $p(y|x)$.

The Tree-structured Parzen Estimator (TPE)~\cite{Bergstra2011,watanabe2026treestructuredparzenestimatorunderstanding} evaluates Eq.~\eqref{eq:Expected_Improvement} by modeling the configurations given the score rather than the score given the configurations. The completed trials are split by their scores into a good set, the top fraction $\gamma$, and a bad set containing the rest, with $y^*$ the splitting threshold, so that $\gamma = P(y \geq y^*)$. A Parzen (kernel-density) estimator is then fit to the configurations of each group,

\begin{equation}
\label{eq:BO_probability_TPE}
    p(x|y)=
    \begin{cases}
        \ell(x), \quad y\geq y^*,\\
        g(x), \quad y<y^* .
    \end{cases}
\end{equation}

Under this model the acquisition of Eq.~\eqref{eq:Expected_Improvement} becomes a monotonically increasing function of the density ratio~\cite{Bergstra2011},

\begin{equation}
    \label{eq:EI_with_density_ratio}
    \mathrm{EI}(x) \propto
    \left(\gamma+(1-\gamma)\frac{g(x)}{\ell(x)}\right)^{-1},
\end{equation}
so that maximizing the acquisition reduces to maximizing $\ell(x)/g(x)$. At each iteration the sampler draws a set of candidates from $\ell(x)$ and proposes the one with the largest ratio, the point where the good configurations concentrate most strongly relative to the bad ones. Replacing the surrogate over $f$ by two kernel-density estimates makes each proposal inexpensive, so that the budget is spent almost entirely on the reservoir simulations themselves.

We use the TPE sampler as implemented in Optuna~\cite{Akiba2019}, setting only the number of startup trials and the random seed and leaving the remaining parameters at their default values. The densities of Eq.~\eqref{eq:BO_probability_TPE} are then Gaussian mixtures fit independently for each parameter, the good set holds the best $\lceil 0.1 n \rceil$ of the $n$ completed trials, so that $\gamma = 0.1$, and $24$ candidates are drawn from $\ell(x)$ at each proposal. A uniform component is included among the kernels, which prevents the densities from collapsing onto the observed points, and the kernels of the older trials are down-weighted, so that the proposals track the most recent evidence. Each study comprises $150$ trials, the first $50$ sampled at random, uniformly in the fields and log-uniformly in $\Delta t$, and the remaining $100$ guided by the ratio of Eq.~\eqref{eq:EI_with_density_ratio}, with every trial run to completion. Studies are run independently for every combination of task, optimization strategy and encoding window $w$, as well as of NARMA order $n$ and Mackey-Glass delay $\tau_{\mathrm{MG}}$ where applicable.

Two features of this scheme are worth keeping in mind when reading the reported optima. Since candidates are drawn from $\ell(x)$, the search concentrates where the good trials already lie, so on a landscape with a broad plateau the sampler can keep proposing configurations within it rather than probing farther regions, and the returned optimum is then an essentially arbitrary member of a wide set of near-equivalent points. In addition, with independent densities the correlations between $h_x$ and $h_z$ are not modeled, which limits the resolution of the parametric search on a two-dimensional landscape whose good region is elongated. Both effects are visible in our results. For the temporal search we can rule them out directly, since the dense scans of Figs.~\ref{fig:STM_scan} and~\ref{fig:NARMA_scan} show that the located optimum coincides with the true maximum of the scanned curve at every window. For the parametric search no such scan is available, and the scatter of the winning fields reported in Tables~\ref{tab:stm_optima}-\ref{tab:sf_optima}, together with the width of the $1\%$-best bands, indicates that the landscape is flat over a broad region of the field plane rather than that a sharp optimum has been missed.

\section{Extended results for the memory tasks}
\label{Appendix_sec:Results_extended}

This Appendix collects extended results for the memory tasks of Sec.~\ref{sec:memory}, complementing the total scores reported in the main text with a more detailed view of how each configuration achieves them.

\subsection{STM task}
\label{appendix_subsec:STM_expansion}

\begin{figure}[t!]
    \centering
    \includegraphics[width=1\linewidth]{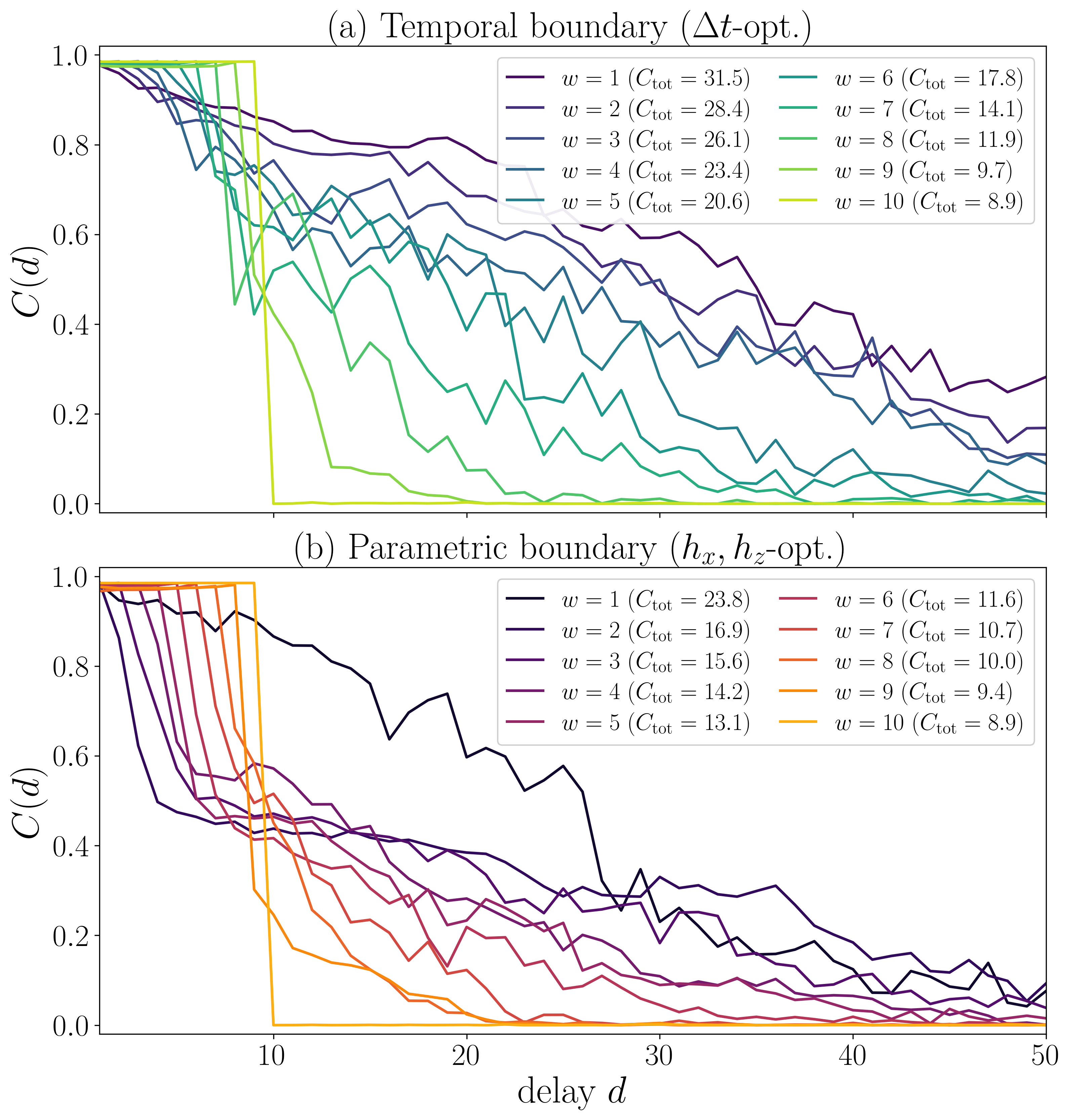}
    \caption{Per-delay memory capacity $C(d)$ of the best configuration found at each encoding window $w$, for the temporal (a) and parametric (b) optimization strategies, with the corresponding total capacities $C_{\mathrm{tot}}$ indicated in the legends. In both panels the capacity is close to one for the delays held explicitly in the encoding window and decays once the input leaves it, with the tail carried by the memory register.}
    \label{fig:stm_appendix}
\end{figure}

We first examine how the total capacity of the STM task is distributed across delays. Fig.~\ref{fig:stm_appendix} shows the per-delay capacity $C(d)$ of the best configuration found by BO at each encoding window, for the temporal (a) and parametric (b) strategies, whose sum over delays gives the totals $C_{\text{tot}}$ reported in Fig.~\ref{fig:STM_results}.

Panel (a) displays the two-part memory profile anticipated in Sec.~\ref{subsec:reservoir}, with near-perfect recall, $C(d)\simeq 1$, for the delays held explicitly in the window, followed by a decaying tail carried by the memory register once the input leaves it. The two parts trade against each other as $w$ varies, and at $w=N$ the profile becomes rectangular, perfect recall of the delays $d=1,\dots,w-1$ and an abrupt drop to zero. This also explains the monotonic decrease of $C_{\text{tot}}$ in the main text, since each qubit moved into the encoding register adds one perfectly recalled delay but shortens a tail worth more than one unit of capacity. Panel (b) shows the same plateau followed by a weaker, more gradual tail, so the parametric strategy preserves the ordering of the totals but reaches a smaller $C_{\text{tot}}$ at every window, the two coinciding at $w=N-1$ and $w=N$ where no memory register remains.

\begin{figure}[t!]
    \centering
    \includegraphics[width=1\linewidth]{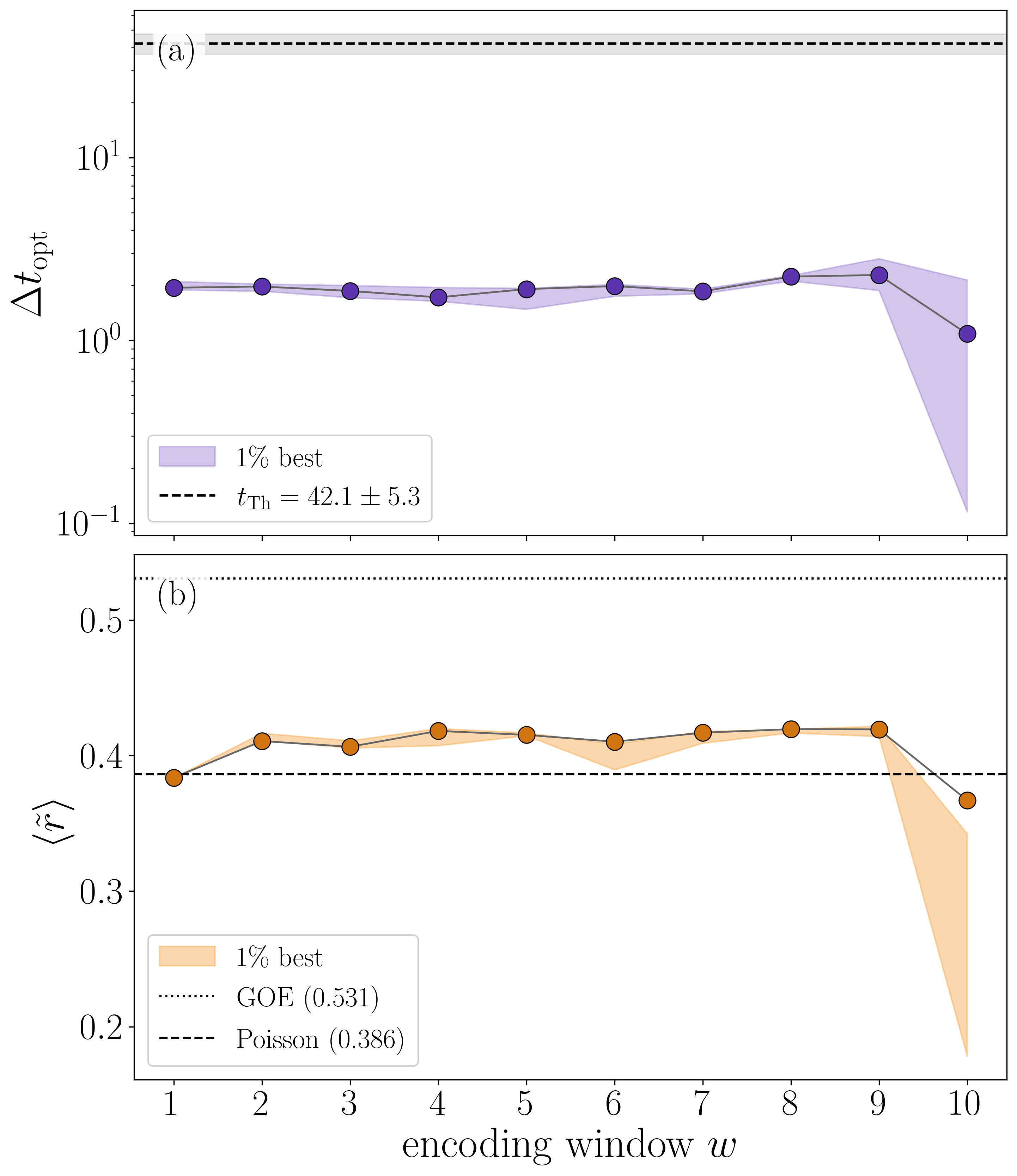}
    \caption{Characterization of the best STM reservoirs at each encoding window $w$ found, with the corresponding configurations listed in Table~\ref{tab:stm_optima}. (a) Optimal evolution time $\Delta t_{\mathrm{opt}}$ of the temporal search, with the shaded band spanning the evolution times whose score lies within $1\%$ of the best. (b) Disorder-averaged gap ratio $\braket{\tilde{r}}$ of the best Hamiltonian found by the parametric search at each window, with the shaded band spanning the mean $\pm \sigma$ of $\braket{\tilde{r}}$ over the configurations whose score lies within $1\%$ of the best, and the Poisson and GOE values marked by the dashed and dotted lines. The gap ratio is computed with uniform per-site disorder of half-width $\epsilon = 0.15$ over $n_r=200$ realizations and averaged over both reflection sectors.}
    \label{fig:stm_topt_rtilde}
\end{figure}

We next characterize the optima found by the BO. Fig.~\ref{fig:stm_topt_rtilde} shows, for every encoding window, the optimal evolution time of the temporal search (a) and the gap ratio of the best Hamiltonian found by the parametric one (b), with the corresponding configuration listed in Table~\ref{tab:stm_optima}. Panel (a) of Fig.~\ref{fig:stm_topt_rtilde} shows that the best evolution time is essentially independent of the window $w$ (for $w<N$), $\Delta t_{\text{opt}}\simeq 2$, more than an order of magnitude below the Thouless time. The shaded band, spanning the evolution times whose score lies within $1\%$ of the best, remains narrow for every $w<N$, so only a fine selection of evolution times reaches the top capacities. At $w=N$ the bands widen showing that the score barely depends on the evolution time, as one can also observe in Fig.~\ref{fig:STM_scan}, where the optimum becomes ill-defined in this QELM limit.

Panel (b) of Fig.~\ref{fig:stm_topt_rtilde} shows that the best Hamiltonians found in the parametric search lie between the Poisson and GOE values, markedly closer to the former. At the long fixed $\Delta t = 70$, a strongly chaotic Hamiltonian would scramble each input toward its Haar-like local equilibrium before the measurement, suppressing the input dependence of the observables, so the highest capacities fall on weakly chaotic dynamics that mix slowly. The parametric search thus reaches through the fields the same moderate effective scrambling that the temporal search attains with a short evolution of a chaotic Hamiltonian, landing at $w=1$ on $(\braket{h_x},\braket{h_z}) = (0.11,3.59)$, nearly switching off the integrability-breaking field and bringing the model close to the TFIM limit. The compensation is only partial, as the parametric totals remain below the temporal ones at every window, and the values reported at $w=N$ carry no significance in this Hamiltonian-independent limit.

\begin{table}[h!]
  \caption{Best STM reservoir configurations found at each encoding window $w$ under the two optimization strategies of Sec.~\ref{subsec:BO_method}, associated with Fig.~\ref{fig:stm_topt_rtilde}. The temporal columns report the optimal evolution time $\Delta t_{\mathrm{opt}}$ found at the fixed chaotic point $(h_x,h_z)=(0.5,1.05)$ and the achieved total capacity. The parametric columns report the winning mean fields at fixed $\Delta t=70$, their disorder-averaged gap ratio $\langle\tilde r\rangle$, computed with uniform per-site disorder of half-width $\epsilon=0.15$ over $n_r=200$ realizations and averaged over both reflection sectors, and the achieved total capacity. The Poisson and GOE reference values are $\langle\tilde r\rangle\approx 0.386$ and $0.531$, respectively.}
  \label{tab:stm_optima}
  \begin{ruledtabular}
  \begin{tabular}{ccccccc}
    \multirow{2}{*}{$w$} & \multicolumn{2}{c}{Temporal} & \multicolumn{4}{c}{Parametric} \\
    \cline{2-3} \cline{4-7}
     & $\Delta t_{\mathrm{opt}}$ & $C_{\mathrm{tot}}$ & $\langle h_x\rangle$ & $\langle h_z\rangle$ & $\langle\tilde r\rangle$ & $C_{\mathrm{tot}}$ \\
    \colrule
    1  & 1.94 & 31.50 & 0.11 & 3.59 & $0.380 \pm 0.053$ & 23.85 \\
    2  & 1.97 & 28.44 & 8.95 & 4.73 & $0.412 \pm 0.016$ & 16.86 \\
    3  & 1.87 & 26.09 & 8.77 & 4.47 & $0.407 \pm 0.018$ & 15.60 \\
    4  & 1.72 & 23.38 & 9.97 & 6.63 & $0.418 \pm 0.015$ & 14.17 \\
    5  & 1.91 & 20.60 & 8.82 & 4.84 & $0.415 \pm 0.016$ & 13.12 \\
    6  & 1.98 & 17.78 & 8.77 & 4.63 & $0.412 \pm 0.016$ & 11.60 \\
    7  & 1.86 & 14.13 & 8.35 & 4.64 & $0.416 \pm 0.017$ & 10.71 \\
    8  & 2.24 & 11.90 & 9.33 & 6.63 & $0.417 \pm 0.016$ & 9.95  \\
    9  & 2.28 & 9.71  & 8.28 & 5.05 & $0.418 \pm 0.016$ & 9.44  \\
    10 & 1.09 & 8.89  & 6.59 & 1.57 & $0.366 \pm 0.019$ & 8.88  \\
  \end{tabular}
  \end{ruledtabular}
\end{table}

\subsection{NARMA-$n$ task}
\label{appendix_subsec:NARMA_expansion}
\begin{figure*}[t!]
    \centering
    \includegraphics[width=1\linewidth]{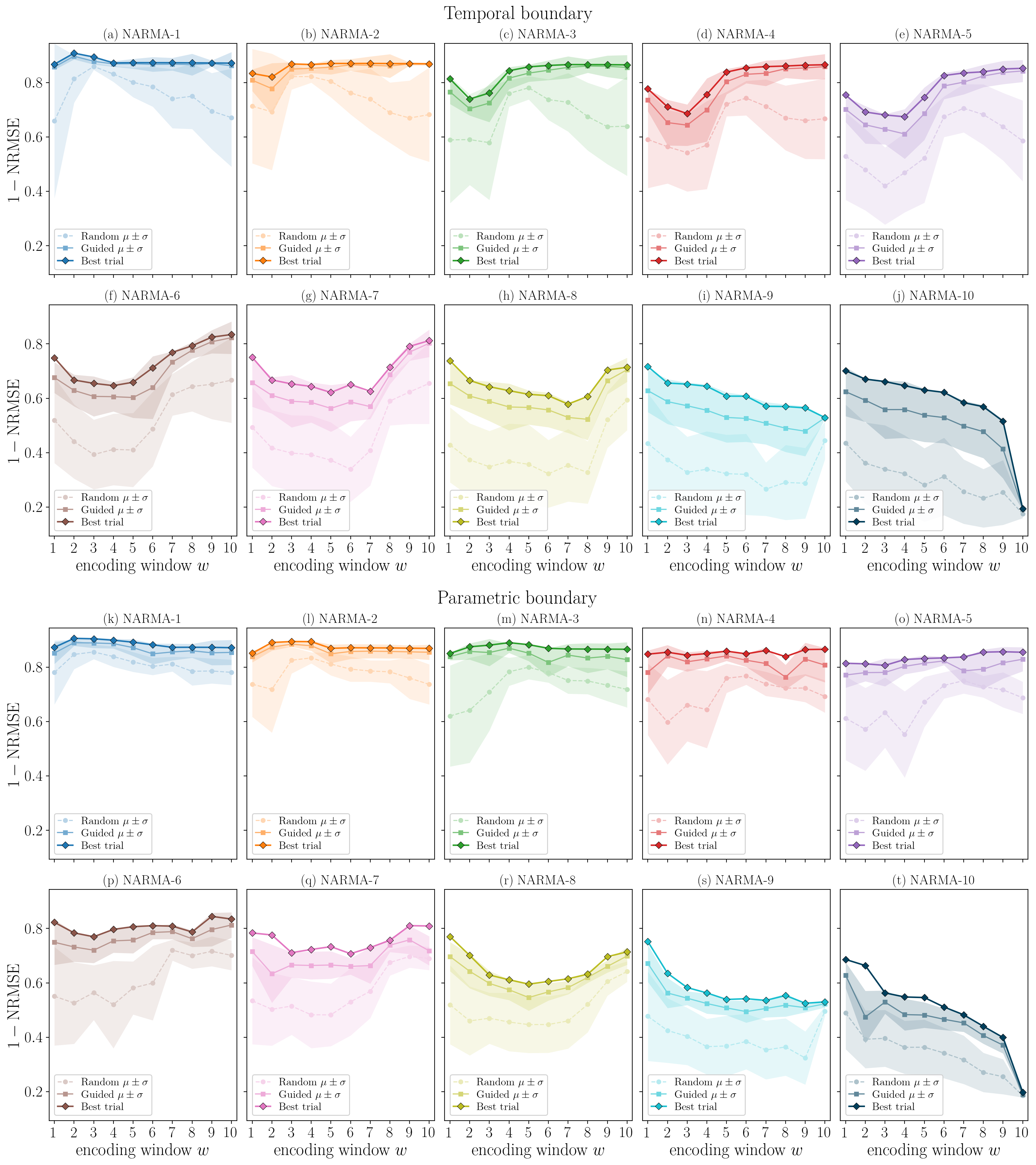}
    \caption{NARMA-$n$ score $1-\mathrm{NRMSE}$ as a function of the encoding window $w$ for every order $n=1$ to $10$, under the temporal (a-j) and parametric (k-t) optimization strategies. Each panel shows the mean $\pm\sigma$ of the $50$ random trials (dashed, light band), the mean $\pm\sigma$ of the $100$ guided trials (solid, darker band) and the best configuration found (diamonds). The guided distributions lie above the random ones throughout, with the largest gains at small windows and high orders, and the distributions converge as $w\to N$. At $n=10$ both strategies collapse at $w=N$, where the required lag is unreachable.}
    \label{fig:narma_appendix_grid}
\end{figure*}

Fig.~\ref{fig:narma_appendix_grid} resolves the heatmaps of Fig.~\ref{fig:NARMA_results} order by order. The guided band lies above the random one throughout, most where the memory qubits must supply the recall, at small windows and high orders. Unlike in the STM task, the dispersion does not collapse as $w\to N$, because even with the lag held explicitly in the window the readout still needs products of the encoded inputs beyond those the register exposes directly, so the mixing by the dynamics remains relevant; a random draw of $\Delta t$ then lands anywhere between the short-time plateau, where the cross-term is directly available through the two-body correlators, and the scrambled long-time regime, and in some cases a moderate evolution helps further, combining the register contents so that additional products reach the readout. The window-dependence tracks the diagonal $n=w$, high and flat at low orders, developing a trough at intermediate windows with maxima at both ends at intermediate orders, where the lag is carried by the $N-1$ memory qubits at $w=1$ or by the register at $w>n$ but by neither in between, and losing the large-$w$ recovery at the highest orders, until $n=10$ falls monotonically to the QELM collapse.

\begin{figure}[t]
    \centering
    \includegraphics[width=1\linewidth]{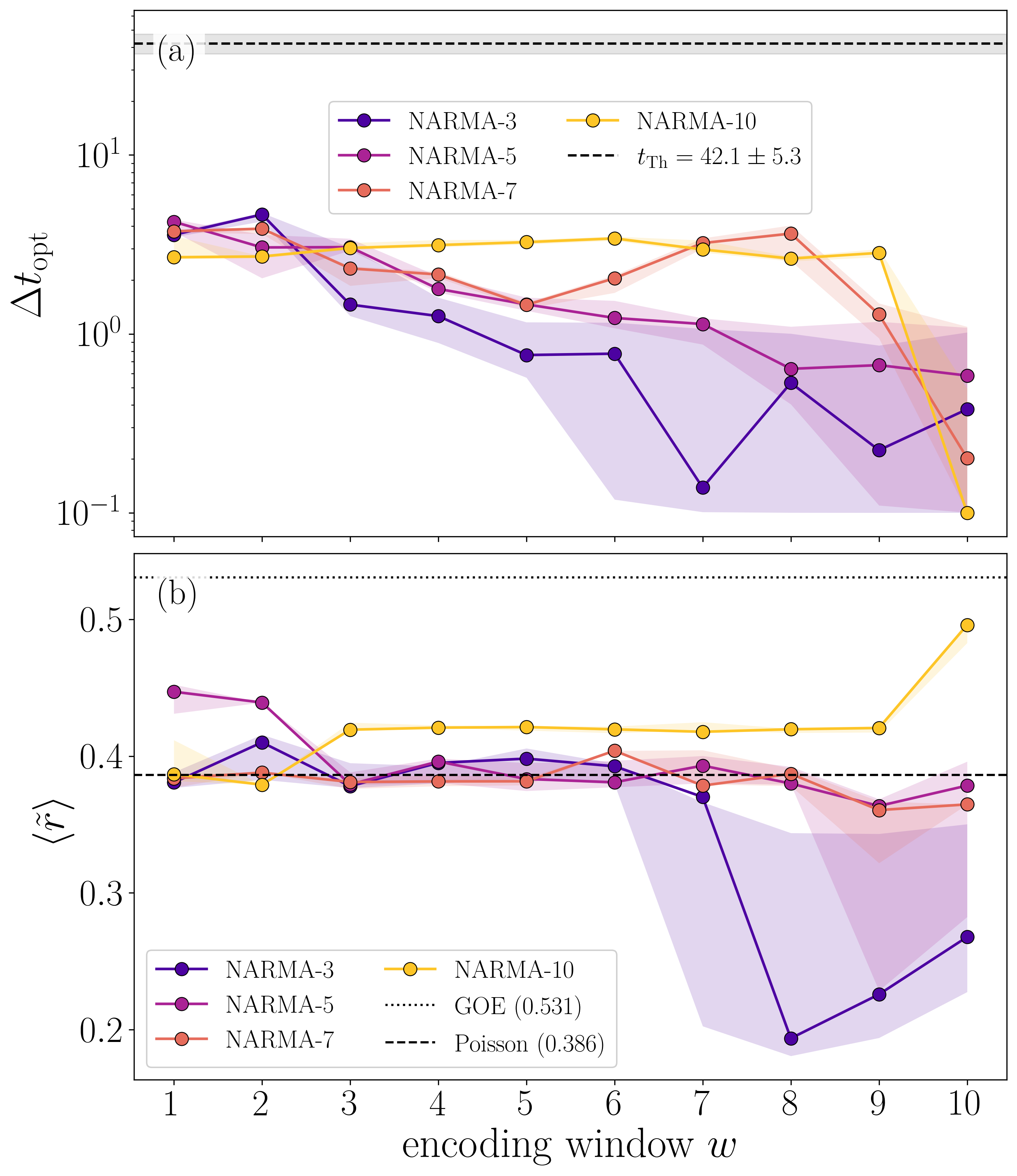}
    \caption{Characterization of the optimal NARMA-$n$ reservoirs at each encoding window $w$ for four representative orders, with the corresponding configurations listed in Table~\ref{tab:narma_optima}. (a) Optimal evolution time $\Delta t_{\mathrm{opt}}$ of the temporal search, with the shaded bands spanning the evolution times whose score lies within $1\%$ of the best. For $n\geq w$ the optimum sits at intermediate times, while once $n<w$ it drops toward the shortest times of the search range and the band widens. (b) Disorder-averaged gap ratio $\langle\tilde r\rangle$ of the winning Hamiltonian of the parametric search, computed as in Fig.~\ref{fig:stm_topt_rtilde}, with the Poisson and GOE values marked.}
    \label{fig:narma_topt_rtilde}
\end{figure}

We next characterize the best configurations, which for the temporal search are the true optima, since the crosses overlap with the stars in Fig.~\ref{fig:NARMA_scan}. Fig.~\ref{fig:narma_topt_rtilde} collects, for four representative orders, the best evolution time (a) and the gap ratio of the best parametric Hamiltonian (b), with the configurations for $n=5,7,10$ in Table~\ref{tab:narma_optima}. Panel (a) splits at the diagonal. For $n\geq w$ the best time sits at intermediate values, roughly independent of the window and slightly above the STM value, consistent with a task that needs somewhat more mixing before the readout can form the target's products. For $n<w$ it drops by an order of magnitude toward the shortest times and the $1\%$-best band widens, since with both cross-term inputs in the window the score is high from the start and further evolution mostly only degrades it.

\begin{table}[t]
  \caption{Best NARMA-$n$ reservoir configuration found at each encoding window $w$ for the orders $n\in\{5,7,10\}$ under the two optimization strategies of Sec.~\ref{subsec:BO_method}, associated with Fig.~\ref{fig:narma_topt_rtilde}. The temporal columns report the best evolution time $\Delta t_{\mathrm{opt}}$ found at the fixed chaotic point $(h_x,h_z)=(0.5,1.05)$ and the achieved score $1-\mathrm{NRMSE}$. The parametric columns report the best mean fields found at fixed $\Delta t=70$, their disorder-averaged gap ratio $\langle\tilde r\rangle$, computed with uniform per-site disorder of half-width $\epsilon=0.15$ over $n_r=200$ realizations and averaged over both reflection sectors, and the achieved score. The Poisson and GOE reference values are $\langle\tilde r\rangle\approx 0.386$ and $0.531$, respectively.}
  \label{tab:narma_optima}
  \begin{ruledtabular}
  \begin{tabular}{cccccccc}
    \multirow{2}{*}{$n$} & \multirow{2}{*}{$w$} & \multicolumn{2}{c}{Temporal} & \multicolumn{4}{c}{Parametric} \\
    \cline{3-4} \cline{5-8}
     & & $\Delta t_{\mathrm{opt}}$ & $1-\mathrm{NRMSE}$ & $\langle h_x\rangle$ & $\langle h_z\rangle$ & $\langle\tilde r\rangle$ & $1-\mathrm{NRMSE}$ \\
    \colrule
    \multirow{10}{*}{5} & 1 & 4.24 & 0.75 & 3.62 & 9.96 & $0.446 \pm 0.015$ & 0.81 \\
     & 2 & 3.04 & 0.69 & 3.43 & 9.99 & $0.436 \pm 0.017$ & 0.81 \\
     & 3 & 3.05 & 0.68 & 0.38 & 8.13 & $0.382 \pm 0.051$ & 0.81 \\
     & 4 & 1.78 & 0.67 & 0.73 & 7.89 & $0.395 \pm 0.046$ & 0.83 \\
     & 5 & 1.46 & 0.75 & 0.20 & 7.93 & $0.381 \pm 0.054$ & 0.83 \\
     & 6 & 1.23 & 0.83 & 0.03 & 8.01 & $0.380 \pm 0.053$ & 0.83 \\
     & 7 & 1.13 & 0.83 & 5.34 & 2.15 & $0.390 \pm 0.017$ & 0.84 \\
     & 8 & 0.64 & 0.84 & 5.72 & 1.97 & $0.381 \pm 0.017$ & 0.85 \\
     & 9 & 0.67 & 0.85 & 6.20 & 1.12 & $0.365 \pm 0.024$ & 0.86 \\
     & 10 & 0.58 & 0.85 & 5.86 & 1.69 & $0.378 \pm 0.020$ & 0.85 \\
    \colrule
    \multirow{10}{*}{7} & 1 & 3.74 & 0.75 & 0.38 & 8.15 & $0.379 \pm 0.050$ & 0.78 \\
     & 2 & 3.88 & 0.67 & 0.02 & 2.78 & $0.387 \pm 0.050$ & 0.78 \\
     & 3 & 2.32 & 0.65 & 0.01 & 9.53 & $0.385 \pm 0.049$ & 0.71 \\
     & 4 & 2.15 & 0.64 & 0.21 & 9.14 & $0.379 \pm 0.048$ & 0.72 \\
     & 5 & 1.45 & 0.62 & 0.02 & 8.95 & $0.379 \pm 0.050$ & 0.73 \\
     & 6 & 2.05 & 0.65 & 0.88 & 8.54 & $0.397 \pm 0.045$ & 0.71 \\
     & 7 & 3.22 & 0.63 & 0.04 & 8.87 & $0.384 \pm 0.048$ & 0.73 \\
     & 8 & 3.64 & 0.71 & 0.22 & 8.57 & $0.384 \pm 0.052$ & 0.76 \\
     & 9 & 1.28 & 0.79 & 6.13 & 1.00 & $0.358 \pm 0.025$ & 0.81 \\
     & 10 & 0.20 & 0.81 & 6.98 & 1.51 & $0.366 \pm 0.019$ & 0.81 \\
    \colrule
    \multirow{10}{*}{10} & 1 & 2.68 & 0.70 & 0.52 & 7.93 & $0.384 \pm 0.048$ & 0.69 \\
     & 2 & 2.71 & 0.67 & 0.00 & 3.93 & $0.376 \pm 0.051$ & 0.66 \\
     & 3 & 3.02 & 0.66 & 9.00 & 7.03 & $0.421 \pm 0.015$ & 0.56 \\
     & 4 & 3.14 & 0.65 & 7.59 & 4.86 & $0.421 \pm 0.015$ & 0.55 \\
     & 5 & 3.26 & 0.63 & 7.69 & 4.89 & $0.421 \pm 0.015$ & 0.55 \\
     & 6 & 3.42 & 0.62 & 8.24 & 4.99 & $0.420 \pm 0.016$ & 0.51 \\
     & 7 & 2.97 & 0.58 & 9.11 & 7.03 & $0.416 \pm 0.016$ & 0.48 \\
     & 8 & 2.64 & 0.57 & 8.84 & 6.90 & $0.423 \pm 0.016$ & 0.44 \\
     & 9 & 2.84 & 0.52 & 7.09 & 5.40 & $0.421 \pm 0.016$ & 0.40 \\
     & 10 & 0.10 & 0.19 & 1.88 & 2.04 & $0.497 \pm 0.012$ & 0.20 \\
  \end{tabular}
  \end{ruledtabular}
\end{table}

Panel (b) shows the same split. The common feature across the whole plane is that the winning fields are much larger than the coupling $J=1$, so that the interaction term, the only source of entanglement between qubits, acts as a perturbation and the effective scrambling rate stays low over the long fixed evolution $\Delta t = 70$. Where the optimizer places the field within that regime, however, changes with the order. For $n \geq w$ it drives $h_x$ to nearly zero while keeping a large transverse field, reducing the chain to the integrable TFIM, with $\braket{\tilde{r}}$ close to the Poisson value, as Table~\ref{tab:narma_optima} shows for $n=7$ at $w\leq 8$. Above the diagonal the search goes further and selects $\braket{\tilde{r}}$ below the Poisson value, now with a dominating longitudinal $h_x$, as the table shows for $n=5$ at $w\geq 8$. In this corner the field is parallel to the coupling and the two share the same eigenbasis, so the evolution is nearly diagonal in $x$ and barely disturbs the encoded register. At the highest order, $n=10$, the optimum never crosses the diagonal and both fields remain large and comparable, giving $\braket{\tilde{r}}\simeq 0.42$, weakly chaotic but far from GOE.

Neither the integrable nor the nearly commuting configurations should be read as inoperative reservoirs. The transverse field does not commute with the coupling, so the near-TFIM dynamics still correlates the qubits and mixes the inputs, through free quasiparticles rather than through many-body scrambling. In the nearly commuting corner the encoded states are not eigenstates of the Hamiltonian either, and although the conserved $\braket{X_i}$ and $\braket{X_iX_j}$ carry only the instantaneous input, the phases the evolution generates between $x$ eigenstates become visible in the transverse components of the feature set, which makes the choice of readout basis essential. Reservoirs of precisely this class, with commuting field and coupling terms read out in a transverse basis, have achieved state-of-the-art experimental performance on the NARMA family~\cite{Hou_2026}. Finally, the configuration reported at $n=10$ and $w=N$, with $\braket{\tilde{r}}$ near the GOE value, carries no significance, since there the task has collapsed for every Hamiltonian and the search returns an arbitrary point of a flat landscape.

\begin{table}[h!]
  \caption{Best Mackey-Glass reservoir configurations found at each encoding window $w$ for the chaotic delays $\tau_{\rm MG}\in\{30,50\}$ under the two optimization strategies of Sec.~\ref{subsec:BO_method}, associated with Fig.~\ref{fig:mg_topt_rtilde}. The temporal columns report the best evolution time $\Delta t_{\rm opt}$ found at the fixed chaotic point $(h_x,h_z)=(0.5,1.05)$ and the achieved forward capacity $C_{\rm tot}^{\rm fut}$ of Eq.~\eqref{eq:forward_capacity}. The parametric columns report the best mean fields found at fixed $\Delta t=70$, their disorder-averaged gap ratio $\langle\tilde r\rangle$, computed with the fields drawn uniformly from a window of half-width $\epsilon=0.15$ around the reported mean values over $n_r=200$ realizations and averaged over both reflection sectors, and the achieved forward capacity. The Poisson and GOE reference values are $\langle\tilde r\rangle\approx 0.386$ and $0.531$,
respectively.}
  \label{tab:mg_optima}
  \begin{ruledtabular}
  \begin{tabular}{cccccccc}
    \multirow{2}{*}{$\tau_{\rm MG}$} & \multirow{2}{*}{$w$} & \multicolumn{2}{c}{Temporal} & \multicolumn{4}{c}{Parametric} \\
    \cline{3-4} \cline{5-8}
     & & $\Delta t_{\mathrm{opt}}$ & $C_{\mathrm{tot}}^{\mathrm{fut}}$ & $\langle h_x\rangle$ & $\langle h_z\rangle$ & $\langle\tilde r\rangle$ & $C_{\mathrm{tot}}^{\mathrm{fut}}$ \\
    \colrule
    \multirow{10}{*}{30} & 1 & 10.51 & 130.2 & 3.94 & 5.57 & $0.490 \pm 0.017$ & 133.4 \\
     & 2 & 5.94 & 130.3 & 5.05 & 8.13 & $0.482 \pm 0.016$ & 131.6 \\
     & 3 & 5.44 & 129.8 & 4.84 & 7.90 & $0.485 \pm 0.017$ & 132.5 \\
     & 4 & 6.10 & 129.6 & 7.13 & 6.77 & $0.446 \pm 0.015$ & 130.6 \\
     & 5 & 4.82 & 127.9 & 8.25 & 7.11 & $0.432 \pm 0.015$ & 130.9 \\
     & 6 & 4.88 & 126.3 & 8.33 & 6.58 & $0.423 \pm 0.016$ & 130.4 \\
     & 7 & 5.05 & 123.4 & 9.47 & 5.50 & $0.417 \pm 0.016$ & 129.4 \\
     & 8 & 3.90 & 120.8 & 3.91 & 2.40 & $0.420 \pm 0.016$ & 126.6 \\
     & 9 & 3.02 & 115.8 & 9.93 & 4.65 & $0.397 \pm 0.017$ & 124.2 \\
     & 10 & 1.65 & 111.9 & 8.04 & 9.88 & $0.470 \pm 0.020$ & 113.3 \\
    \colrule
    \multirow{10}{*}{50} & 1 & 3.26 & 89.2 & 4.38 & 3.11 & $0.427 \pm 0.016$ & 107.9 \\
     & 2 & 4.53 & 91.5 & 5.29 & 2.30 & $0.394 \pm 0.018$ & 105.0 \\
     & 3 & 4.76 & 88.0 & 9.29 & 7.07 & $0.416 \pm 0.014$ & 102.0 \\
     & 4 & 4.74 & 84.8 & 9.61 & 6.88 & $0.416 \pm 0.014$ & 102.1 \\
     & 5 & 4.73 & 80.5 & 9.80 & 5.91 & $0.419 \pm 0.015$ & 103.1 \\
     & 6 & 3.18 & 77.5 & 5.92 & 2.32 & $0.391 \pm 0.019$ & 99.2 \\
     & 7 & 5.55 & 72.2 & 9.94 & 2.68 & $0.363 \pm 0.020$ & 95.4 \\
     & 8 & 5.94 & 68.2 & 8.36 & 4.13 & $0.405 \pm 0.018$ & 84.7 \\
     & 9 & 0.32 & 63.0 & 9.92 & 4.07 & $0.382 \pm 0.018$ & 82.5 \\
     & 10 & 0.18 & 61.7 & 6.96 & 0.02 & $0.131 \pm 0.026$ & 61.4 \\
  \end{tabular}
  \end{ruledtabular}
\end{table}

\section{Extended results for the forecasting tasks}
\label{Appendix_sec:Results_extended_forecasting}

\subsection{Mackey-Glass task}
\label{appendix_subsec:MG_expansion}

The Mackey-Glass system of Eq.~\eqref{eq:mackey_glass} is a delay-differential equation, so its state at time $t$ is the whole history segment over $[t-\tau_{\rm MG}, t]$. We integrate it with a fixed-step fourth-order Runge-Kutta scheme of step $h=0.05$, chosen so that the delay spans an integer number of steps for every $\tau_{\rm MG}$ considered and the delayed value falls on the stored grid. Starting from a constant initial history $x(t') = 1.2$, we discard a transient of $5000$ time units and sample the settled trajectory at unit spacing, giving the input sequences after the rescaling to $[0,1]$. As the delay grows the dynamics transition from a stable limit cycle to a chaotic attractor of increasing dimension, a complexity that can be measured through the Kaplan-Yorke dimension $D_{\rm KY}$~\cite{Kaplan1979}, estimated from the Lyapunov spectrum that we compute with the Benettin algorithm adapted to delay-differential equations~\cite{Benettin1980}. For this system $D_{\rm KY}$ grows roughly linearly with the delay, from $D_{\rm KY} = 1$ at $\tau_{\rm MG} = 7$ and $15$, through $2.14$ at the onset $\tau_{\rm MG} = 17$, to $3.50$ and $5.27$ at $\tau_{\rm MG} = 30$ and $50$, so that the delay acts as a single dial on the difficulty of the forecasting task.

\begin{figure}[t!]
    \centering
    \includegraphics[width=1\linewidth]{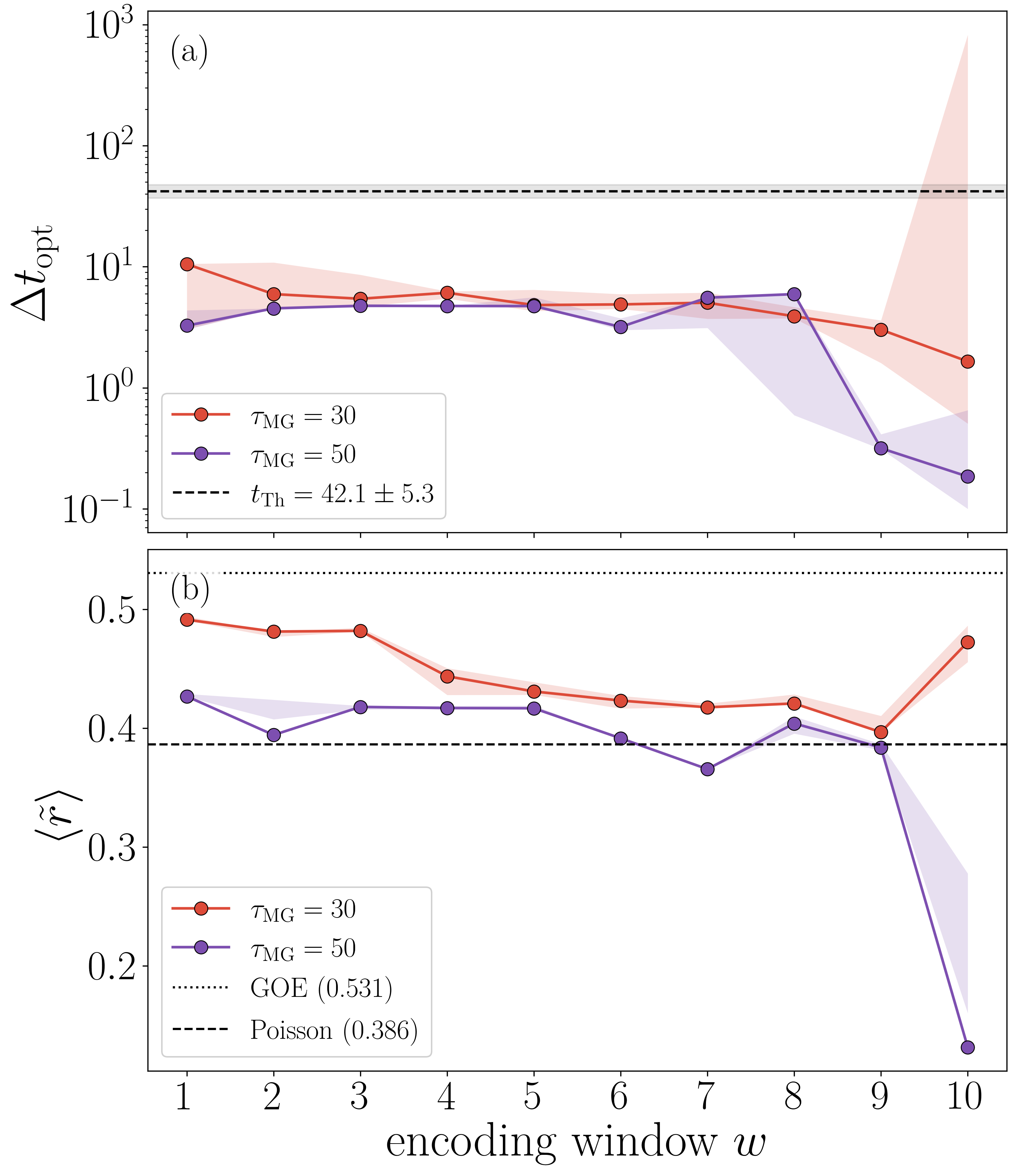}
    \caption{Characterization of the best Mackey-Glass reservoirs found at each encoding window $w$ for the two chaotic delays $\tau_{\rm MG} = 30$ and $50$, with the corresponding configurations listed in Table~\ref{tab:mg_optima}. (a) Best evolution time $\Delta t_{\rm opt}$ of the temporal search, with the shaded band spanning the evolution times whose score lies within $1\%$ of the best. It sits at intermediate values for most windows, far below the Thouless time (dashed), and drops toward the shortest times of the search range as the encoding register fills the chain, most sharply at $\tau_{\rm MG} = 50$. (b) Disorder-averaged gap ratio $\langle\tilde r\rangle$ of the best Hamiltonian found by the parametric search, with the Poisson and GOE values marked. The winners at $\tau_{\rm MG} = 50$ lie systematically below those at $\tau_{\rm MG} = 30$, at or near the Poisson value for several windows, and in both panels the bands widen at the largest windows, where the score barely depends on the configuration.}
    \label{fig:mg_topt_rtilde}
\end{figure}

Fig.~\ref{fig:mg_topt_rtilde} characterizes the best configurations at the two chaotic delays, with parameters in Table~\ref{tab:mg_optima}. The best evolution time (a) sits at intermediate values $\Delta t_{\rm opt} \simeq 3$--$10$ over most windows for both delays, below the Thouless time and close to the memory-task values, and stays roughly independent of $w$ until the register nearly fills the chain, where at $\tau_{\rm MG} = 50$ it drops by an order of magnitude at $w=9,10$ as the register already holds the recent history. The parametric gap ratios (b) are ordered by delay, lower at $\tau_{\rm MG} = 50$ (between $\simeq 0.36$ and $0.43$) than at $\tau_{\rm MG} = 30$ ($\simeq 0.40$--$0.49$), so forecasting the longer-memory attractor pushes the search toward more weakly scrambling dynamics that preserve the encoded history over the long fixed $\Delta t = 70$. The $w=N$ values, most visibly $\langle\tilde r\rangle = 0.13$ at $\tau_{\rm MG}=50$, carry limited significance on the nearly flat QELM landscape.

\subsection{Santa Fe task}
\label{Appendix_subsec:results_Santa_Fe}

\begin{figure}[!t]
    \centering
    \includegraphics[width=1\linewidth]{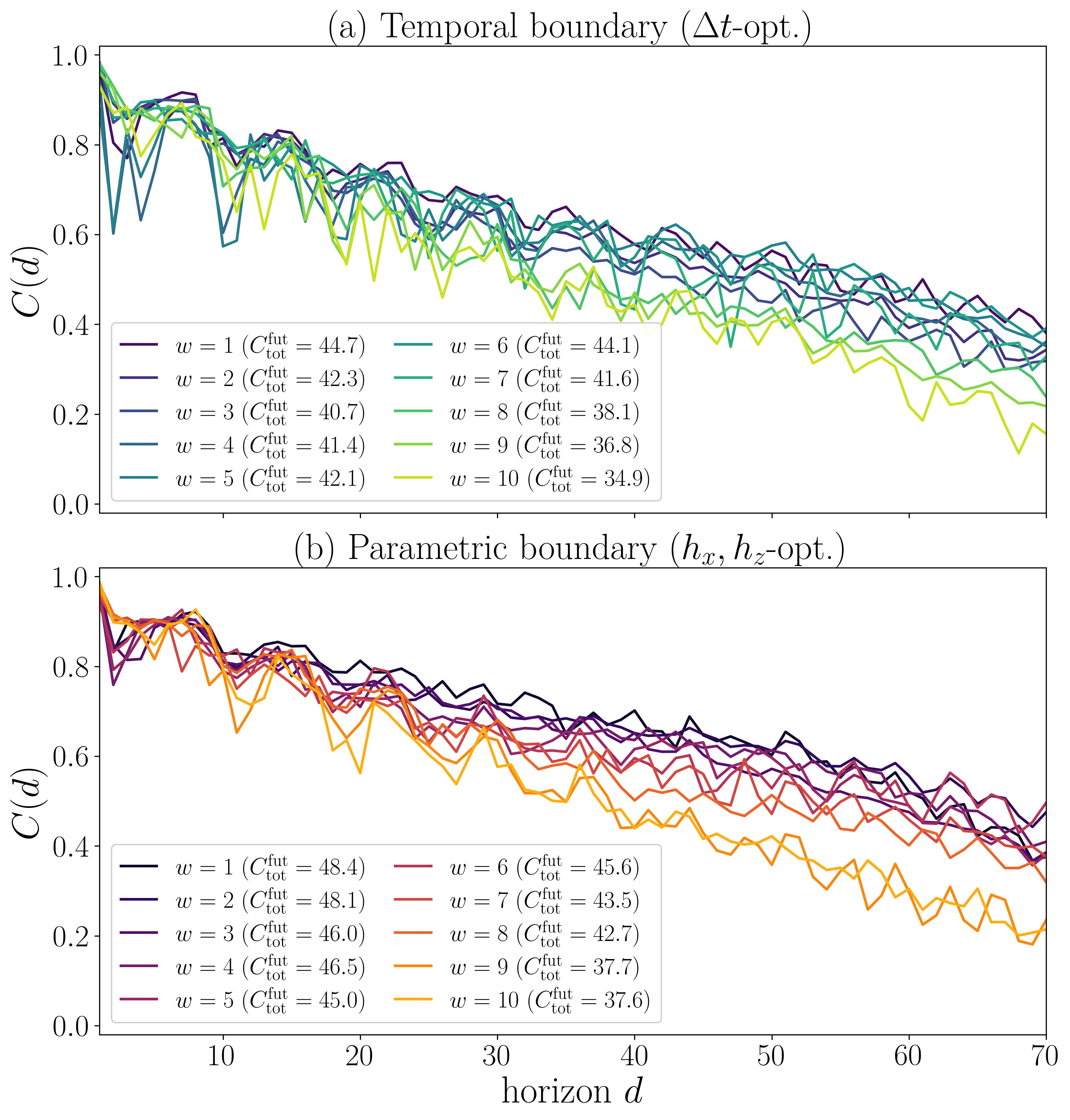}
    \caption{Per-horizon forward capacity $C^{\text{fut}}(d)$ of the best Santa Fe configuration found at each encoding window $w$, for the temporal (a) and parametric (b) optimization strategies, with the corresponding total capacities $C_{\rm tot}^{\rm fut}$ indicated in the legends.}
    \label{fig:C_tot_vs_delay_santa_fe}
\end{figure}

Fig.~\ref{fig:C_tot_vs_delay_santa_fe} shows the per-horizon forward capacity $C^{\text{fut}}(d)$ of the best configuration at each window, for the temporal (a) and parametric (b) strategies. At short horizons the windows are indistinguishable, the recent inputs in the register already carrying what these horizons require, and they separate only in the tail, where the small windows sustain a higher capacity and the largest, $w=9,10$, fall below with little recurrent memory left to supply the older history. The QELM limit thus stays close to the best except at the longest predictions, so the moderate drop reported in the main text is concentrated in the tail, reflecting the short-ranged correlations of the signal.

\begin{figure}[!t]
    \centering
    \includegraphics[width=1\linewidth]{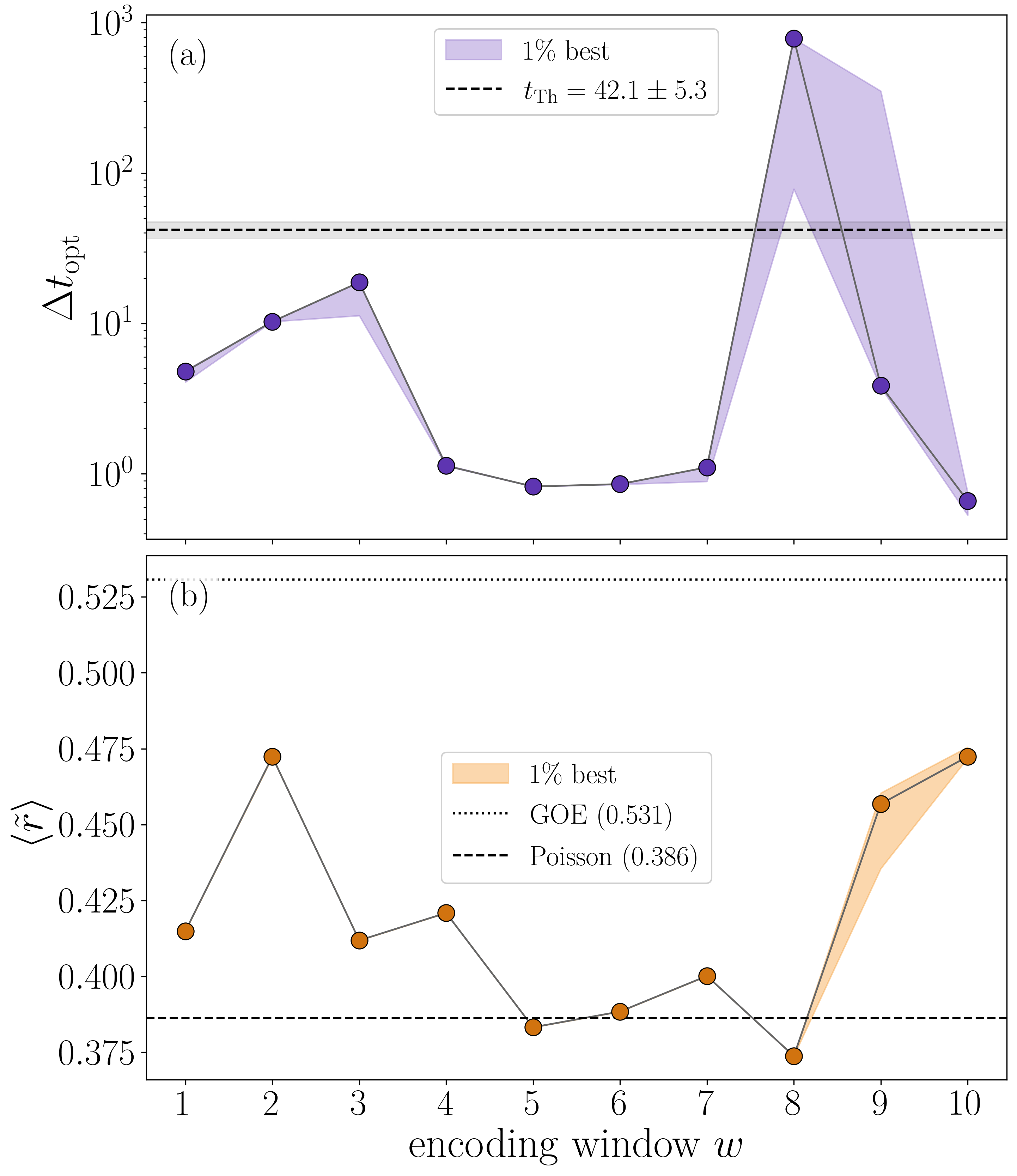}
    \caption{Characterization of the best Santa Fe reservoirs found at each encoding window $w$, with the corresponding configurations listed in Table~\ref{tab:sf_optima}. (a) Best evolution time $\Delta t_{\rm opt}$ of the temporal search, with the shaded band spanning the evolution times whose score lies within $1\%$ of the best. At the smallest windows the optimum sits at intermediate values, from $w=4$ to $7$ it drops toward the shortest times of the search range, and from $w = 8$ the band spans  several decades and the optimum becomes ill-defined, since the register already holds the short memory signal and the score barely depends on the evolution time. (b) Disorder-averaged gap ratio $\langle\tilde r\rangle$ of the best Hamiltonian found by the parametric search, with the Poisson and GOE values marked.}
    \label{fig:sf_topt_rtilde}
\end{figure}

Fig.~\ref{fig:sf_topt_rtilde} characterizes the best configurations, with parameters in Table~\ref{tab:sf_optima}. With no dense scan of the evolution time available here, the $1\%$-best bands are our main handle on the landscape. At the smallest windows $w=1$--$3$ the band is narrow and the optimum sits at intermediate $\Delta t_{\rm opt} \simeq 5$--$19$, as for the memory tasks and Mackey-Glass, the reservoir extracting from the recurrent memory the history the register does not hold. Beyond $w=3$ the register already covers the short correlations of the signal and the landscape depends only weakly on $\Delta t$, the search settling at the short end for $w=4$--$7$ and scattering over several decades from $w=8$ on. We read this as a nearly flat landscape, though without a dense scan we cannot dismiss structure the search missed. The extreme case $\Delta t_{\rm opt} = 784$ at $w=8$, far beyond the Heisenberg time, scores well only because exact expectation values let the linear readout exploit an arbitrarily small noiseless signal, a regime that would not survive finite-shot estimation (Appendix~\ref{Appendix_sec:Separability}).

\begin{table}[h!]
  \centering
  \caption{Best Santa Fe forecasting reservoir configurations found at each encoding window $w$ under the two optimization strategies of Sec.~\ref{subsec:BO_method}, associated with Fig.~\ref{fig:sf_topt_rtilde}. The temporal columns report the best  evolution time $\Delta t_{\rm opt}$ found at the fixed chaotic point $(h_x,h_z)=(0.5,1.05)$ and the achieved forward capacity $C_{\rm tot}^{\rm fut}$ of Eq.~\eqref{eq:forward_capacity}. The parametric columns report the best mean fields found at fixed $\Delta t = 70$, their disorder-averaged gap ratio $\langle\tilde r\rangle$, computed with the fields drawn uniformly from a window of half-width $\epsilon = 0.15$ around the reported mean values over $n_r = 200$ realizations and averaged over both reflection sectors, and the achieved forward capacity. The Poisson and GOE reference values are $\langle\tilde r\rangle \approx 0.386$ and $0.531$.}
  \label{tab:sf_optima}
  \begin{ruledtabular}
  \begin{tabular}{ccccccc}
    \multirow{2}{*}{$w$} & \multicolumn{2}{c}{Temporal} & \multicolumn{4}{c}{Parametric} \\
    \cline{2-3} \cline{4-7}
     & $\Delta t_{\rm opt}$ & $C_{\rm tot}^{\rm fut}$ & $\langle h_x\rangle$ & $\langle h_z\rangle$ & $\langle\tilde r\rangle$ & $C_{\rm tot}^{\rm fut}$ \\
    \colrule
    1  & 4.79   & 44.66 & 1.44 & 9.56 & $0.413 \pm 0.040$ & 48.36 \\
    2  & 10.29  & 42.28 & 7.53 & 9.11 & $0.472 \pm 0.022$ & 48.14 \\
    3  & 18.79  & 40.66 & 6.00 & 3.04 & $0.412 \pm 0.018$ & 46.01 \\
    4  & 1.13   & 41.45 & 6.42 & 3.84 & $0.421 \pm 0.017$ & 46.54 \\
    5  & 0.82   & 42.15 & 7.09 & 2.66 & $0.383 \pm 0.016$ & 45.00 \\
    6  & 0.85   & 44.09 & 3.20 & 0.46 & $0.388 \pm 0.021$ & 45.64 \\
    7  & 1.10   & 41.56 & 2.98 & 0.76 & $0.400 \pm 0.018$ & 43.55 \\
    8  & 784.30 & 38.13 & 7.07 & 2.20 & $0.374 \pm 0.018$ & 42.74 \\
    9  & 3.87   & 36.85 & 0.42 & 2.45 & $0.457 \pm 0.024$ & 37.72 \\
    10 & 0.66   & 34.86 & 2.49 & 4.79 & $0.472 \pm 0.016$ & 37.64 \\
  \end{tabular}
  \end{ruledtabular}
\end{table}

\section{The densely connected model as a quantum reservoir}
\label{Appendix_sec:dense_model}

In this Appendix we characterize the densely connected model used in the comparison of Sec.~\ref{subsec:comparison_models}, mirroring the analysis of Sec.~\ref{sec:The 1D Mixed-Field Ising Model as Quantum Reservoir} for the chain. 

The Hamiltonian is given by Eq.~\eqref{eq:H_dense_main}, with $N = 10$ qubits, uniform local fields $(h_x, h_z)$, and couplings $J_{ij} \sim \mathcal{U}[-1,1]$ drawn independently for every pair. The model replaces the nearest-neighbor coupling of Eq.~\eqref{eq:mfim} by a random all-to-all connected interaction, while keeping the fields uniform so that the connectivity is the only difference between the two models.

\begin{figure}[!t]
    \centering
    \includegraphics[width=1\linewidth]{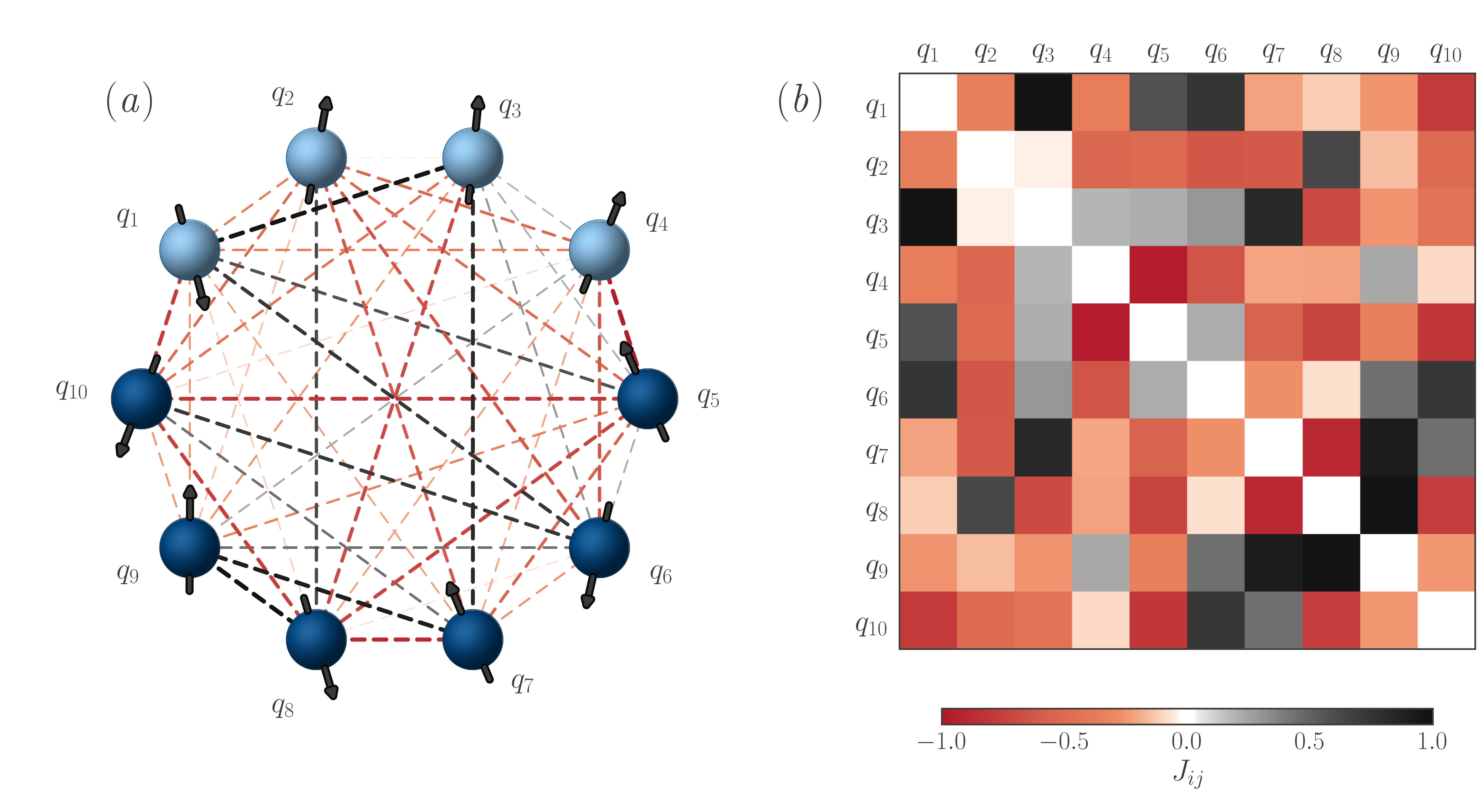}
    \caption{The densely connected model. (a) Sketch of the $N = 10$ densely connected reservoir, with the explicit register (light blue) and the recurrent memory (dark blue) as in Fig.~\ref{fig:setup}(b), and every pair of qubits coupled by $J_{ij} X_i X_j$, the dashes colored by the coupling strength. (b) The fixed realization of the couplings $J_{ij} \sim \mathcal{U}[-1,1]$ used in the optimization experiments, with the color scale shared with (a).}
    \label{fig:dense_model}
\end{figure}

Since the couplings are random, the spectral and dynamical characterization below averages over realizations of $J_{ij}$ at fixed fields, which replaces the field-disorder average used for the chain. However, the optimization experiments admit no such average. A Bayesian search needs a definite score per trial, so the searches of Sec.~\ref{subsec:comparison_models} for the dense model run on one fixed realization, shown in Fig.~\ref{fig:dense_model}, which plays the role that the coupling $J=1$ plays in the chain. As throughout this work, every Hamiltonian is normalized by its own root-mean-square spectral width $\tilde{H}_{\text{dense}} = H_{\text{dense}}/\sigma_H$, which absorbs the realization-dependent energy scale of the couplings and makes a given evolution time physically comparable across draws, strategies, and models. The couplings are never optimization parameters, and the two strategies act, as for the linear chain, on the evolution time and the uniform local fields. Once each search has returned its best configuration, we re-evaluate it on $10$ additional realizations of the couplings.

\begin{figure}[t!]
    \centering
    \includegraphics[width=\linewidth]{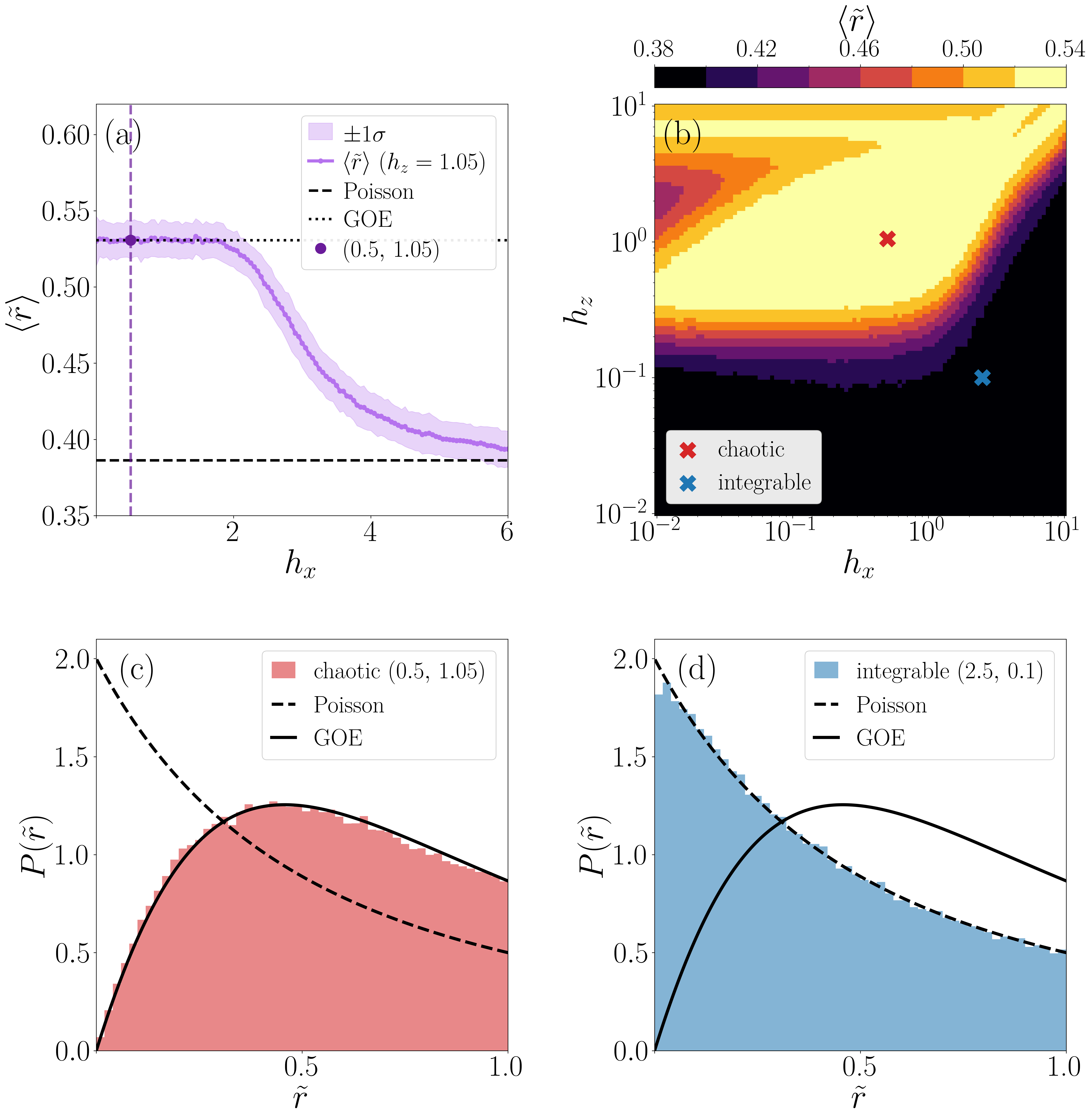}
    \caption{Spectral characterization of the densely connected model for $N = 10$, with the couplings $J_{ij} \sim \mathcal{U}[-1,1]$ drawn independently for each of the $n_r = 200$ realizations at fixed fields. (a) Average gap ratio $\langle\tilde r\rangle$ versus $h_x$ at fixed $h_z = 1.05$, resolved within the sectors of the parity $\prod_i Z_i$, with the shaded band showing $\pm 1\sigma$ over realizations and the dashed and dotted lines marking the Poisson ($0.386$) and GOE ($0.531$) values. (b) Heatmap of $\langle\tilde r\rangle$ over the field plane, evaluated over the full spectrum, with the chaotic $(0.5, 1.05)$ and integrable $(2.5, 0.1)$ reference points marked by the red and blue crosses. (c,\,d) Gap-ratio distributions $P(\tilde r)$ at those two points, compared with the Poisson and GOE predictions.}
    \label{fig:dense_spectral}
\end{figure}

Fig.~\ref{fig:dense_spectral} shows the spectral characterization of Sec.~\ref{subsec:static} applied to the densely connected model, with the average over $n_r = 200$ realizations of the couplings at fixed fields. For this model, the random $J_{ij}$ break the reflection symmetry of the chain, and for finite fields no symmetry remains, so the gap ratio of the heatmap in Fig.~\ref{fig:dense_spectral}(b) is evaluated over the full spectrum, $\mathcal{D} = 2^N = 1024$. The exception is the parity $\mathcal{P}=\prod_i Z_i$, which is conserved for $h_x = 0$ and only perturbatively broken while $h_x$ remains small compared to the couplings and $h_z$. In that region the two parity sectors stay effectively unmixed, and superposing them biases the full-spectrum statistics toward Poisson, so the departures from GOE near the left edge of Fig.~\ref{fig:dense_spectral}(b), wide on the logarithmic axis but confined to perturbative $h_x$, reflect the unresolved quasi-symmetry rather than a loss of chaos. The cut of Fig.~\ref{fig:dense_spectral}(a) confirms this, resolving the parity sectors and showing that $\langle\tilde r\rangle$ holds the GOE value from $h_x = 0$ up to $h_x \simeq 2$, falling toward Poisson only once the longitudinal field dominates the couplings and shares an eigenbasis with them. The chaotic reference point of the chain, $(h_x, h_z) = (0.5, 1.05)$,
remains well inside the chaotic region (red cross in Fig.~\ref{fig:dense_spectral}(b)), so we keep the same local fields for the chaotic Hamiltonian, while the integrable reference of the chain is now chaotic and we take $(2.5, 0.1)$, in the field-dominated region, in its place (blue cross). The gap-ratio distributions at the two points (Fig.~\ref{fig:dense_spectral}(c,\,d)) follow the GOE surmise and the Poisson form, respectively.

\begin{figure}[h!]
    \centering
    \includegraphics[width=\linewidth]{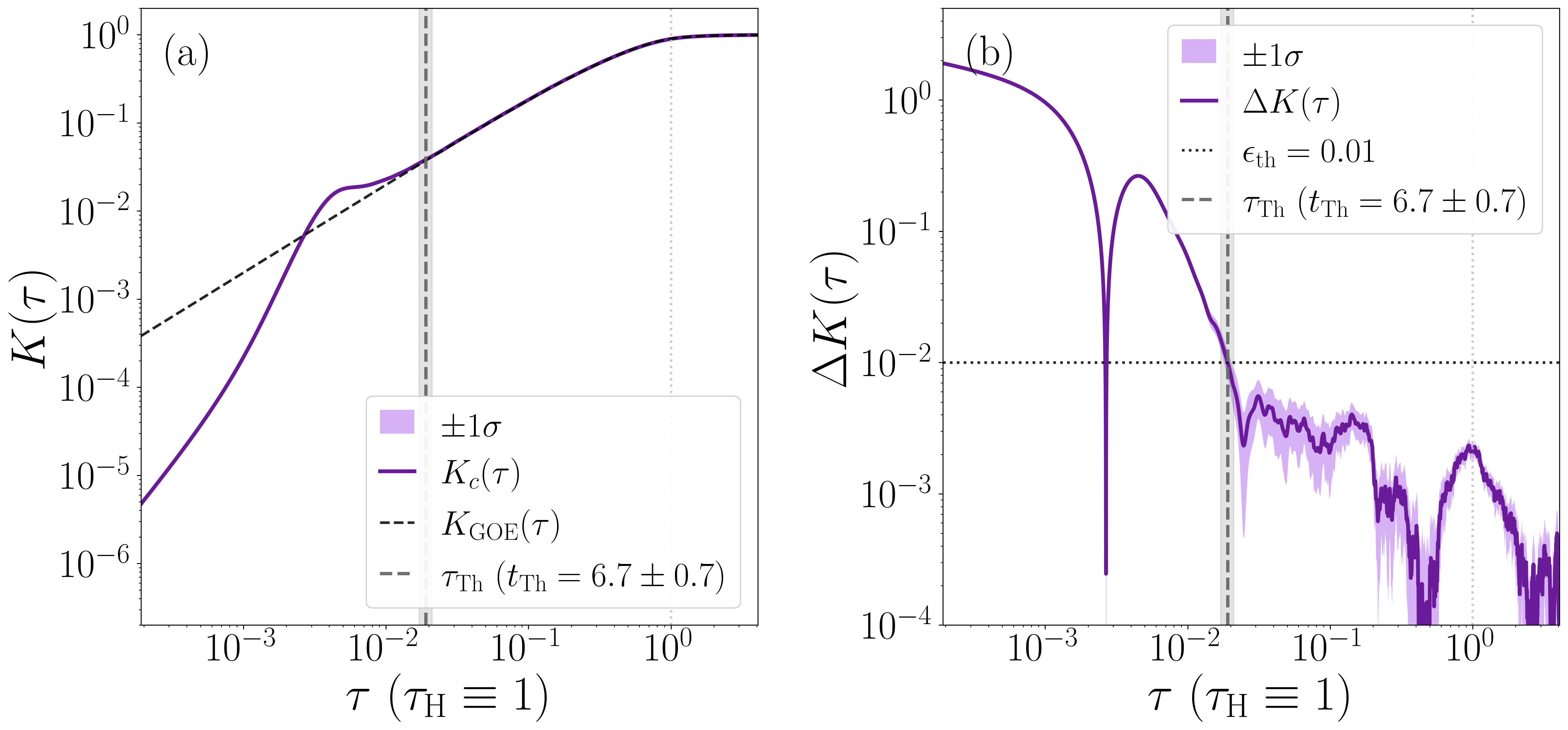}
    \caption{Thouless time of the dense model at the chaotic point $(h_x, h_z) = (0.5, 1.05)$, averaged over $n_r = 10^4$ realizations of the couplings $J_{ij} \sim \mathcal{U}[-1,1]$ at fixed fields. (a) Connected spectral form factor $K_c(\tau)$ versus the rescaled time $\tau = t/t_H$ (solid), compared with the GOE prediction of Eq.~\eqref{eq:SFFGOE} (dashed), saturating to the plateau at $\tau_H = 1$ (light dotted vertical line). (b) Deviation $\Delta K(\tau)$ of Eq.~\eqref{eq:deltaK}, with the threshold $\epsilon_{\mathrm{th}} = 0.01$ (dotted horizontal line). In both panels the vertical gray band marks $\tau_{\mathrm{Th}}$ with its uncertainty, and shaded bands show $\pm 1\sigma$ over the coupling realizations.}
    \label{fig:dense_thouless}
\end{figure}

The characteristic timescales shift with the connectivity of the system~\cite{Gharibyan_2018}. At the chaotic point $(h_x, h_z) = (0.5, 1.05)$, the Heisenberg time is obtained from Eq.~\eqref{eq:tH_def}. With no symmetry sectors to resolve, the full dimension $\mathcal{D} = 2^N = 1024$ enters the calculation instead of the $\mathcal{D} = 528$ of the reflection sector used for the chain, and the spectral variance is now evaluated on the same full spectrum used to normalize, so that $\Gamma_0 = 1$ exactly and $t_H = \chi\mathcal{D} = 349.5$. Repeating the SFF analysis of Appendix~\ref{Appendix_sec:Detailed calculation of the Thouless time} over the coupling ensemble, with the same unfolding, Gaussian filter, and threshold crossing, gives $t_{\mathrm Th} = 6.7\pm 0.7$ (Fig.~\ref{fig:dense_thouless}), approximately six times shorter than the chain value $t_{\mathrm{Th}} = 42.1 \pm 5.3$. This reduction comes from the connectivity of the system. Chaotic correlations develop only once a local perturbation has spread over the whole system. In the linear chain this spreading proceeds qubit by qubit along the lattice, and the Thouless time carries that cost, while in the dense model every qubit couples to all others, the spreading is no longer limited by propagation, and the universal GOE regime of the SFF is reached almost an order of magnitude earlier. This early onset, however, raises the statistical cost of locating the crossing. The noise floor of $\Delta K(\tau)$ decreases as $1/\sqrt{n_r}$, and with the $n_r = 2000$
realizations used for the chain it grazed the threshold $\epsilon_{\mathrm{th}} = 0.01$, so the backward search intermittently locked onto residual fluctuations rather than onto the true departure from the GOE ramp. We therefore average over $n_r = 10^4$ realizations of the couplings, which lowers the floor well below the threshold.
 
Finally, we verify that the dense model fulfills the essential reservoir properties established in Sec.~\ref{subsec:reservoir} and analyzed in Appendix~\ref{Appendix_sec:ESP_and_input_dependence} for the linear chain. 

\begin{figure}[t!]
    \centering
    \includegraphics[width=\linewidth]{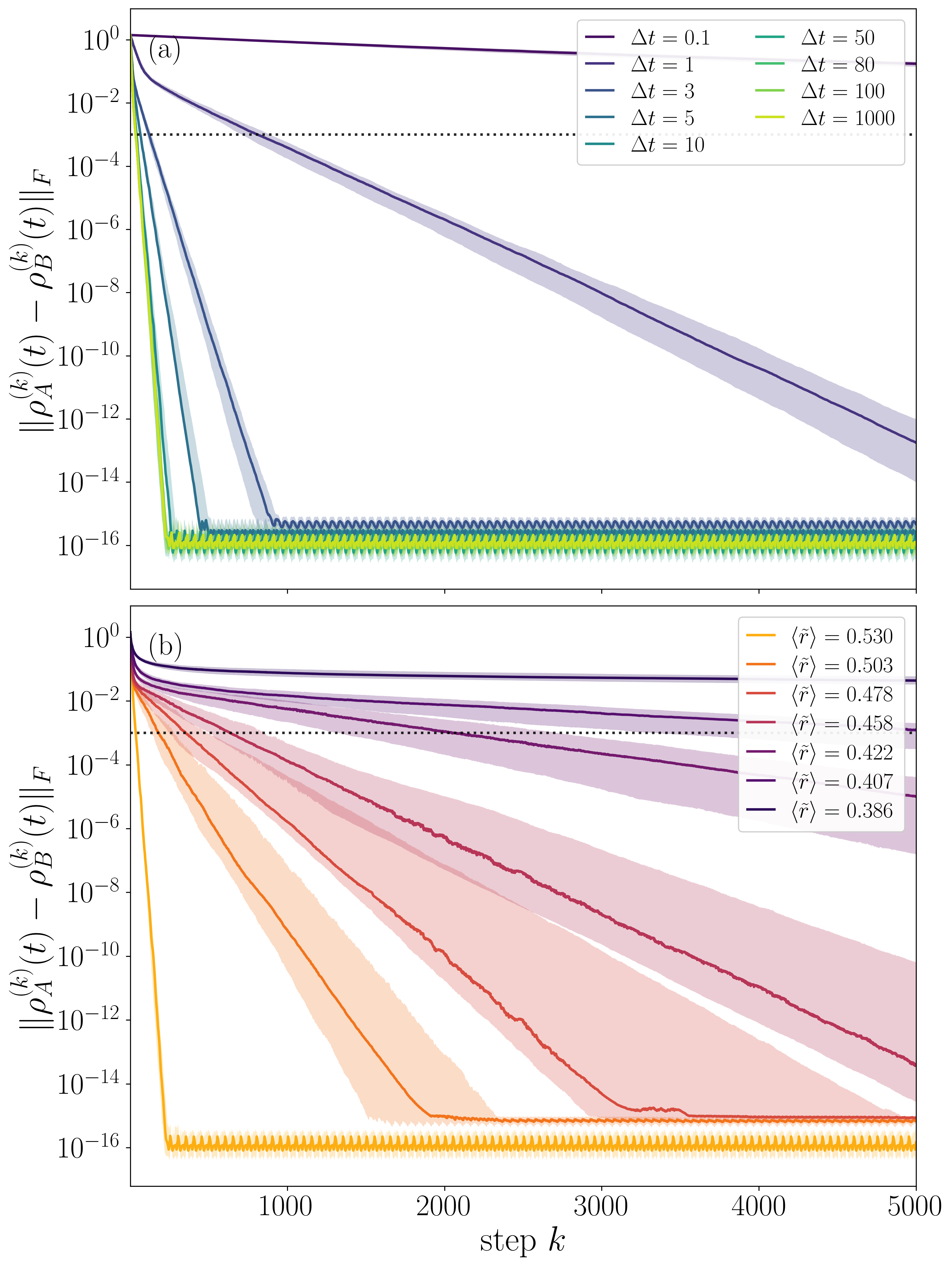}
    \caption{Convergence to the echo-state property in the dense model, measured by the Frobenius distance between two reservoir copies initialized in the reference and in a random product state, driven by the same input sequence at $w = 1$. Curves show the median over 10 realizations of the couplings with 20 random state pairs each, 200 trajectories in total, with shaded bands spanning the interquartile range and the dotted line marking the washout threshold $\epsilon = 10^{-3}$. (a) Chaotic point $(h_x, h_z) = (0.5, 1.05)$ for different evolution times $\Delta t$. (b) Fixed $\Delta t = 70$ for Hamiltonians across the integrable-to-chaotic crossover, labeled by their gap ratio $\langle\tilde r\rangle$; the decay is exponential in every case, with the rate collapsing toward the field-dominated integrable corner.}
    \label{fig:dense_esp}
\end{figure}

For the ESP, one copy of the reservoir starts from the reference state $\rho_A^{(0)} = \ket{0}\bra{0}^{\otimes N}$ and the other from a random product state $\rho_B^{(0)}$, both driven by the same input sequence at $w = 1$, the worst case for the erasure, with $20$ state pairs for each of the $10$ realizations of the couplings $J_{ij}$. Figure~\ref{fig:dense_esp} tracks the Frobenius distance between the two copies, reporting the median over the $200$ trajectories with the bands covering the interquartile range. We use the median here because the mean is pulled by the slowest realizations of the couplings, which lag orders of magnitude behind the rest on the logarithmic scale. Fig.~\ref{fig:dense_esp}(a) fixes the chaotic point and sweeps the evolution time, while Fig.~\ref{fig:dense_esp}(b) fixes $\Delta t = 70$ and scans Hamiltonians across the integrable-to-chaotic crossover. In both panels the decay is exponential and only its rate changes, slowing at short evolution times and toward the integrable region, where the longitudinal field shares the $X$ eigenbasis with the couplings and the nearly diagonal evolution delays the erasure. As in the chain, the dynamics set the rate of the washout rather than whether it occurs. Over the range of evolution times and Hamiltonians relevant to the optimization, the distance reaches the threshold well before $N_{\mathrm{wo}} = 3000$ steps, so the washout length used for the chain remains valid for the dense model.

\begin{figure}[h!!]
    \centering
    \includegraphics[width=\linewidth]{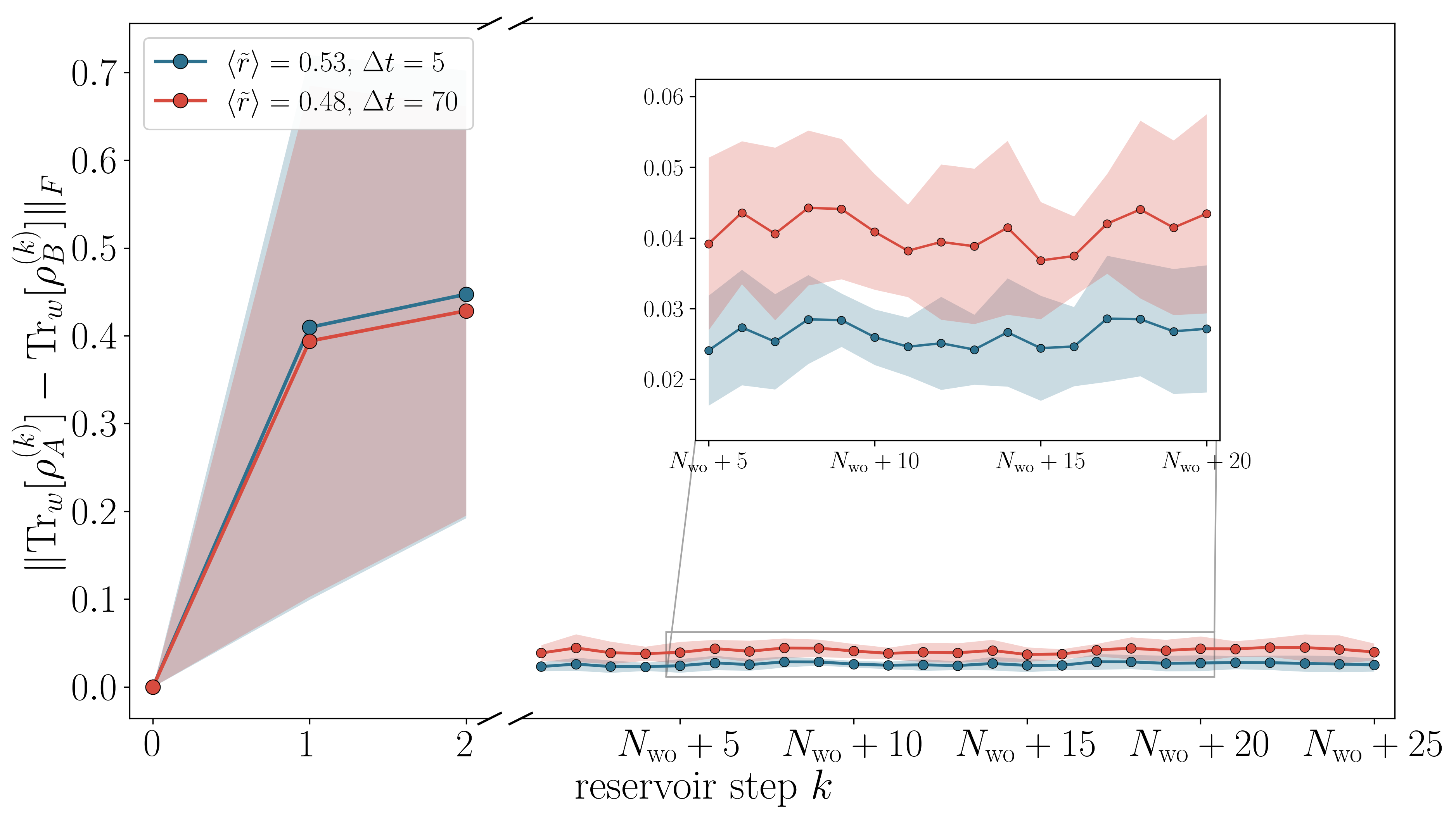}
    \caption{Verification of input separability in the dense model, measured by the Frobenius distance $D_k^{\mathrm{mem}}$ of Eq.~\eqref{eq:sep_distance} between the memory-qubit reduced states of two reservoir copies initialized in the same reference state and driven by different input sequences at $w = 1$. Curves show the mean $\pm 1\sigma$ over the $10$ realizations of the couplings, with the same pair of input sequences shared across configurations. Two of them are shown, the chaotic point $(h_x, h_z) = (0.5, 1.05)$ at $\Delta t = 5$ (blue) and a weakly chaotic Hamiltonian with $\langle\tilde r\rangle = 0.48$ at $\Delta t = 70$ (red). The left segment shows the growth of the distance over the first steps and the right segment the post-washout window, with the inset zooming into the stationary regime, where both configurations keep fluctuating around a finite value instead of decaying, in contrast with the collapse of Fig.~\ref{fig:dense_esp}.}
    \label{fig:dense_input_sep}
\end{figure}

For the input separability, two copies of the same reservoir start from the reference state and are driven by independent input sequences, tracking $D_k^{\mathrm{mem}}$ of Eq.~\eqref{eq:sep_distance} at $w = 1$ over the $10$ realizations of the couplings, with the same pair of input sequences for all of them. Fig.~\ref{fig:dense_input_sep} shows the chaotic point at $\Delta t = 5$ and a weakly chaotic Hamiltonian with $\langle\tilde r\rangle = 0.48$ at $\Delta t = 70$, representative of the two optimization strategies. In both the distance grows from zero and settles around $2.8 \times 10^{-2}$ and $4.3 \times 10^{-2}$ respectively, far above the numerical floor of the ESP test, and the inset shows that it keeps fluctuating rather than converging to a constant, as each new input drives the two copies apart. The values are nonetheless smaller than in the chain, $5 \times 10^{-2}$ at the same chaotic point and evolution time and of order $10^{-1}$ for the weakly chaotic one, because the dense couplings spread the injected input over all ten spins instead of leaving it partially localized near the encoding site, diluting the trace that any local observable retains. The same mixing that erases the initial condition faster also thins the per-site signature of the input, tightening the measurement budget discussed in Sec.~\ref{subsec:reservoir}.

\section{Extended results for the dense model}
\label{app:extended_results_dense_model}

This Appendix collects the extended results of the comparison of Sec.~\ref{subsec:comparison_models}, showing for the dense model the same analysis that Appendices~\ref{Appendix_sec:Results_extended} and~\ref{Appendix_sec:Results_extended_forecasting} present for the chain. All results refer to the fixed realization $J_{ij}^{\text{fix}}$ of Fig.~\ref{fig:dense_model}. 

\begin{figure}[t!]
    \centering
    \includegraphics[width=\linewidth]{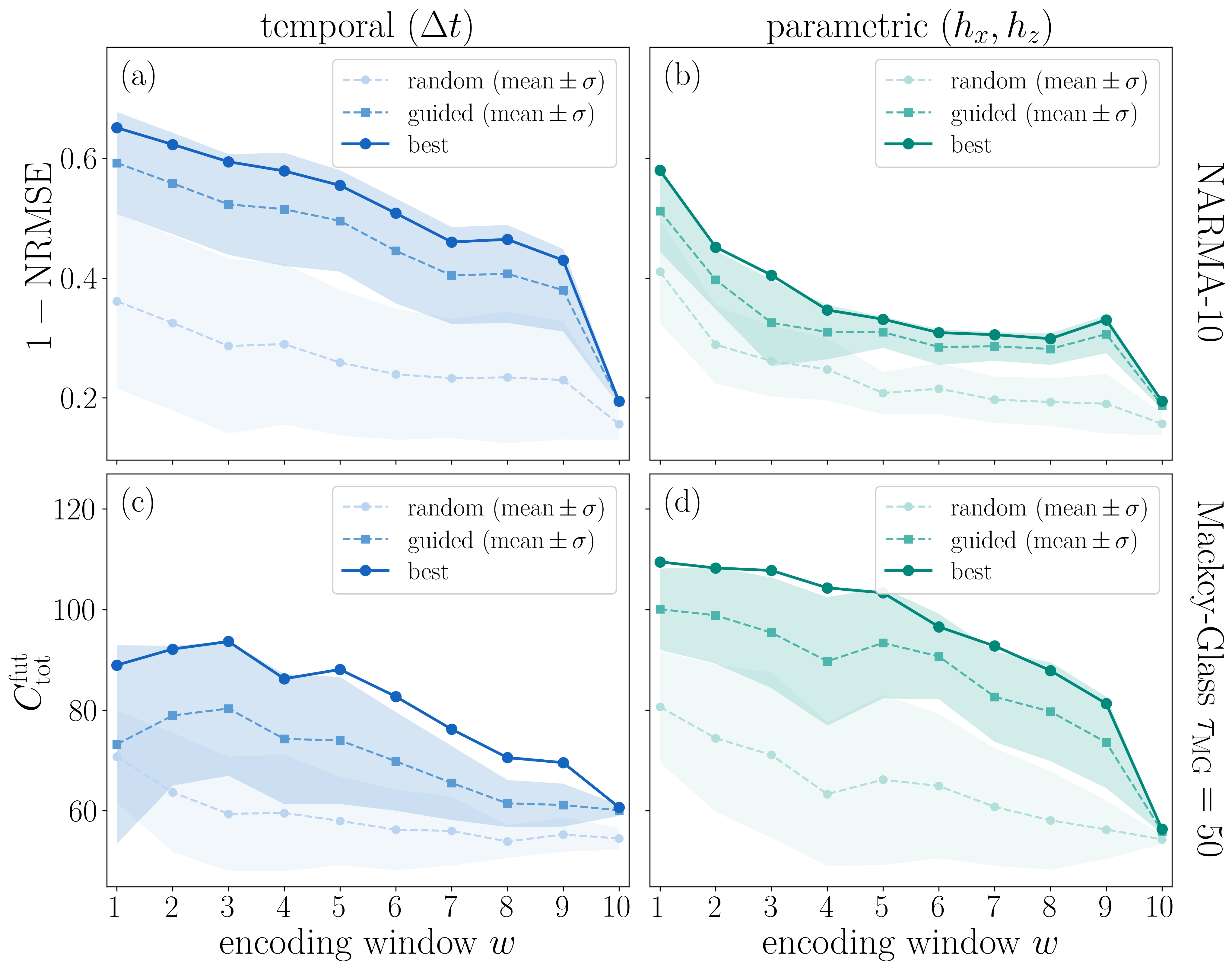}
    \caption{Score of the dense model versus encoding window $w$ at the fixed realization $J_{ij}^{\text{fix}}$, for NARMA-10 (a,\,b) and Mackey-Glass at $\tau_{\rm MG} = 50$ (c,\,d), under the temporal (left) and parametric (right) optimization strategies. Light markers with band show the mean $\pm\sigma$ over the random trials, dark markers over the guided (TPE) trials, and the solid line the best configuration found, following the conventions of Fig.~\ref{fig:STM_results}.}
    \label{fig:dense_bands}
\end{figure}

Fig.~\ref{fig:dense_bands} shows the score of the random, guided, and best configurations as a function of the encoding window, for NARMA-10 (top) and Mackey-Glass at $\tau_{\rm MG} = 50$ (bottom) under the two optimization strategies. The guided band lies above the random one throughout, as for the chain, so BO successfully exploits the structure of the landscape in every case. On NARMA-10 the temporal best decreases monotonically from $0.65$ at $w = 1$ to $0.43$ at $w = 9$ before the collapse at $w = N$, while the parametric best falls much faster, dropping below $0.35$ already at $w = 4$ and staying near $0.3$ thereafter. On Mackey-Glass the pattern of the chain is reproduced instead, with the parametric search reaching higher capacities than the temporal one across the whole window range and both decreasing toward the QELM limit.

\begin{figure}[h!]
    \centering
    \includegraphics[width=\linewidth]{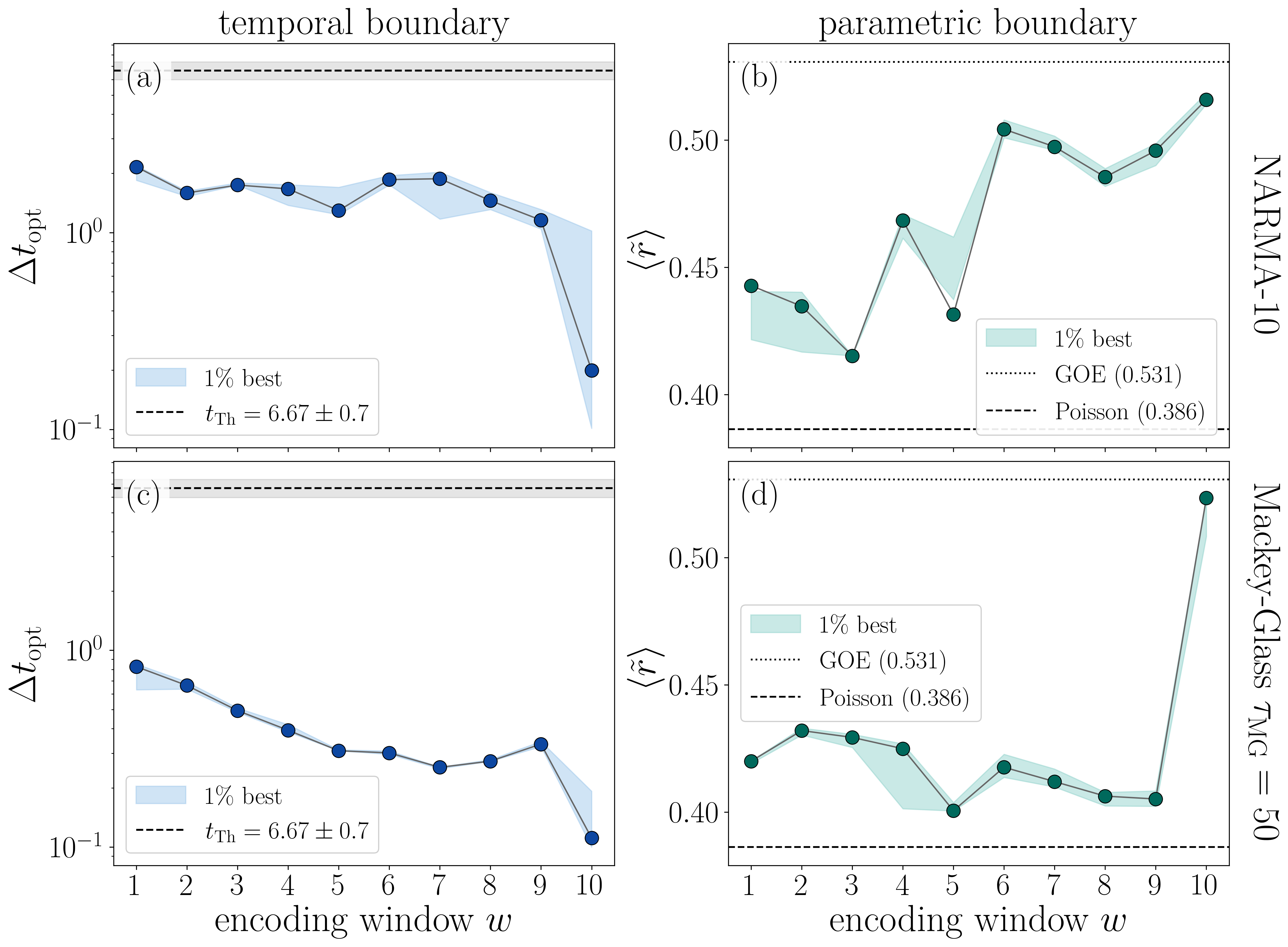}
    \caption{Characterization of the best dense-model configurations at each encoding window $w$, for NARMA-10 (a,\,b) and Mackey-Glass at $\tau_{\rm MG} = 50$ (c,\,d), with the corresponding parameters listed in Table~\ref{tab:dense_optima}. Left panels show the best evolution time $\Delta t_{\rm opt}$ of the temporal search, with the shaded band spanning the evolution times whose score lies within $1\%$ of the best and the dashed line marking the Thouless time $t_{\rm Th} = 6.7 \pm 0.7$ of the dense model. Right panels show the gap ratio $\langle\tilde r\rangle$ of the best Hamiltonian found by the parametric search, computed over $n_r = 100$ realizations of the couplings, with the Poisson and GOE values marked.}
    \label{fig:dense_topt}
\end{figure}

Fig.~\ref{fig:dense_topt} characterizes the best configurations, with the corresponding parameters listed in Table~\ref{tab:dense_optima}. The best evolution times sit below the Thouless time $t_{\rm Th} = 6.7 \pm 0.7$ of the dense model, at $\Delta t_{\rm opt} \simeq 1$-$2$ for NARMA-10 and shorter still, $\Delta t_{\rm opt} \simeq 0.1$-$0.8$, for Mackey-Glass, so the temporal search selects evolutions shorter than the Thouless time as it does for the chain. The best fields of the parametric search are large compared to the couplings, as for the chain, and their gap ratios lie between the Poisson and GOE values in every case. On Mackey-Glass they stay flat at $\langle\tilde r\rangle \simeq 0.40$-$0.43$ across the windows, while on NARMA-10 they rise from $0.42$ toward the GOE value as the register fills the chain; the values at $w = N$, where the task has collapsed and the landscape is flat, carry no significance, as in the chain. The $\braket{\text{score}}_{J_{ij}}$ columns of Table~\ref{tab:dense_optima} report the best configuration re-evaluated, without re-optimization, on the $10$ additional realizations of the couplings, exposing how much of the optimized performance survives on couplings the search never saw. On NARMA-10 the temporal winners carry over essentially unchanged while the parametric ones lose up to $0.06$, and on Mackey-Glass the pattern reverses, with the parametric winners robust within a few percent while the temporal ones lose up to ten units at intermediate windows, consistent with their very short optimal evolutions probing the particular coupling pattern rather than its ensemble-typical dynamics. In every case the spread leaves the comparison with the chain unchanged, so no conclusion depends on the particular draw. Several parametric winners sit near the upper edge of the search domain, with $h_x$ close to $10$. Extending the domain might push them further, but we keep the same bounds used for the chain so that the comparison stays on equal footing, and bounded control ranges are in any case the situation of any real device.

\begin{table*}[t!]
  \caption{\label{tab:dense_optima}
  Best configurations of the dense model at each encoding window $w$ for NARMA-10 (top) and Mackey-Glass at $\tau_{\mathrm{MG}} = 50$ (bottom), under the two optimization strategies of Sec.~\ref{subsec:BO_method} and associated with Fig.~\ref{fig:dense_topt}. The temporal columns report the best evolution time $\Delta t_{\mathrm{opt}}$ found at the fixed chaotic point $(h_x, h_z) = (0.5, 1.05)$ and the achieved score; the parametric columns report the best fields found at fixed $\Delta t = 70$ and their gap ratio $\langle\tilde r\rangle$, computed over $n_r = 100$ realizations of the couplings with uncertainties of $\pm 0.01$ throughout. The score, evaluated on the working realization $J_{ij}^{\text{fix}}$, is $1-\mathrm{NRMSE}$ for NARMA-10 and the forward capacity $C_{\mathrm{tot}}^{\mathrm{fut}}$ of Eq.~\eqref{eq:forward_capacity} for Mackey-Glass. The columns $\braket{\text{score}}_{J_{ij}}$ report the best configuration re-evaluated, without re-optimization, on the $10$ additional realizations of the couplings, as mean $\pm$ standard deviation. The Poisson and GOE reference values are $\langle\tilde r\rangle \approx 0.386$ and $0.531$, respectively.}
  \begin{ruledtabular}
  \begin{tabular}{cccccccccc}
    & \multirow{2}{*}{$w$} & \multicolumn{3}{c}{Temporal} & \multicolumn{5}{c}{Parametric} \\
    \cline{3-5} \cline{6-10}
    & & $\Delta t_{\mathrm{opt}}$ & $\text{score}_{J_{ij}^{\text{fix}}}$ & $\braket{\text{score}}_{J_{ij}}$ & $h_x$ & $h_z$ & $\langle\tilde r\rangle$ & $\text{score}_{J_{ij}^{\text{fix}}}$ & $\braket{\text{score}}_{J_{ij}}$ \\
    \colrule
    \multirow{10}{*}{\rotatebox{90}{NARMA-10}}
    & 1  & 2.16 & 0.65 & $0.65 \pm 0.01$ & 8.85 & 3.70 & 0.443 & 0.58 & $0.55 \pm 0.03$ \\
    & 2  & 1.59 & 0.62 & $0.62 \pm 0.01$ & 9.97 & 4.00 & 0.435 & 0.45 & $0.47 \pm 0.03$ \\
    & 3  & 1.74 & 0.59 & $0.58 \pm 0.01$ & 4.49 & 1.22 & 0.415 & 0.41 & $0.35 \pm 0.02$ \\
    & 4  & 1.67 & 0.58 & $0.56 \pm 0.02$ & 6.92 & 3.31 & 0.468 & 0.35 & $0.32 \pm 0.01$ \\
    & 5  & 1.29 & 0.55 & $0.55 \pm 0.02$ & 6.96 & 2.59 & 0.431 & 0.33 & $0.31 \pm 0.01$ \\
    & 6  & 1.86 & 0.51 & $0.51 \pm 0.03$ & 8.19 & 5.24 & 0.504 & 0.31 & $0.30 \pm 0.02$ \\
    & 7  & 1.88 & 0.46 & $0.46 \pm 0.03$ & 6.45 & 3.79 & 0.497 & 0.31 & $0.27 \pm 0.02$ \\
    & 8  & 1.45 & 0.46 & $0.46 \pm 0.03$ & 9.99 & 5.55 & 0.485 & 0.30 & $0.29 \pm 0.01$ \\
    & 9  & 1.16 & 0.43 & $0.41 \pm 0.04$ & 9.99 & 6.05 & 0.496 & 0.33 & $0.28 \pm 0.04$ \\
    & 10 & 0.20 & 0.19 & $0.19 \pm 0.01$ & 8.56 & 6.27 & 0.516 & 0.19 & $0.18 \pm 0.01$ \\
    \colrule
    \multirow{10}{*}{\rotatebox{90}{Mackey-Glass, $\tau_{\mathrm{MG}}=50$}}
    & 1  & 0.83 & 89.0 & $89.0 \pm 2.8$ & 4.67 & 1.40 & 0.420 & 109.5 & $103.2 \pm 3.5$ \\
    & 2  & 0.66 & 92.1 & $90.1 \pm 2.9$ & 7.93 & 3.00 & 0.432 & 108.3 & $102.4 \pm 1.6$ \\
    & 3  & 0.49 & 93.7 & $89.4 \pm 5.0$ & 2.78 & 0.66 & 0.429 & 107.8 & $102.3 \pm 2.8$ \\
    & 4  & 0.39 & 86.2 & $85.5 \pm 4.8$ & 3.41 & 0.88 & 0.425 & 104.3 & $100.2 \pm 3.4$ \\
    & 5  & 0.31 & 88.1 & $80.0 \pm 5.7$ & 7.19 & 1.66 & 0.401 & 103.3 & $97.6 \pm 2.5$ \\
    & 6  & 0.30 & 82.7 & $75.0 \pm 5.2$ & 9.23 & 3.15 & 0.418 & 96.6  & $95.3 \pm 1.3$ \\
    & 7  & 0.25 & 76.2 & $66.8 \pm 6.4$ & 7.48 & 2.29 & 0.412 & 92.8  & $92.4 \pm 3.4$ \\
    & 8  & 0.27 & 70.6 & $65.0 \pm 4.0$ & 7.14 & 1.89 & 0.406 & 87.9  & $86.2 \pm 3.1$ \\
    & 9  & 0.33 & 69.6 & $60.6 \pm 2.6$ & 7.58 & 2.02 & 0.405 & 81.3  & $76.0 \pm 6.0$ \\
    & 10 & 0.11 & 60.7 & $60.4 \pm 0.2$ & 6.14 & 7.72 & 0.523 & 56.4  & $55.4 \pm 0.4$ \\
  \end{tabular}
  \end{ruledtabular}
\end{table*}

\end{document}